\documentclass[11pt, draftclsnofoot, onecolumn]{IEEEtran}
\usepackage{microtype}
\usepackage{graphicx}
\usepackage{subfigure}
\usepackage{booktabs} % for professional tables
\usepackage{amsmath,amsfonts,amssymb,amsbsy,bm,paralist,theorem,ifthen,color}
\usepackage{algorithmic}
\usepackage{algorithm}% http://ctan.org/pkg/algorithms
\usepackage{hyperref}
\usepackage{multirow}
\usepackage[table]{xcolor}  
\newcommand\KK{\ensuremath{\kappa}}

\newcommand\Dc{\ensuremath{\mathcal{D}}}

\newcommand\Sc{\ensuremath{\mathcal{S}}}
\newcommand\Vc{\ensuremath{{\mathcal{V}}}}

\newcommand\Lc{\ensuremath{{\mathcal{L}}}}

\newcommand\Ac{\ensuremath{{\mathcal{A}}}}

\newcommand\Uc{\ensuremath{{\mathcal{U}}}}

\newcommand\xb{\ensuremath{{\bf x}}}

\newcommand\yb{\ensuremath{{\bf y}}}

\newcommand\ub{\ensuremath{{\bf u}}}

\newcommand\bb{\ensuremath{{\bf b}}}

\newcommand\cb{\ensuremath{{\bf c}}}
\newcommand\db{\ensuremath{{\bf d}}}

\newcommand\Ibb{\ensuremath{{\mathbb{I}}}}

\newcommand\vb{\ensuremath{{\bf v}}}

\newcommand\zb{\ensuremath{{\bf z}}}
\newcommand\mub{\ensuremath{{\bm \mu}}}

\newcommand\thetab{\ensuremath{{\bm \theta}}}

\newcommand\nub{\ensuremath{{\bm \nu}}}
\newcommand\lambdab{\ensuremath{{\bm \lambda}}}

\newcommand\zerob{\ensuremath{{\bm 0}}}

\newcommand\E{\ensuremath{{\mathbb{E}}}}

\newcommand\oneb{\ensuremath{{\bf 1}}}

\newcommand\Prob{\ensuremath{{\rm Prob}}}

\newcommand\Var{\ensuremath{{\rm Var}}}

\newcommand\Rbb{\ensuremath{{\mathbb{R}}}}

\newcommand{\wt}{\widetilde}

\newcommand{\ol}{\overline}
\newcommand{\ul}{\underline}

\newtheorem{Lemma}{Lemma}

\newtheorem{Theorem}{Theorem}

\newtheorem{Corollary}{Corollary}

\newtheorem{Rmk}{Remark}
\newtheorem{assumption}{Assumption}

\newcommand{\tabincell}[2]{\begin{tabular}{@{}#1@{}}#2\end{tabular}}

\graphicspath{{fig/}}

\ifCLASSOPTIONcompsoc
  \usepackage[nocompress]{cite}
\else
  \usepackage{cite}
\fi
\ifCLASSINFOpdf
\else
\fi
\allowdisplaybreaks[4]
\begin{document}
%\title{Federated Clustering via Matrix Factorization Models: From Model Averaging to Gradient Sharing}
%\title{Federated Matrix Factorization with Application to Data Clustering}
\title{Towards Fast Personalized Semi-Supervised Federated Learning in Edge Networks: Algorithm Design and Theoretical Guarantee}

\vspace{-0.3cm}
\author{Shuai~Wang, ~Yanqing~Xu,
	~Yanli~Yuan, % <-this % stops a space
	~and~Tony~Q.~S.~Quek% <-this % stops a space
	\IEEEcompsocitemizethanks{
		\IEEEcompsocthanksitem Shuai Wang and Yanli Yuan are with the Information Systems Technology and Design, Singapore University of Technology and Design, Singapore 487372 (e-mail: shuai\_wang@sutd.edu.sg,~yanliyuan@bit.edu.cn).
		\IEEEcompsocthanksitem Tony Q. S. Quek is with the Information Systems Technology and Design, Singapore University of Technology and Design,  Singapore 487372,  and also with the Department of Electronic Engineering, Kyung Hee University, Yongin 17104, South Korea (e-mail: tonyquek@sutd.edu.sg).
		\IEEEcompsocthanksitem   Yanqing Xu is with the Shenzhen Research Institute of Big Data and School of Science and Engineering, The Chinese University of Hong Kong, Shenzhen 518172, China (e-mail:  xuyanqing@cuhk.edu.cn).}
}
\maketitle

		\vspace{-1.3cm}
	\begin{abstract}
		Recent years have witnessed a huge demand for artificial intelligence and machine learning applications in wireless edge networks to assist individuals with real-time services. Owing to the practical setting and privacy preservation of federated learning (FL), it is a suitable and appealing distributed learning paradigm to deploy these applications at the network edge. Despite the many successful efforts made to apply FL to wireless edge networks, the adopted algorithms mostly follow the same spirit as FedAvg, thereby heavily suffering from the practical challenges of label deficiency and device heterogeneity. These challenges not only decelerate the model training in FL but also downgrade the application performance. In this paper, we focus on the algorithm design and address these challenges by investigating the personalized semi-supervised FL problem and proposing an effective algorithm, namely FedCPSL. In particular, the techniques of pseudo-labeling, and interpolation-based model personalization are judiciously combined to provide a new problem formulation for personalized semi-supervised FL. The proposed FedCPSL algorithm adopts novel strategies, including adaptive client variance reduction, local momentum, and normalized global aggregation, to combat the challenge of device heterogeneity and boost algorithm convergence. Moreover, the convergence property of FedCPSL is thoroughly analyzed and shows that FedCPSL is resilient to both statistical and system heterogeneity, obtaining a sublinear convergence rate. Experimental results on image classification tasks are also presented to demonstrate that the proposed approach outperforms its counterparts in terms of both convergence speed and application performance.
	\end{abstract}
%	\vspace{-3mm}
	\begin{IEEEkeywords}
		Wireless edge networks, heterogeneous edge devices, federated learning, semi-supervised learning, personalized federated learning
\end{IEEEkeywords}
%}


% is not selected <OR> if conference mode is selected - because compsoc
% conference papers position the abstract like regular (non-compsoc)
% \IEEEdisplaynontitleabstractindextext

% \fi
\IEEEpeerreviewmaketitle

%
%\ifCLASSOPTIONcompsoc
%\IEEEraisesectionheading{\section{Introduction}\label{sec:introduction}}
%\else
\vspace{-0.3cm}
\section{Introduction}
\label{sec:introduction}
The proliferation of distributed sensitive data at the network edge and the increasing computational capability of network edge devices have motivated the research on deploying artificial intelligence (AI) and machine learning (ML) applications in wireless edge networks to provide diverse customized and heterogeneous mobile services for customers\cite{Tariq2020WC}. This promising direction has attracted significant interest from researchers who pursue high-quality wireless AI services in a distributed edge network \cite{Tong2022WC,Letaief2019CM}. Such a network usually consists of a central server which needs to collaborate with a massive number of edge devices to train a ML model for application scenarios where privacy protection and resource utilization are critical. Motivated by this, federated learning (FL), as an emerging distributed learning paradigm, has become increasingly appealing in wireless edge networks in recent years as a result of its similar and practical configuration to edge computing, as well as its inherent nature for privacy preservation \cite{Nguyen2021CST,Khan2021CST}. In particular, FL runs under the orchestration of the central server without the need of knowing the raw
private data of each edge device.

Different from the conventional centralized learning, FL requires frequent model exchanges between the central server and edge devices via wireless links. Therefore, it is important to investigate the impacts of wireless aspects on the performance of the FL algorithms. In view of this, recent research on FL in wireless networks can be categorized into two main directions. The first direction focuses on the wireless resource allocation, such as bandwidth allocation, power control and computation resource allocation, during the FL process to improve the algorithm convergence \cite{Yang2020TCOM,ChenTWC2021,ZhuTWC2020,Elbir2020CL,Zheng2022TSP,Elbir2022TWC}. For example, \cite{Yang2020TCOM} investigated the edge device scheduling problem to improve the convergence rate of the FL algorithms under the constraint of the limited radio resources of each device. \cite{ChenTWC2021} studied the joint learning and radio resource allocation design to minimize the cost function. \cite{Elbir2020CL} investigated the hybrid beamforming design via FL, and \cite{Zheng2022TSP,Elbir2022TWC} proposed to use FL to estimate the wireless channel of massive MIMO systems. However, these works are generally based on simple FL frameworks. Among them, federated averaging (FedAvg) \cite{FedAvg_noniid_2019} is the most frequently adopted FL algorithm because of its simple-yet-good empirical performance. Roughly, it follows a computation and aggregation protocol where the central server updates the global model based on the local ML models received from and computed by edge devices, and then broadcast it to the devices. The process continues in a round-by-round fashion until convergence. The other important direction focuses on advanced FL algorithm designs \cite{shiTWC2022,millsTPDS2021,ZhaoTWC2022,Zhang2021IoT,Zhang2022IoT}. For example, 
\cite{shiTWC2022} proposed the dynamic batch sizes assisted FL (DBFL) algorithm, which has been shown to converge faster than the FedAvg algorithm. \cite{millsTPDS2021} proposed a novel multi-task personalized FL framework which also outperformed FedAvg.

Despite the successful progress of FL, it still faces many practical challenges. Especially when applied to wireless edge networks, there are two main challenges that cannot be well addressed by traditional simple FL algorithms, e.g., FedAvg. Let us elaborate on the challenges in detail and explain why the adopted FedAvg algorithm fails to handle them.
Firstly, the majority of data in edge devices is typically unlabeled, and each edge device only processes low-quality, inadequately labeled data. This realistic phenomenon arises from a severe shortage of experts to perform correct labeling, as well as the huge cost of labeling the massive data \cite{FedSSL} stored in these edge devices.
For instance, in a smart healthcare system, there is a scarcity in the amount of labeled data in each medical institute, and it is daunting for doctors to label them manually. However, FedAvg and its variants are incapable of addressing this issue because they assume that all data in edge devices are properly labeled. It is acknowledged that these algorithms perform poorly with merely a small amount of labeled data. This naturally necessitates the urgent demand for algorithmic development for FL with a deficiency of labeled data, namely semi-supervised FL (SSFL), where both labeled and unlabeled data are exploited for model training \cite{Zhang2021IoT,Zhang2022IoT}.

The second challenge is the presence of massive heterogeneous edge devices. The high degree of device heterogeneity decelerates the model training in FL and downgrades the application performance that the FL algorithms can finally reach. In particular, these heterogeneous edge devices have both \emph{statistical} and \emph{system heterogeneity}. 
\begin{itemize}
	\item[\emph{Statistical heterogenity}:] In wireless edge networks, the local datasets owned by edge devices are usually unbalanced and non-i.i.d. (non-independent and identically distributed) across these devices \cite{Ma2021JSAC,Zhao2022TWC}. It has been shown that FedAvg is inefficient in dealing with the non-i.i.d. data \cite{SCAFFOLD_2020}. The presence of unbalanced and non-i.i.d. data could corrupt the algorithm convergence speed and even lead to model divergence \cite{FedMAJ_2020}. On the other hand, as the conventional FL requires all edge devices to agree on a global model, the resulting global model could perform arbitrarily poorly once applied to the local dataset of each device because of the non-i.i.d. data \cite{APFL_2020}. 
	A large body of literature has overcomed this issue by adopting the technique of model personalization which aims to achieve a device-specific or personalized model to fit the dataset of each device for good generalization performance \cite{PerFedAvg_2020,FedAMP_2021,APFL_2020}. This motivates the need for personalized FL in wireless edge networks.
	\item[\emph{System heterogeneity}:]  In addition to non-i.i.d. data, edge devices may have varied system constraints including different hardwares (CPU, GPU,  and memory size), network configurations (wireless channels, bandwidth, and transmission power), battery states, etc. Indeed, these heterogeneous characteristics can cause unpredictable behaviors of edge devices as well as the straggler problem in FL \cite{FedToE2022,AsyncFedOpt_2020,LocalSGD_2019,FLANP_2022}.
	Specifically, different computational capabilities and the unavailability of edge devices can adversely impact the algorithm convergence. Besides, the delay caused by the straggler (the slowest edge device) can significantly slow down the model training because FL requires waiting for all edge devices to communicate with the server in each round.
\end{itemize}

The statistical heterogeneity and system heterogeneity can be widely seen in practical wireless edge communication systems. Considering a distributed multiple-input multiple-output (MIMO) system at the network edge, there are many distributed access points (APs) to serve multiple users. Recently, the FL algorithms have been applied to such network for signal processing (SP) tasks, e.g, channel estimation (CE) and beamforming, to overcome the disadvantages of high communication and computation costs of tradition SP algorithms \cite{ Zheng2022TSP,Elbir2022TWC}. Take the uplink CE problem for example, each AP is viewed as an edge device and the dataset of each edge device is the received pilot signals from its associated users. There is a central sever who coordinates the CE of all APs to approach the centralized performance based on the FL algorithms. However, applying these algorithms to such a system is faced with the aforementioned challenges. For example, i) since each AP may serve different users, the datasets (received pilot signals) of the APs could be unbalanced and non-i.i.d.; ii) the APs may be equipped with different system hardwares, and thus have varied computational capabilities; iii) the wireless channels between the APs to the central sever are dramatically different due to the small-scale fading and shadowing.

It is worth-noting that the interplay of these two challenges could exacerbate their respective negative effects. For example, the statistical heterogeneity among edge devices increases in the presence of unlabeled data since the unlabeled data in each device may contain most of the same patterns as those of other devices \cite{FedSSL}. As a result, using merely labeled data for training often leads to not only algorithm inefficiency but also a severe performance degradation of FL algorithms. Due to the limitations of FedAvg in addressing the two main challenges, our work focuses on designing an advanced FL algorithm to overcome these challenges in wireless edge networks.

\vspace{-0.2cm}
\subsection{Related works}

We notice that, while many successful efforts have been made to address the aforementioned challenges in wireless edge networks, most of them merely take into account one challenge and might not well address it. Few have simultaneously considered these two issues and offered effective solutions. 

Recent FL works \cite{lin2021semifed, FlexMatch_2021,FedMatch_2021, FedSSL} attempted to address the challenge of label deficiency by studying the SSFL problem. They mostly relied on the technique of pseudo-labeling and performance-promoting regularization, where the unlabeled data samples are respectively assigned artificial labels, and both labeled and unlabeled data are then considered together with specific regularizations to improve the training performance. The two works \cite{FedSemL2022, SSFL_TMI2022} respectively applied similar strategies to the applications of human activity recognition and travel model identification while accounting for the resource constraints in wireless edge networks. Nevertheless, these works adopted or followed the same spirit as FedAvg for model training, and thus, as mentioned previously, suffer from the adverse effect of device heterogeneity.

%However, they either address the edge device drift problem caused by non-i.i.d data or allow heterogeneous local updates to account for computational diversity among edge devices. Few of them have simultaneously considered both of these two factors and offer effective solutions. 

%The applications of federated learning to wireless edge network has attracted great attentions recently and many successful progresses has been made \cite{Yang2020TCOM,ChenTWC2021,ZhuTWC2020,Zheng2022TSP,Elbir2022TWC}. For example,  \cite{Yang2020TCOM} investigated the edge device scheduling problem to improve the convergence rate of federated learning algorithm under the constraint of limited radio resource of each edge device. \cite{ChenTWC2021} studied the joint learning and radio resource allocation design to minimize the loss function of the learning algorithm. \cite{Elbir2020CL} investigated the hybrid beamforming design via federated learning and \cite{Zheng2022TSP,Elbir2022TWC} proposed to use federated learning to estimate the wireless channel of massive multiple-input-multiple-output systems. 

%However, the aforementioned works all assumed that the data are well labeled. In view of this, federated semi-supervised learning method was proposed in \cite{lee2013pseudo,lin2021semifed} which used the pseudo labeling technique, by assigning a psedo label to the unlabeled data, to improve the performance of trained model. 

To address the challenge of device heterogeneity, researchers have proposed methods such as FedProx\cite{FedProx_2018}, SCAFFOLD \cite{SCAFFOLD_2020}, and FedDyn \cite{FedDyn_2021}, which penalized model updates with appropriate regularizations in order to stabilize the algorithm convergence. However, they either fail to address the challenge or generalize badly for many edge devices' local datasets because of the lack of personalization. To deal with the personalized FL problem, recent works proposed algorithms such as PerFedAvg \cite{PerFedAvg_2020}, FedAMP\cite{FedAMP_2021}, Ditto\cite{Ditto_2021}, pFedMe\cite{pFedMe_2020} and APFL\cite{APFL_2020}. In particular, the authors of APFL \cite{APFL_2020} adopted model interpolation to provide a device-specific model for each edge device through a convex mixture of a local model and the global model. On the other hand, it is noticed that, instead of model personalization, some works \cite{DRFL_2020, DRFLM_2022, FedACL_2019, CE-DRDL_2022} mitigated this issue with the distributionally robust FL (DRFL).  Unlike the former, DRFL aims to pursue a global model which performs well over the worst-case combination of all local data distributions, and thus its generalization performance will be resilient to the non-i.i.d. data. However, the preceding works all relied solely on labeled data for model training, thereby performing poorly in practical scenarios where label deficiency is prevalent.

%Personalized federated learning method can combat the impacts from edge devices heterogeneity and has been studied recently. In particular, \cite{kairouz2021advances} classified the related method to three main categories, i.e., the meta-learning based approach \cite{jiang2019improving,khodak2019adaptive}, model regularization approach \cite{t2020personalized,shen2020federated} and model interpolation approach \cite{mansour2020three}. Specifically, an online convex optimization based meta-learningapproach is proposed in \cite{khodak2019adaptive}, which can be used in federated learning for better personalization.In \cite{t2020personalized}, personalized learning is realized by using the regularization between local and global mode. While \cite{shen2020federated} propose a knowledgedistillation method to achieve personalization, where the regularization on the predictions between local model and global model is applied. 
%For the model interpolation approach, the local model and global model are trained and then a convex combination the two models is used as the local interference model \cite{mansour2020three}. 

More recently, a few works \cite{bettini2021personalized,Yu2021TMC,tashakori2022semipfl} have emerged to investigate the problem of personalized SSFL. Specifically, the works\cite{bettini2021personalized}\cite{Yu2021TMC} focused on the application of human activity recognition and utilized active learning and label propagation for data labeling and transfer learning for personalization. The authors in \cite{tashakori2022semipfl} proposed to generate a personalized autoencoder using a hyper-network over the labeled and unlabeled data for multi-sensory classification. Unfortunately, these works still did not address the challenge of device heterogeneity as the adopted algorithms again are similar to FedAvg and thus affected by heterogeneous edge devices. Besides, none of these works theoretically analyzed the convergence of their proposed algorithms.

\vspace{-0.2cm}
\subsection{Contribution}

In this paper, we study the personalized semi-supervised FL (PSSFL) problem in wireless edge networks, aiming at proposing an efficient algorithm with both high-quality application performance and solid theoretical guarantee. In particular, we propose an algorithm named FedCPSL, which leverages novel strategies to address the aforementioned two challenges and outperforms its existing counterparts in terms of both convergence speed and application performance. The contributions can be expounded from the following aspects:

{\bf 1) Problem formulation:} We propose a general problem formulation for the PSSFL problem in wireless edge networks which combines the techniques of pseudo-labeling, regularization, and model interpolation to model the PSSFL problem in a novel way. Unlike APFL\cite{APFL_2020}, we introduce a new localized model for each device and consider the mixture of the outputs of the localized model and the global model as that of the personalized model, which enjoys better theoretical justification. To the best of our knowledge, this is the first work considering both the interpolation-based model personalization and semi-supervised FL. The resulting problem formulation is novel for modeling the PSSFL problem. 

{\bf 2) Algorithm design:} We propose the FedCPSL algorithm, which ensures fast and simultaneous algorithm convergence with respect to both the global and localized model parameters by addressing the challenge of device heterogeneity. Specifically, the global and localized model parameters are simultaneously trained. FedCPSL adopts the strategy of adaptive client variance reduction to combat the statistical heterogeneity among edge devices. For system heterogeneity, it allows edge devices to perform heterogeneous local updates across rounds based on their system constraints and leverages the strategy of global normalized aggregation to alleviate the resulting solution bias. Besides, the momentum approach is introduced into the updates of both the global and localized model parameters to further boost the algorithm convergence.

{\bf 3) Convergence analysis :} We establish the convergence property of the FedCPSL algorithm. In contrast to the existing analysis of FL algorithms with one block of variables, we justify the convergence of both the global and localized model parameters in the presence of the above novel strategies. In particular, under mild assumptions, they all have a sublinear convergence rate for nonconvex objectives, which matches the state-of-the-art lower bounds. More importantly, the theoretical analysis shows that the convergences of both the global and localized model parameters are resilient to the challenge of device heterogeneity, which causes the speedup of FedCPSL over its counterparts. Such convergence results for the PSSFL problem are new and have never appeared in the literature.

{\bf 4) Experiments:} We evaluate the performance of the proposed FedCPSL algorithm by applying it to the image classification tasks on the MNIST and CIFAR-10 datasets. The experimental results demonstrate that FedCPSL outperforms benchmark algorithms under different experimental settings. It not only enjoys a faster convergence but also achieves a better application performance.

%We conduct extensive simulations on the MNIST and CIFAR-10 data set to examine the FedPCSL algorithm under different settings. The experiments demonstrate that FedPCSL  can improve the generalization capability of the local model for unlablled data. Besides, compared with benchmark algorithms, the  FedPCSL has a faster convergence rate.

{\textbf{Synopsis:}} Section \ref{sec: problem formulation} introduces the proposed system model and problem formulation of PSSFL  in wireless edge networks. Section \ref{sec: FedPCSL algorithm} presents the proposed FedCPSL algorithm to solve the formulated problem, and the associated theoretical analysis is given in Section \ref{sec: convergence analysis}. The experiment results are presented in Section \ref{sec: simulation}. Section \ref{sec: conclusion} concludes this paper.

{\section{Problem formulation } \label{sec: problem formulation}
Consider a wireless network where a central server coordinates $N$ edge devices to learn a ML model. Without loss of generality, we focus on the classic data classification task, and denote $\Dc_i$ as the local dataset that collects the data generated or sensed by edge device $i$. In this work, we consider a semi-supervised setting where the local dataset $\Dc_i$ consists of a set of labeled data $\Lc_i$ and a set of unlabeled data $\Uc_i$, i.e.,
\begin{align}
	\Dc_i = \Lc_i  \cup \Uc_i, ~\Lc_i = \{(\xb_{i, k}, \yb_{i, k})\}_{k = 1}^{n_i},~\Uc_i = \{\ub_{i, k}\}_{k=1}^{m_i},
\end{align}
where $(\xb_{i, k}, \yb_{i, k}) \in \Rbb^{S} \times \Rbb^C$ is the $k$-th input and output pair in $\Lc_i$ with $S$ denoting the dimension of the input data and $C$ signifying the number of data classes;
$\ub_{i, k}$ is the $k$-th input of $\Uc_i$,
and $n_i$ and $m_i$ are the sizes of the labeled and unlabeled data at edge device $i$, respectively.
Since the generated or sensed data by edge devices are biased to the applications and environment, the datasets have the following characteristics: i) the local datasets, $\Dc_i, i \in [N]$, could be non-i.i.d. and their sizes could be unbalanced; ii) the datasets may follow different data distributions; iii) the number of labeled data could be much smaller than the unlabeled data, i.e., $n_i < m_i$. These edge devices  $i \in [N]$ may have varied system constraints including different hardwares (CPU, GPU, and memory size), network configurations (wireless channels, bandwidth, and transmission power), battery states, etc.

Our objective is to learn an accurate and personalized ML model for each edge device by utilizing their local datasets $\Dc_i$. To this end, one needs to address the two challenges, i.e., the label deficiency and device heterogeneity. Since the traditional FL algorithms, e.g., FedAvg, cannot deal with these two challenges well, we are interested in developing an advanced FL algorithm by considering the challenges into the problem formulation and the algorithm design, respectively. Specifically, to handle the challenge of label deficiency, we consider to use semi-supervised learning and interpolation-based model personalization to formulate the learning problem. Meanwhile, to deal with the adverse impacts of device heterogeneity on the algorithm convergence, a sophisticated and effective FL algorithm is developed (see Sec. \ref{sec: FedPCSL algorithm}).

To fully exploit the unlabeled data $\Uc_i, \forall i \in [N]$, the pseudo-labeling strategy is applied. In this strategy, each unlabeled data sample is initially assigned an artificial label, called pseudo label. These pseudo labels of unlabeled data are then optimized and used during the training process to improve the application performance of the ML model \cite{lin2021semifed}.
Specifically, let $h(\thetab; \cdot)$ represent the ML model with model parameter $\thetab$, and introduce $\nub_i =[\nub_{i, 1}, \nub_{i, 2}, \ldots, \nub_{i, m_i}]$ with $\nub_{i, k} \in \Rbb^C$ being a probability vector as the pseudo labels for the unlabeled data $\Uc_i$. 
Then, we define 
\begin{align}
	\xb_i =[\xb_{i, 1}, \xb_{i, 2},\ldots, \xb_{i, n_i}], 	\yb_i =[\yb_{i, 1}, \yb_{i, 2}, \ldots, \yb_{i, n_i}],
	\ub_i = [\ub_{i, 1}, \ub_{i, 2},\ldots, \ub_{i, m_i}],
\end{align}
and consider to train a global model $\thetab$ by minimizing a joint cost function as follows.
\begin{align}
	\min_{\thetab, \nub_{i} } \sum_{i=1}^{N}\omega_i f_i(\thetab, \nub_{i}), ~{\rm s.t.} ~\nub_{i} \in \Vc_i,
\end{align}where $\Vc_i = \{\nub_{i} |\nub_{i, k} \geq 0, \oneb^\top \nub_{i, k} = 1, \forall k \in [m_i]\}$ collects all the possible pseudo labels, and 
\begin{align} \label{def: f_i}
	f_i(\thetab, \nub_{i}) \triangleq &l(h(\thetab; \xb_i), \yb_i) + \alpha_p l(h(\thetab; \ub_i), \nub_i) + \alpha_r R(h(\thetab; \ub_i), \nub_{i})
\end{align} is the local cost function for each edge device $i$; $\omega_i$ is the weight associated with edge device $i$ satisfying $\sum_{i=1}^{N}\omega_i = 1$; for example, $\omega_i = \frac{n_i + m_i}{n + m}$ with $n$ and $m$ representing the total number of labeled and unlabeled data of all edge devices, respectively; $l$ is the loss function for the ML model $h$ w.r.t. a set of data samples.
%{\blue $R(\cdot)$ is the regularizer to control the pseudo labels}.
One can see that the local cost $ f_i(\thetab, \nub_{i})$ penalizes the output of the model  $h(\thetab, \cdot)$ for being dissimilar to the target labels with respect to both the labeled and unlabeled datasets.
The model $h(\thetab, :)$ could be any ML classifier, such as decision tree, SVM, and deep neural network, which accepts data inputs and then respectively outputs their classes or class probability distributions. The regularizer, $R(h(\thetab; \ub_i), \nub_{i})$, on the model parameter and pseudo labels can be well-defined to avoid incorrect estimation of pseudo labels and thus benefits the learning process from unlabeled data.

As mentioned previously, model interpolation is an effective way to combat the negative effects caused by non-i.i.d. data by convexly combining the global model and local model \cite{APFL_2020}. This motivates us to design a personalized model under the semi-supervised setting for each edge device with the help of a localized model and the global model.
Specifically, we assume that, for any input $\xb$, the output of the personalized model $h(\thetab_{i, p}; \cdot)$ can be represented as a convex combination of the outputs of a localized model $h(\thetab_{i, lc}; \cdot)$ and the global model $h(\thetab; \cdot)$, i.e.,
\begin{align} \label{eqn: mixture relation}
	h(\thetab_{i, p}; \xb) = \beta_ih(\thetab_{i, lc}; \xb) + (1-\beta_i)h(\thetab; \xb), \forall \xb,
\end{align}where $\beta_i \in [0, 1]$ is a predefined mixture coefficient. Then, the problem of personalized semi-supervised FL is to find a localized model which makes \eqref{eqn: mixture relation} be the minimizer of the local cost function. To be concrete, one can write the problem as
\begin{subequations} \label{eqn: PSFL}
	\begin{align}
		\min_{\substack{\thetab_{i, lc}, i \in [N]}} &~ \sum_{i=1}^{N} \omega_i  F_i(\thetab_{i, lc}, \thetab, \nub_i), \label{eqn: PSFL_obj}\\
		{\rm s.t.} 
		&~(\thetab, \{\nub_i\}_{i = 1}^{N}) = \arg\min_{\thetab^\prime, \nub_{i}^\prime \in \Vc_i} \sum_{i=1}^{N}\omega_i f_i(\thetab^\prime, \nub_{i}^\prime),\label{eqn: PSFL_c1}
	\end{align}where
	\begin{align}\label{eqn: local loss}
		F_i(\thetab_i^{lc}, \thetab, \nub_i) \triangleq 
		& l( \beta_ih(\thetab_{i, lc}) + (1-\beta_i)h(\thetab), \yb_i) + \alpha_p l(\beta_ih(\thetab_{i, lc}) + (1-\beta_i)h(\thetab), \nub_i) \notag \\
		&+ \alpha_r R(\beta_ih(\thetab_{i, lc}) + (1-\beta_i)h(\thetab), \nub_{i}),
	\end{align} and $f_i(\cdot, \cdot)$ is defined in \eqref{def: f_i}. Here we use $h(\thetab)$ to represent $h(\thetab; \cdot)$ for ease of presentation. It is worth-noting that, in contrast to the traditional model interpolation, we construct the personalized model by taking a convex combination of the global model and the localized model in \eqref{eqn: local loss}, rather than combining their parameters. This particularly aligns with the theoretical justification of the model interpolation technique in \cite[Theorem 1 and Corollary 3]{APFL_2020}.
\end{subequations}

One notable feature of the problem formulation \eqref{eqn: PSFL} is that the optimization process for the localized models can be completely decoupled from the training of the global model. This is because problem  \eqref{eqn: PSFL} has a form of two-phase optimization, with global model training and model personalization as two key ingredients. To be specific, problem \eqref{eqn: PSFL} can be solved using a two-phase algorithm where in the first phase, the global model parameter $\thetab$ and pseudo labels $\{\nub_i\}_{i = 1}^{N}$ are trained by solving the subproblem \eqref{eqn: PSFL_obj}, and then, with $\thetab$ and $\{\nub_i\}_{i = 1}^{N}$ as inputs, each edge device $i$ solves \eqref{eqn: PSFL_obj} in the second phase to obtain a localized model parameter $\thetab_{i,lc}$. In the first phase, since problem \eqref{eqn: PSFL_obj} is a SSFL problem, it is natural to follow the model averaging (MA) strategy \cite{FedMAJ_2020} adopted in the FL algorithms to solve problem \eqref{eqn: PSFL_obj} \cite{Shuai_FedMA_2021}. The second phase is responsible for model personalization, and lastly, the output of a new input data $\xb$ of edge device $i$ is generated by following \eqref{eqn: mixture relation}. 
	
%where the server collaborates with the $N$ edge devices to learn a shared model and pseudo labels by following the model averaging strategy used in FedAvg. 
	
%	Naively, problem  \eqref{eqn: PSFL} can be solved by a two-phase algorithm. In particular, in the first phase, \eqref{eqn: PSFL_c1} is solved by the collaboration between the server and all edge devices; while, in the second phase, each edge device $i$ minimizes $F_i$ in \eqref{eqn: PSFL_obj} to obtain the localized model. 
	
Nevertheless, problem \eqref{eqn: PSFL} is much more challenging to solve than that in the literature \cite{CE_DDNN_2017,FedProx_2018,EDDL_DModelAvg_2018, Parallel_RSGD_2019,FedAvg_noniid_2019} as it combines the tasks of global model training, pseudo label prediction, and model personalization into a unified formulation which is possibly non-convex and non-smooth, and involves three different blocks of optimization variables. The aforementioned two-phase optimization algorithm is not preferable for solving problem \eqref{eqn: PSFL}  in wireless edge networks because of its low efficiency, the lack of considering the statistical and system heterogeneity, and its incapability of handle them. For example, to achieve desirable performance, the global model and the pseudo labels should be optimized for high accuracy and then fed back to edge devices for the training of the localized models. This process demands a significant amount of training time.
Therefore, it needs careful considerations on how to effectively design efficient and high-performance algorithms for problem \eqref{eqn: PSFL}.

%\begin{Rmk} \rm
%	{It is worthy to pointing out that, although the cost functions in \eqref{eqn: PSFL_obj} and \eqref{eqn: PSFL_c1} take summation over all edge clients, it doesn't mean that all the clients should be participated in the model training in each communication round. Actually, in the next section, we will proposed a novel FL algorithm to solve problem \eqref{eqn: PSFL}, which permits only part of the edge devices to communicate with the central sever. This scheme not only overcomes the challenge of bandwidth limitation but also allows flexible clients selections in case that the clients who are experiencing deep channel fading are selected  for model training.}
%\end{Rmk}
%

\begin{Rmk}\rm
	It is indicated in \cite[Theorem 1 and Corollary 3]{APFL_2020} that, by carefully choosing the mixture coefficient $\beta_i, \forall i \in [N]$, the joint prediction model in \eqref{eqn: mixture relation} with an optimal solution $(\thetab_{i,lc}^\star, \thetab^\star)$ to problem \eqref{eqn: PSFL} generalizes well on the associated local data distribution. Specifically, the value of $\beta_i$ controls the contribution of the global model to model personalization. The setting of $\beta_i  = 1$ suggests no FL while $\beta_i = 0$ implies no model personalization. Thus, the choice of $\beta_i$ for edge device $i$ should match the dissimilarity (the degree of non-i.i.d. data) between its own data distribution and the average of all edge devices' data distributions. When the dissimilarity is large, a large $\beta_i$ is preferred to incorporate more of the localized model into the personalized model of edge device $i$. On the contrary, a small $\beta_i$ is suggested such that the personalized model of edge device $i$ benefits more from the global model.
\end{Rmk}

\section{Proposed FedCPSL Algorithm} \label{sec: FedPCSL algorithm}

In this section, we develop an efficient FL algorithm for problem \eqref{eqn: PSFL}, termed FedCPSL. The proposed FL algorithm allows simultaneous training of all the optimization variables to spead up the learning process, and overcomes the challenge of device heterogeneity in wireless edge networks with advanced algorithm strategies. Furthermore, an intuitive explanation on the advantage of FedCPSL in dealing with the challenge is presented.

\subsection{Algorithm design}

Instead of using the two-phase optimization, we aim to optimize the global model and the localized models simultaneously, such that one can obtain desirable personalized models when an accurate global model is reached. Notice that the decoupling structure of \eqref{eqn: PSFL} makes it possible to update the global model and pseudo labels similarly to the existing FL algorithms while parallelly optimizing the localized models with low complexity. In particular, since problem \eqref{eqn: PSFL} involves three kinds of variables (model parameters), i.e., $\thetab$, $\nub_{i}$, and $\thetab_{i, lc}$, we consider leveraging the strategies of alternating minimization (AM) and MA in FL to sequentially update these model parameters round by round until they converge.

As per the above idea, one can design an iterative algorithm as follows. For round $r=1,2,\ldots$, the server randomly samples a small, fixed-size subset of edge devices (denoted by $\Ac^r$ with size $|\Ac^r| = m \ll N$) and broadcasts the global model parameter $\thetab^r$ to them. Each of the selected edge devices $i \in \Ac^r$ sequentially obtains approximate solutions to the following subproblems of \eqref{eqn: PSFL}:
\begin{subequations}
	\begin{align}
		\nub_i^{r+1}&=\arg\min_{\substack{\nub_i \in \Vc_i}} ~f_i(\thetab^{r}, \nub_i), \label{eqn: prob_pseudo_labels}\\
		\thetab_i^{r+1}&=\arg\min_{\thetab_i} ~f_i(\thetab_i, \nub_i^{r+1}), \label{eqn: prob_local_copy}\\
		\thetab_{i, lc}^{r+1} &= \arg\min_{\thetab_{i,lc}} ~F_i(\thetab_{i, lc}, \thetab_i^{r+1},\nub_i^{r+1}).\label{eqn: prob_localized_model}
	\end{align}
\end{subequations}
Then, the server collects and takes certain average of $\thetab_i^{r+1}, \forall i \in \Ac^r$ to produce $\thetab^{r+1}$ for the next round of updates. In practice, \eqref{eqn: prob_pseudo_labels} may have a closed-form solution; otherwise, it will be approximated by one-step gradient descent (GD). Besides, it is sufficient to employ the local SGD \cite{CE_DDNN_2017} to approximate the solutions of \eqref{eqn: prob_local_copy} and \eqref{eqn: prob_localized_model}. In particular, each edge device $i \in \Ac^r$ performs $Q_i^r \geq 1$ consecutive steps of SGD with respect to $\thetab_{i}$ and $\thetab_{i, lc}$, i.e., for $t=0, \ldots, Q_i^r-1$,\begin{subequations}
	\begin{align}
		&\thetab_i^{r, t+1}= \thetab_i^{r, t} - \eta g_i(\thetab_i^{r,t}, \nub_i^{r+1}), \label{eqn: update_theta_i} \\
		& \thetab_{i, lc}^{r, t+1} = \thetab_{i, lc}^{r,t} -\eta_{c} G_i(\thetab_{i, lc}^{r,t}, \thetab_i^{r, t}, \nub_{i}^{r+1}), \label{eqn: update_theta_loc}
	\end{align}
\end{subequations}where  $\eta >0$ and $\eta_{c} >0$ are two step sizes; $g_i(\thetab_i, \nub_i)$ denote the SGD of $f_i(\cdot, \cdot)$ over a mini-batch $\Sc = B_l \cup B_u$ which consists $S_l$ samples i.i.d. drawn from $\Lc_i$ and $S_u$ samples i.i.d. drawn from $\Uc_i$; $G_i(\cdot, \nub_i^{r+1})$ is the SGD of $F_i$ over $\Sc$;
\begin{align}
	&\thetab_i^{r, 0}=\thetab^{r},  \thetab_{i, lc}^{r, 0}=  \thetab_{i, lc}^{r - 1, Q_i^r - 1};\label{eqn: update_theta_init}
\end{align}
Then, $\thetab_i^{r+1}$ and $\thetab_{i, lc}^{r+1}$ are set as $\thetab_i^{r+1}=\thetab_i^{r, Q_i^r},  \thetab_{i, lc}^{r+1} =  \thetab_{i, lc}^{r, Q_i^r}$,$ \forall i \in \Ac^r$.

In contrast to \cite{FedSSL}, we leverage the scheme of partial client participation (PCP) in which only a small set of edge devices are sampled to perform the local update and communicate with the server in each round. It is practically preferred because federated training of a ML model demands huge communication resources and reliable connections, which are not affordable for wireless edge networks where bandwidth is limited and communication links are usually unreliable. In particular, radio resource limited wireless edge networks cannot support the participation of all edge devices in every communication round, which leads to network congestion if full participation is required. The scheme of PCP can greatly alleviate such problem by adjusting the size of participants according to the available bandwith. Besides, the poor link quality in wireless edge networks may raise the problem of transmission outage \cite{QFL_Outage_2022} because part of edge devices experience deep channel fading and thus unsuccessful transmission of the model parameters to the server. PCP also can mitigate this issue through modeling the randomly offline edge devices.  It is worth-noting that, the algorithm convergence could be further accelerated by PCP if advanced device selection and frequency bandwidth allocation schemes are incorporated \cite{Yang2020TCOM,ChenTWC2021,ZhuTWC2020}.

On the other hand, as mentioned, the edge devices in wireless edge networks may have heterogeneous system constraints including computational capabilities and network configurations. The heterogeneity in the computational speeds results in large variations in the number of local update steps performed by the participating devices. This would cause large model training delay if the participating devices are restricted to perform the same amount of local computations in each round \cite{FedProx_2018}. To resolve this issue, we allow time-varying heterogeneous
local updates (HLU) in which each active edge device $i$ performs $Q_i^r$ local update steps at round $r$. It should be remarked that the scheme of time-varying HLU across rounds is highly desired in wireless edge computing scenarios where low-latency ML services are demanded. The benefit of such scheme lies in that each device can flexibility decide the amount of its local update steps according to its current system state in each round to reduce the training delay. For example, at one round, the mobile devices with full battery and good wireless connections can contribute more to the training process than those in poor system conditions.

Nevertheless, the above update rules \eqref{eqn: update_theta_i}\eqref{eqn: update_theta_loc}\eqref{eqn: update_theta_init} have the following disadvantages: 1) The scheme of local SGD usually leads to relatively slow algorithm convergence; 2) the non-i.i.d. data can cause the client drift problem\cite{SCAFFOLD_2020}; 3) using time-varying HLU can result in the issue of objective inconsistency \cite{FedNova_2020}.
To address these issues and boost the efficiency, three specific strategies are introduced in the proposed algorithm to deal with the above three issues, respectively. The details of the adopted strategies are stated as follows:
\begin{itemize}
	\item {\bf Accelerated local SGD with momentum}.  
	To speed up the algorithm convergence, the momentum technique is applied to the local updates \eqref{eqn: update_theta_i} and \eqref{eqn: update_theta_loc}. Specifically, in the momentum-based SGD, two local momentum buffers are introduced for each edge device and initialized as zero at the beginning of each round. In each round, the local momentum buffers dynamically accumulate the historical gradients in the current round, and are then used for updating \eqref{eqn: update_theta_i} and \eqref{eqn: update_theta_loc} respectively. By taking the gradients in all the previous steps into consideration, the momentum-based SGD can speed up the algorithm convergence, which is verified by the numerical results in Section \ref{sec: simulation}.
	\item {\bf Adaptive client variance reduction}. As the non-i.i.d. data causes high gradient variance among edge devices, which leads to the client drift problem and slows down the convergence of the global model parameter $\thetab$, the technique of client variance reduction is desired. In parallel, the introduction of momentum and  HLU presents new challenges that traditional variance reduction schemes cannot address. To remedy this issue, we adopt an approach which follows similar spirits as client variance reduction but is adapted to the momentum-based local SGD and  HLU. In particular, two kinds of control variables (one for the server and the other one for edge devices) are introduced to correct the local update directions. Unlike traditional client variance reduction schemes \cite{SCAFFOLD_2020}, we update these control variables in a novel way that not only retains the capability of client variance reduction but also accommodates the challenges brought by these two schemes. Therefore, it can benefit the stable and fast convergence of the proposed algorithm; See more details in Section \ref{sec: algorithm implement}. 
	\item {\bf Normalized global aggregation}. Using time-varying HLU may result in the issue of objective inconsistency, which downgrades the application performance \cite{FedNova_2020}. Following the same spirit as FedNova in \cite{FedNova_2020}, we employ the strategy of normalized global aggregation to account for (time-varying) HLU. However, unlike FedNova, we utilize a flexible step-size $\eta_g$ for global aggregation, which not only benefits the theoretical analysis but also leads to faster algorithm convergence \cite{SCAFFOLD_2020}.
\end{itemize}

\subsection{Algorithm implementation and explanation}
\label{sec: algorithm implement}
\begin{algorithm}[t!] 
	\caption{Proposed FedCPSL Algorithm} %: Federated clustering by model averaging}
	\label{alg: FedPCSL}{
		\begin{algorithmic} \footnotesize 
			\STATE {\bfseries Input:} initial values of $\thetab_i^0 = \thetab$, $\cb^0 = \cb_i^0 = 0, \forall i \in [N], \eta > 0, \eta_v > 0$.
			\FOR{round $r=0$ {\bfseries to} $R - 1$}
			\STATE {\bfseries \underline{Server side:}} select a set of edge devices $\Ac^r$ (with size $|\Ac^r| = m$) by uniform sampling without replacement, and broadcast 
			$\thetab^{r}$ and $\cb^r$ to all edge devices in $\Ac^r$.
			\STATE {\bfseries \underline{edge device side:}}
			\FOR{edge device $i=1$ {\bfseries to} $N$ in parallel} 
			%	\FOR{edge device $p = 1$ {\bfseries to} $P$ in parallel}  
			\IF{edge device $i \notin \Ac^r$} \STATE Set $\thetab_i^{r+1}=\thetab^{r}, \thetab_{i, lc}^{r+1}= \thetab_{i, lc}^{r}$.
			\ELSE 
			\STATE Compute $\nub_i^{r+1}$ via \eqref{eqn: prob_pseudo_labels} or \vspace{-0.3cm}
			\begin{align}
				\nub_{i}^{r+1} = \nub_{i}^r - \eta_v \nabla_{\nub_i} f_i(\thetab^r, \nub_{i}^r). \label{alg: update_nu}
			\end{align} \vspace{-0.6cm}
			\STATE Set $\thetab_i^{r, 0}=\thetab^{r },  \thetab_{i, loc}^{r, 0}= \thetab_{i, loc}^{r-1, Q_i^{r-1}}, \mub_i^{r, 0} = \zerob, \mub_{i, lc}^{r, 0} = \zerob$.
			\FOR{ $t = 0$ {\bfseries to} $Q_i^r - 1$}
			\STATE \vspace{-0.8cm} \begin{align}
					&\mub_i^{r, t+1} = \gamma \mub_i^{r, t} + g_i(\thetab_i^{r,t}, \nub_i^{r+1}) - \cb_i^r + \cb^r,  \label{alg: update_momentum}\\
					&\thetab_i^{r, t+1}= \thetab_i^{r, t} - \eta \mub_i^{r, t+1}, \label{alg: update_theta_i}\\
					& \mub_{i, lc}^{r, t+1} = \gamma \mub_{i, lc}^{r, t} + G_i(\thetab_{i, lc}^{r,t}, \thetab_i^{r, t}, \nub_{i}^{r+1}),  \label{alg: update_momentum_lc}\\
					& \thetab_{i, lc}^{r, t+1} =  \thetab_{i, lc}^{r,t} -\eta_{c} \mub_{i, lc}^{r, t+1}. \label{alg: update_theta_loc}
				\end{align}
%			\end{subequations}
			\vspace{-0.6cm}
			\ENDFOR
			\STATE Set $\thetab_i^{r+1}=\thetab_i^{r, Q_i^r}, \thetab_{i, lc}^{r+1} = \thetab_{i, lc}^{r, Q_i^r}$, and compute %\vspace{-0.2cm}
			\begin{align}
				&\wt Q_{i}^r = \frac{Q_i^r}{1-\gamma} -\frac{\gamma(1-\gamma^{Q_i^r})}{(1-\gamma)^2}, \label{alg: update_eff_Q}\\
				&\cb_i^{r+1} =\cb_i^r - \bigg(\cb^r + \frac{ \thetab_i^{r+1}-\thetab^{r} }{\eta \wt Q_{i}^r }\bigg). \label{alg: update_edge device_ctl}
			\end{align}\vspace{-0.5cm}
			\STATE Send $\thetab_i^{r+1} - \thetab^r$ and $\wt Q_{i}^r $ to the server.
			\ENDIF
			\ENDFOR
			\STATE {\bfseries \underline{Server side:}} Compute 
			\STATE  \vspace{-0.6cm}
			\begin{align} 
				\thetab^{r+1} =& \thetab^r + \eta_g \frac{N}{m}\sum_{i\in \Ac^r} \omega_i \frac{\thetab_i^{r+1} - \thetab^r}{\wt Q_{i}^r }, \label{alg: update_theta}\\
				\cb^{r+1} =& \cb^r - \sum_{i \in \Ac^r} \omega_i \bigg(\cb^r + \frac{ \thetab_i^{r+1}-\thetab^{r} }{\eta \wt Q_{i}^r }\bigg) , \label{alg: update_server_ctl}
			\end{align}
			%\vspace{-0.5cm}
			\ENDFOR
	\end{algorithmic}}%\vspace{-1mm} 
\end{algorithm}

With the above three strategies, the details of the proposed FedCPSL algorithm are summarized in Algorithm \ref{alg: FedPCSL}. During the process of Algorithm \ref{alg: FedPCSL}, the server maintains two variables: the global model parameter $\thetab$ and the server control variable $\cb$, while each edge device $i$ maintains the localized model parameter $\thetab_{i, lc}$, the local copy of global model parameter $\thetab_i$, the client control variable $\cb_i$, the pseudo label $\nub_{i}$, and the local momentum buffers $\mub_i$ and $\mub_{i, lc}$. In each communication round $r$, the server randomly selects a set of active edge devices $\Ac^r$ by uniform sampling without replacement, and broadcasts $\thetab^r$ and $\cb^r$ to the devices in $\Ac^r$. Given $\thetab^r$ as the initial point, each active edge device $i \in \Ac^r$ computes $\nub_i^{r+1}$ via \eqref{eqn: prob_pseudo_labels}. If \eqref{eqn: prob_pseudo_labels} cannot be solved analytically, one step of GD will be adopted, i.e.,
\begin{align}
	\nub_{i}^{r+1} = \nub_{i}^r - \eta_v \nabla_{\nub_i} f_i(\thetab^r, \nub_{i}^r).
\end{align}
One feature of the proposed FedCPSL algorithm is that each edge device $i \in \Ac^r$ updates $\thetab_i$ using the local SGD with momentum for fast convergence. In each local step $t$, it employs a local momentum buffer $\mub_i$ which is initialized as zero at the beginning of each round and updated by geometrically accumulating all past variance-reduced SG's in the current round as
\begin{align}
		\mub_i^{r, t+1} =&  \gamma\mub_i^{r, t} +  g_i(\thetab_i^{r, t}, \nub_{i}^{r+1}) - \cb_i^r + \cb^r \notag \\
	=&\sum_{k = 0}^{t} \gamma^{t - k}(g_i(\thetab_i^{r, t}, \nub_{i}^{r+1}) - \cb_i^r + \cb^r),
\end{align}where $\gamma \in [0, 1)$ is the momentum factor.  It then  updates $\thetab_i$ by applying one step of SGD with $\mub_i^{r,t+1}$, i.e., \eqref{alg: update_theta_i}. Note that, similar to traditional client variance reduction schemes \cite{SCAFFOLD_2020}, the proposed FedCPSL algorithm incorporates $\cb - \cb_i$ to the local update of $\thetab_i$ to perform client variance reduction. However, owing to the introduction of momentum $\mub_i$ and time-varying local update steps $Q_i^r$, the updates of the control variables, $\cb_i$ and $\cb$, are respectively adjusted by \eqref{alg: update_edge device_ctl} and \eqref{alg: update_server_ctl} to accommodate these changes. Analogously, a local momentum buffer $\mub_{i, lc}$ is adopted for the update of the localized model $\thetab_{i, lc}$. Specifically, it is initialized as zero at the beginning of round $r$, updated by \eqref{alg: update_momentum_lc} and then used to update $\thetab_{i, lc}$ via \eqref{alg: update_theta_loc}. After that, these active edge devices upload the local updates $\thetab_i^{r+1} - \thetab^r$ and $\wt Q_i^r$ to the server, which will be used for global aggregation through \eqref{alg: update_theta}. Lastly, the next round starts with the newly generated $\thetab^{r+1}$ and $ \cb^{r+1}$.

Let us see how $\cb$ and $\cb_i$'s help to reduce the gradient variance among edge devices and correct the corresponding local update directions. In round $r$, if edge device $i$ is active, one can derive from \eqref{alg: update_momentum} and \eqref{alg: update_theta_i} that
\begin{align}
		&\thetab_i^{r+1} - \thetab^r = -\eta \wt Q_i^r \sum_{t = 0}^{Q_i^r-1} \frac{(\bb_i^r)^t} {\wt Q_i^r}(g_i( \thetab_i^{r,t}, \nub_i^{r+1})  + \cb^r - \cb_i^r), \label{eqn: local update}
\end{align}where $\bb_i^r = [1 - \gamma^{Q_i^r}, 1 - \gamma^{Q_i^r-1}, \ldots, 1 - \gamma] / (1-\gamma)$ and $\wt Q_i^r = \|\bb_i^r\|_1$. As a result, $\cb_i^{r+1}$ can be rewritten as
\begin{align}
	\cb_i^{r+1} &= \cb_i^r - \bigg( \cb^r - \sum_{t = 0}^{Q_i^r-1}\frac{(\bb_i^r)^t}{\wt Q_i^r} (g_i(\thetab_i^{r,t}, \nub_i^{r+1})  + \cb^r - \cb_i^r)\bigg) \notag\\
	&= \sum_{t = 0}^{Q_i^r-1}\frac{(\bb_i^r)^t}{\wt Q_i^r}g_i( \thetab_i^{r,t}, \nub_i^{r+1}). \label{eqn: edge device_ctl_detail}
\end{align}
Intriguingly, the above results \eqref{eqn: local update}\eqref{eqn: edge device_ctl_detail} suggest that the update of $\thetab_i$ generalizes the variance-reduction techniques used in traditional client variance reduction schemes where the local update is performed along an estimated global gradient direction considering the data of all edge devices. To be concrete, suppose $\rho = 1, \gamma = 0$ , $Q_i^r = Q, \nub_{i}^r = \nub_{i}^0, \forall i, r$, then we obtain $\mub_i^{r,t} = 0$ and 
\begin{align}
		&\thetab_i^{r, t+1}= \thetab_i^{r, t} - \eta\bigg (g_i( \thetab_i^{r,t}, \nub_i^{r+1}) - \cb_i^r + \sum_{i=1}^{N}\omega_i \cb_i^r\bigg),\\
		& \cb_i^{r+1} = \frac{1}{Q}\sum_{t = 0}^{Q-1}g_i( \thetab_i^{r,t}, \nub_i^{r+1}).
\end{align}
Clearly, these resulting update rules of $\thetab_{i}$ and $\cb_{i}$ are identical to that of traditional client variance reduction schemes. Thus, the update rules of $\thetab_i$, $\cb_i$ and $\cb$ in Algorithm \ref{alg: FedPCSL} can be seen as an extension of client variance reduction to the local SGD with momentum in the presence of heterogeneous edge devices.

It is worth noting that, the server control variable $\cb$ is updated only relying on the local updates $\thetab_i^{r+1} - \thetab^r$ and $\wt Q_{i}^r$ from active edge devices. In contrast, traditional client variance reduction schemes \cite{SCAFFOLD_2020} usually require transmitting back and forth both the local update and the server control variable. Thus,  \eqref{alg: update_server_ctl} enables a significant reduction of communication cost as the uplink channel is typically much slower than the downlink and the downlink cost is usually negligible compared to the uplink cost. Furthermore, to remedy the objective inconsistency, we utilize the technique of the local update normalization in the stage of global aggregation where the local update $\thetab_i^{r+1} - \thetab^r$ is normalized before aggregation.

\section{Convergence analysis} \label{sec: convergence analysis}
Before proceeding, we make the following assumptions, which are commonly adopted in the FL literature, e.g., \cite{Yang2020TCOM,ChenTWC2021,APFL_2020}.

%\vspace{-0.1cm}
\begin{assumption}[\textbf{Lower bounded}] \label{assumption: lower-bounded}
	All local cost functions $f_i(\cdot, \cdot)$ in problem \eqref{eqn: PSFL} are non-convex and lower bounded, i.e., 
	\begin{align}
		&f_i(\thetab_i, \nub_i) \geq \underline{f} > {-\infty},~\forall~\thetab_{i},  \nub_i\in \Vc_i.
	\end{align}
\end{assumption}

\begin{assumption}[\textbf{Convexity}] \label{assumption: nu-convexity}
	The cost function $l(\cdot, \cdot)$ in problem \eqref{eqn: PSFL}  is bi-convex and $f_i(\cdot, \nub_{i})$ is strongly convex with modulus $\mu$, i.e, $\forall \nub_i, \nub_i^\prime \in  \Vc_i$,
	\begin{align}
	&f_i(\thetab_i, \nub_i) \geq f_i(\thetab_i, \nub_i^\prime) + \langle \nabla_{\theta} f_i(\thetab_i, \nub_i^\prime), \nub_i - \nub_i^\prime\rangle + \frac{\mu}{2} \|\nub_i - \nub_i^\prime\|^2.
	\end{align}
Let $L_i(\zb_i, \nub_{i}) \triangleq l(\zb_i, \yb_i) + \alpha_p l(\zb_i, \nub_i) + \alpha_r R(\thetab, \nub_{i})$, then both $L_i(\cdot, \nub_{i})$ and $L_i(\zb_i, \cdot)$ are strongly convex with modulus $\mu$.
\end{assumption}
%Assumption \ref{assumption: nu-convexity} certainly holds as the KL divergences in $r_1(\cdot)$ and $r_2(\cdot)$ are strongly convex and the entropy loss is linear with respect to $\nub$.

\begin{assumption}[\textbf{Smoothness}] \label{assumption: L-smooth}
	Each local cost function $f_i(\cdot, \cdot)$ in problem \eqref{eqn: PSFL}  is $L$-smooth with respect to $\thetab_i$ and $\vb_i$. i.e, $\forall (\thetab_i, \nub_i), (\thetab_i^\prime, \nub_i^\prime) \in \Rbb^n \times  \Vc_i$,
	\begin{align}
		\|\nabla f_i(\thetab_i, \nub_i) - \nabla f_i(\thetab_i^\prime, \nub_i^\prime)\| \leq L \|(\thetab_i, \nub_i)-(\thetab_i^\prime, \nub_i^\prime)\|.
	\end{align}
Meanwhile, the local cost function $F_i(\cdot, \cdot, \cdot)$ is $L_F$-smooth. The function $L_i(\cdot, \cdot)$ (resp. $h(\cdot)$) is  $\wt L$-smooth (resp. $L_h$-smooth). 
\end{assumption}

\begin{assumption}[Bounded variance] \label{assumption: SGD_variance}
	For a data samples $d_i$ uniformly sampled at random from $\Lc_i$ or $\Uc_i$, the resulting stochastic gradients (SG) for problem  \eqref{eqn: PSFL} are unbiased and have bounded variances, i.e.,
	\begin{align}
		& \E[\nabla_{\theta} f_i(\thetab_i, \nub_i; d_i)] = \nabla_{\thetab} f_i(\thetab_i, \nub_i), \\
		&\E[\|\nabla_{\theta} f_i(\thetab_i, \nub_i; d_i) - \nabla_{\thetab} f_i(\thetab_i, \nub_i)\|^2] \leq \sigma^2,\\
		& \E[\nabla_{\thetab_{i, lc}}F_i(\thetab_{i, lc}, \thetab_{i}, \nub_i; d_i)] = \nabla_{\thetab_{i, lc}} F_i(\thetab_{i, lc}, \thetab_{i},\nub_i), \\
		&\E[\|\nabla_{\thetab_{i, lc}} F_i(\thetab_{i, lc}, \thetab_{i},\nub_i; d_i) - \nabla_{\thetab_{i, lc}} F_i(\thetab_{i, lc}, \thetab_{i}, \nub_i)\|^2] \leq \sigma^2,
	\end{align}where $\sigma > 0$ is a constant.
\end{assumption}
Then, $g_i(\thetab_i, \nub_i)$ is unbiased with variance bounded by $\frac{\sigma^2}{S}$ where $S = S_l + S_u$.  Analogously, $G_i(\thetab_{i,lc}, \thetab_{i}, \nub_i)$ is also unbiased with variance bounded by $\frac{\sigma^2}{S}$. 

\begin{assumption} [Model discrepancy]\label{assumption: model_distance} For any two data inputs $\xb_1, \xb_2$, the distance between their model outputs is bounded, i.e.,
	\begin{align}
		\|h(\xb_1) - h(\xb_2)\| \leq \Gamma.
	\end{align}
\end{assumption}
In practice, Assumption \ref{assumption: model_distance} is mild as it is known that the output of $h(\xb)$ usually lies in a bounded space. For example, for a classification task, $h(\xb) \in [0, 1]^n$, and thus $\Gamma \leq 2$.

Then, we start to build the convergence properties of the proposed FedCPSL algorithm. It demands the convergence of the global model parameter $\thetab$, the pseudo labels $\nub$, and the localized model parameter $\thetab_{i, lc}$ to a solution of problem \eqref{eqn: PSFL}. 
To achieve this, we first establish the convergence of the sequence  $\{(\thetab^r, \nub^r)\}$ generated by FedCPSL to a stationary point of problem \eqref{eqn: PSFL_c1}. Then, we build the convergence of  $\{(\thetab_{i,lc}^r, \nub^r)\}$ to a stationary point of problem \eqref{eqn: PSFL}. 

We define the following term as the optimality gap between $(\thetab^r, \nub^r)$ and a stationary solution of problem \eqref{eqn: PSFL_c1}.
\begin{align}
	&G(\thetab^r, \nub^r) 
	\triangleq \| \nabla_{\thetab}f(\thetab^r, \nub^{r})\|^2 +\frac{31L}{64}\sum_{i=1}^{N}\omega_i\| \nub_i^{r+1}- \nub_i^r\|^2. 
\end{align}Note that if $G(\thetab^r, \nub^r)=0$, then $(\thetab^r, \nub^r)$ is a stationary solution of problem \eqref{eqn: PSFL_c1}. The following theorem establishes the convergence of $\{(\thetab^r, \nub^r)\}$. 

\begin{Theorem}\label{thm: FedGAPD}
	Let $\Ac^r (|\Ac^r| = m < N)$ be a set of edge devices obtained by uniform sampling without replacement in round $r$. In Algorithm \ref{alg: FedPCSL}, suppose $\nub_{i}$ is updated via \eqref{alg: update_nu}, the stepsize $\eta_v \leq \frac{1}{4L}$, the stepsize $\eta$ satisfy
		\begin{align}
			& \eta \leq \min\bigg\{\frac{m}{48\eta_gNL},\frac{1}{8 L\ol Q},\frac{3\eta_g N}{100mL\ol Q^2}, \frac{m}{32\eta_gNL} \bigg(1 + \frac{2N}{ m}\bigg)^{-\frac{1}{2}} \bigg\}, \label{eqn: condition_eta}
		\end{align}where $\ol Q = \max\limits_{i, r} \wt Q_i^r$. Then, under Assumption  \ref{assumption: lower-bounded}, \ref{assumption: nu-convexity}, \ref{assumption: L-smooth} and \ref{assumption: SGD_variance}, the generated sequence $\{(\thetab^r, \nub^r)\}$ of  FedCPSL satisfies
		\begin{align}
			\frac{1}{R}\sum_{r = 0}^{R-1}\E[G(\thetab^r, \nub^r)] \leq  \frac{4(P^0 - \ul f)}{\eta \eta_g R}+ \bigg(\frac{259}{64} + \frac{ m}{N} + \frac{5\eta mL\ol Q^2}{24\eta_g N}\bigg)\frac{12\eta \eta_g NL\sigma^2}{mS}.\label{thm1: FedGAPD_bd}
		\end{align}where  $P^0$ is independent of $R, \sigma, S$, and it is defined in \eqref{eqn: def_potential}.
\end{Theorem}

%{\bf Proof:} 
\begin{IEEEproof}
Unlike the existing works \cite{FedSSL}\cite{SCAFFOLD_2020}, we consider a non-convex SSFL problem with the presence of heterogeneous edge devices. To reduce the negative effects of non-i.i.d. data and time-varying HLU and accelerate the algorithm convergence, we adopt schemes of local SGD with momentum and novel client variance reduction, which makes Theorem \ref{thm: FedGAPD} much more challenging to prove. Following the analysis framework in \cite{SCAFFOLD_2020}, we attempt to build a potential function which descends as $(\thetab_{i}, \thetab, \nub_{i})$ proceeds in Algorithm \ref{alg: FedPCSL}  by analyzing the one-round progress of the cost function $f$. To this end, we develop some new techniques to overcome the challenges brought by the novel update of $\thetab_{i}$ in Algorithm \ref{alg: FedPCSL}. Details are presented in Appendix \ref{appdix: theorem1}.
\end{IEEEproof}

One can see from Theorem \ref{thm: FedGAPD} that the convergence of the global model parameter $\thetab$ is resilient to heterogeneous edge devices. In particular, it converges with much less influence caused by non-i.i.d. data compared to FedAvg \cite{FedAvg_noniid_2019} and allows time-varying HLU. The following corollary demonstrates the sublinear convergence rate of $\thetab$ and $\nub_{i}$ with a carefully chosen constant step size $\eta$. The convergence rate matches the known lower bounds for nonconvex FL with client variance reduction \cite{SCAFFOLD_2020}. However, with the inclusion of local momentum, the convergence of $\thetab$ and $\nub_{i}$ is provably faster in practice. This point will be further examined via numerical experiments in Section \ref{sec: simulation}.

\begin{Corollary} \label{corly: uniform_conv_rate}
	Let $\eta  = \mathcal{O}\bigg(\frac{\sqrt{m}}{\sqrt{\eta_gN R}}\bigg)$, then under the same setting as Theorem \ref{thm: FedGAPD}, we obtain a sublinear convergence rate of the sequence $\{(\thetab^r, \nub_{i}^r)\}$, i.e.  $\mathcal{O}\bigg(\frac{\sqrt{N}}{\sqrt{m \eta_g R}}\bigg)$.
\end{Corollary}

Next, let us move to the convergence of the localized model parameter $\thetab_{i, lc}$. To build the convergence condition, we employ the term $\sum_{i=1}^{N}\omega_i\|\nabla_{\thetab_{i, lc}} F_i(\thetab_{i, lc}^r,\thetab^{r}, \nub_{i}^{r})\|^2$ as the optimality gap. When $\sum_{i=1}^{N}\omega_i\|\nabla_{\thetab_{i, lc}} F_i(\thetab_{i, lc}^r,\thetab^{r}, \nub_{i}^{r})\|^2 = 0$, $\thetab_{i, lc}^r, i \in [N]$ is a stationary point of problem \eqref{eqn: PSFL}. 

\begin{Theorem} \label{thm: localized model}
	Consider the same setting of $\rho$ and the update of $\nub_{i}$ as Theorem \ref{thm: FedGAPD}, and suppose the stepsizes $\eta$, $\eta_c$ and $\beta_i, i \in [N]$ satisfy
		\begin{align}
			 \eta &\leq \min\bigg\{\frac{m}{48\eta_gNL},\frac{m}{48\eta_gND_0}, \frac{1}{8 L\ol Q},\frac{3\eta_g N}{100mL\ol Q^2},  \frac{m}{32\eta_gNL} \bigg(1 + \frac{2N}{ m}\bigg)^{-\frac{1}{2}} \bigg\}, \\
			\eta_{c} &\leq \min\bigg\{ \frac{1}{2\ol Q L_F}, \frac{\wt \eta N L^2}{\ul Q m L_F^2}, \frac{\wt \eta N }{\ul Q m}\bigg\}, \\
			D_0 &\leq \frac{11L}{8} +  \frac{11m}{2N}, 
		\end{align}where $D_0 \triangleq \frac{1}{4}\sum_{i=1}^{N}\omega_i (1-\beta_i)(4\beta_i\wt LL_h \Gamma + 5L)$, $\ul Q = \min\limits_{i, r} \wt Q_i^r$, and $\ol Q$ are defined in Theorem 1. Then, under Assumption  \ref{assumption: lower-bounded}, \ref{assumption: nu-convexity}, \ref{assumption: L-smooth} and \ref{assumption: SGD_variance}, the generated sequence $\{\thetab_{i, c}^r\}$ of  FedCPSL satisfies
	\begin{align}
		&\frac{1}{R}\sum_{r = 0}^{R- 1}\sum_{i=1}^{N}\omega_i \E[\|\nabla_{\thetab_{i, lc}} F_i(\thetab_{i, lc}^r,\thetab^{r}, \nub_{i}^{r})\|^2] \leq  \frac{3N}{\eta_{c} \ul Q m R} (f(\thetab^0, \nub_{i}^0)-\ul f) + \frac{11N }{2\eta_{c}\ul QmR}(P^0 - \ul f)   \notag \\
		& \qquad + \underbrace{ \frac{3\eta_c L_F(\eta_c L_F + 1)\ol Q \sigma^2}{2S} }_{\triangleq \rm (a)}
		+ \underbrace{\frac{3\eta^2 \eta_g^2 N^2\sigma^2}{2\eta_{c}\ul Qm^2S}}_{\triangleq \rm (b)} +  \underbrace{\bigg(\frac{259}{64} + \frac{ m}{N} + \frac{5\eta mL\ol Q^2}{24\eta_g N}\bigg)\frac{66\eta^2 \eta_g^2  N^2L\sigma^2}{\eta_{c} \ul Q m^2S}}_{\triangleq \rm (c)} \notag \\
		& \qquad + \underbrace{\frac{3\wt L^2 \Gamma^2}{2\eta_{c} \ul Q L}\sum_{i=1}^{N}\omega_i\beta_i^2}_{\triangleq \rm (d)}   +  \underbrace{\frac{3N\Gamma^2}{\eta_{c}\ul Q m}\sum_{i=1}^{N}\omega_i\beta_i(1-\beta_i)(\beta_i(\wt L-\mu) + \wt L)}_{\triangleq \rm (e)}  , \label{thm: localized model_bd}
	\end{align}where  $P^0$ is independent of $R, \sigma, S$, and it is defined in \eqref{eqn: def_potential}.
\end{Theorem}

%{\bf Proof:} 
\begin{IEEEproof}
In contrast to APFL \cite{APFL_2020}, we study the convergence of the localized model parameter $\thetab_{i, lc}$ instead of the mixed model $\beta_ih(\thetab_{i, lc}) + (1-\beta_i)h(\thetab)$. The challenge arises from the fact that the optimization of $\thetab_{i, lc}$ partially incorporates the global model $h(\thetab)$, which brings difficulties in characterizing the evolution of $\thetab_{i, lc}$ as the algorithm proceeds. To establish the convergence of $\thetab_{i, lc}$, we follow the analysis framework of local SGD within each edge device $i$, while studying the contribution of the dynamic $\thetab$ and $\nub_{i}$ to the update of $\thetab_{i, lc}$. To be specific, we analyze the one-round progress of $F_i$ by considering not only $\thetab_{i, lc}$ but also the effects of $\thetab$ and $\nub_{i}$ from the proof of Theorem \ref{thm: FedGAPD}. The bound of the optimality gap is then derived. More details can be found in Appendix \ref{appdix: proof_theorem2} . 
\end{IEEEproof} 

Theorem \ref{thm: localized model} shows the convergence of $\thetab_{i, lc}$ with respect to four kinds of error terms. The term $\rm (a)$ in the RHS of \eqref{thm: localized model_bd} is caused by SGD in the update of $\thetab_{i, lc}$ while $\rm (b)(c)$ are due to SGD in the update of $\thetab$. Both of them are obviously $0$ if $\sigma = 0$ or can vanish to $0$ in a sublinear manner with a carefully chosen step size $\eta$ and mini-batch size $S$. The term $\rm (d)$ (resp. $\rm (e)$) in the RHS of \eqref{thm: localized model_bd} stems from the incorporation of $\nub_{i}$ (resp. $\thetab$) into the update of \eqref{thm: localized model_bd}, which implies that the dynamic $\thetab$ and $\nub_{i}$ can deteriorate the convergence of $\thetab_{i, lc}$, even though they gradually converge. According to Corollary \ref{corly: localized model}, we obtain that the sequence  $\{\thetab_{i, lc}^r\}$ sublinearly converges towards a ball with the center being a stationary point and radius being $\rm (d) + (e)$.  It is worth-noting that, when $\beta_i \rightarrow 0, i \in [N]$, the terms $\rm (d)(e)$ will approximate $0$, and $\beta_i \rightarrow 1, i \in [N]$ also means that $\rm (e) \rightarrow 0$. This implies that, with proper choices of $\eta_{c}$ and $\beta_i$, both $\rm (d)$ and $\rm (e)$ are relatively small and $\{\thetab_{i, lc}^r\}$ converges within a neighborhood of the solutions.

\begin{Corollary} \label{corly: localized model}
		Let $\eta  = \mathcal{O}\bigg(\frac{\sqrt{m}}{\sqrt{\eta_gN R}}\bigg)$ and $S = \mathcal{O}(\sqrt{R})$, then under the same setting as Theorem \ref{thm: localized model}, the following convergence result of the sequence $\{\thetab_{i, lc}^r\}$ holds.
		\begin{align}
				&\frac{1}{R}\sum_{r = 0}^{R- 1}\sum_{i=1}^{N}\omega_i \E[\|\nabla_{\thetab_{i, lc}} F_i(\thetab_{i, lc}^r,\thetab^{r}, \nub_{i}^{r})\|^2] \notag \\
			\leq& \mathcal{O}\bigg(\frac{N}{m R} + \frac{\sigma^2}{\sqrt{R}}\bigg) +{\rm (d) + (e)},
			\end{align}where $\rm (d)$ and $\rm (e)$ are defined in Theorem \ref{thm: localized model}.
\end{Corollary}

From Theorem \ref{thm: localized model} and Corollary \ref{corly: localized model}, we have some important observations and insights which are summarized in Remark \ref{remark: Robust to heterogeneous edge devices}, Remark \ref{remark: Trade-off between accuracy and efficiency} and Remark \ref{remark: update of pseudo-label}. 

\begin{Rmk} \label{remark: Robust to heterogeneous edge devices}
	\rm
	 (Robustness to heterogeneous edge devices) As seen from Theorem \ref{thm: localized model}, unlike \cite{APFL_2020}, the convergence of $\thetab_{i, lc}$ does not suffer from the non-i.i.d. data.  This is understandable as the effect of non-i.i.d. data on $\thetab_{i, lc}$ originates from the update of $\thetab$ where we, however, employ client variance reduction to remedy that issue. In addition, we adopts the technique of momentum to boost the convergence of $\thetab_{i, lc}$. Thus, it is believed to be faster than existing algorithms using merely local SGD without client variance reduction and momentum \cite{APFL_2020}.
\end{Rmk}

\begin{Rmk}\label{remark: Trade-off between accuracy and efficiency}
{\rm(Trade-off between accuracy and efficiency) One can notice that the terms $\rm (d)$ and $\rm (e)$ in the RHS of \eqref{thm: localized model_bd} exist because of the simultaneous training of $\thetab, \nub_{i}$ and $\thetab_{i, lc}$. They will disappear if $\thetab$ and $\nub_{i}$ are constant in the optimization of $\thetab_{i, lc}$. However, the joint optimization of $\thetab, \nub_{i}$ and $\thetab_{i, lc}$ in Algorithm \ref{alg: FedPCSL} would greatly improve the algorithm efficiency as we can obtain these desirable parameters simultaneously. Thus, there is a trade-off between convergence performance and algorithm efficiency. Fortunately, this can be easily resolved by simply running a few more rounds for updating $\thetab_{i, lc}$ when $\thetab$ and $\nub_{i}$ have negligible changes. By this way, the proposed FedPCSL can offer both high efficiency and model accuracy.}
\end{Rmk}

\begin{Rmk} \label{remark: update of pseudo-label}
	{\rm The above theoretical results are all based on one-step GD update rule 	in \eqref{alg: update_nu} for $\nub_{i}, i \in [N]$. Similar theoretical guarantees can also be easily obtained when the exact solution of \eqref{eqn: prob_pseudo_labels} is pursued to update 
	$\nub_{i}, i \in [N]$ at each round. In particular, following the same spirit in \cite{FedSSL}, we can utilize the strong convexity of $f_i(\thetab, \cdot)$ and $F_i(\thetab_{i, lc}, \thetab, \cdot)$ to establish similar descent lemmas with respect to $\nub_{i}$ to obtain Lemma \ref{lem: diff_f} in the proof of Theorem \ref{thm: FedGAPD}. Due to space limitations, we omit this and refer \cite{FedSSL} for more details.}
\end{Rmk}

\begin{Rmk} {\rm 
 	The choice of $\beta_i$ for each edge device plays an important rule in the generalization performance of FedCPSL, and a particular setup of it would complement FedCPSL. Nevertheless, the task of effectively estimating the optimal $\beta_i, i \in [N]$ is non-trivial. By \cite[Theorem 1 and equation (3)]{APFL_2020}, the optimal value of $\beta_i$ depends on numerous factors which are remarkably difficult to compute or estimate. Direct incorporation of existing techniques such as the similarity scores  \cite{ClusterFL2020, UserFL2021} between local data distributions of any two devices would yield incorrect estimations. The strategy of regarding $\beta_i$ as an optimization variable and updating it periodically also cannot be straightforwardly applied because of the complexity of problem \eqref{eqn: PSFL}. It will be an interesting direction to study the choices of $\beta_i$ with theoretical justification in the future.}
\end{Rmk}

\section{Experiment Results} \label{sec: simulation}

In this section, we will examine the performance of the proposed FedCPSL algorithm against the state-of-the-art algorithms on real datasets.

\subsection{Experiment setup}

{\bf Cost function $f_i$ and $F_i$:} We consider the problem formulation in \cite{FedSSL} for the local cost function $f_i$ in problem \eqref{eqn: PSFL_c1} with
\begin{align}
		l(h(\thetab; \xb_i), \yb_i) =& - \frac{1}{n_i}\sum_{k=1}^{n_i} \langle \yb_{i, k}, \log(h(\thetab;\xb_{i,k})), \\
		l(h(\thetab; \ub_i), \nub_i)=&- \frac{1}{m_i}\sum_{k=1}^{m_i} \langle \nub_{i,k}, \log(h(\thetab;\ub_{i,k})), \\
		R(h(\thetab; \ub_i), \nub_{i})=& \frac{1}{m_i}\sum_{k=1}^{m_i} KL(\nub_{i,k}, \db) + \frac{1}{m_i}\sum_{k=1}^{m_i}KL(h(\thetab;\ub_{i,k}), \db), 
\end{align}where  $KL(\cdot, \cdot)$ is the Kullback-Leibler divergence; $\db \triangleq \frac{\oneb}{C}$ with $\oneb$ being all-one vector. The cost function $F_i$ is obtained by replacing $h(\thetab; \cdot))$ in $f_i$ with $\beta_ih(\thetab_{i, lc}; \cdot) + (1-\beta_i)h(\thetab; \cdot)$.

{\bf Datasets and models $h$:} The popular MNIST  \cite{website_MNIST} and CIFAR-10 \cite{CIFAR10_Krizhevsky09} datasets are considered for evaluation both of which are widely used in previous FL works. Specifically, the MNIST dataset contains 60K training images and 10K test ones while the CIFAR-10 dataset has 50K training images of handwritten digits and 10K test ones. We respectively adopt a CNN model for the CIFAR-10 image classification task and a fully-connection neural network model for the MNIST image classification task. The former is similar to that in \cite{FedAvg_noniid_2019} which consists of two convolutional layers and two fully connected layers while the latter is the same as that in \cite{FedDyn_2021}.

We simulate the PSSFL process by distributing the training samples of each dataset to $N = 20$ edge devices in the fashion of {\bf Non-IID}. To obtain the {\bf non-IID} distributed data, we follow the heterogeneous data partition method in \cite{FedAvg_noniid_2019} where each edge device is allocated data samples of only a few class labels. In particular, in \cite{FedAvg_noniid_2019}, the training data is sorted by the label and then partitioned into shards. Each edge device is assigned a small and fixed number of shards uniformly at random. This leads to highly non-i.i.d. datasets among edge devices. The number of class labels in each edge device determines the non-i.i.d. degree. Note that the  method in \cite{FedAvg_noniid_2019} is, as far as we know, one of the most widely used data distribution method in the FL literature, and is acknowledged to be sufficient for performance evaluation. Then, we randomly extract $20\%$ of data samples from each edge device as its test data for performance evaluation. Similar to \cite{FedSSL}, in each edge device, we vary a ratio $\epsilon$ $ (0 \leq \epsilon \leq 1)$ of unlabeled data to divide the rest $80\%$ data samples into the unlabeled data $\Uc_i$ and the labeled data $\Lc_i$. $\epsilon = 0.5$ means that half of the training data of each edge device have no labels. By default, $\epsilon$ is set as $0.9$ in the experiment to simulate the real scenario where each edge device only has a few labeled data samples.

{\bf Baseline algorithms for comparison:} We evaluate the proposed FedCPSL algorithm by comparing it with the following four baseline FL algorithms. For all these methods, we report test accuracy as a function of communication rounds. The reported results are the average of the performances of the corresponding models over all edge devices' test data. 
\begin{itemize}
	\item FedSHVR: The algorithm proposed in \cite{FedSSL} for semi-supervised FL but which only supports full participation.
	\item FedSHVRP: The algorithm implemented by adapting FedSHVR to the case of PCP and using the technique of client variance reduction in \cite{SCAFFOLD_2020}.
	\item APFL\footnote{For fair comparison, we modify the original APFL by considering the mixture of the global model and the localized model as the personalized model for each edge device. }: The algorithm proposed in \cite{APFL_2020} for personalized FL by adopting model interpolation, but which only leverages the labeled data for training.
	\item APFSL: The algorithm implemented by adapting APFL to our proposed formulation so as to handle the semi-supervised FL with both labeled and unlabeled data.
\end{itemize}

{\bf Parameter setting:} For all algorithms under test,  the mini-batch size $B_l = 32$ for labeled data and $B_u = 32$ for unlabeled data. In each communication round, we only sample $10\%$ or $20\%$ of the total edge devices ($|\Ac^r| = 2$ or $4$) and all active edge devices perform $E = 2$ local epochs. To evaluate FedCPSL in the presence of heterogeneous edge devices, we also consider (time-varying) {\bf HLU} in which the number of local update steps of each participating edge device is randomly sampled from a fixed range in each communication round. In particular, if {\bf HLU} is considered, we choose the number of local epochs at random from $[1, 5]$ at each communication round. The learning rate $\eta$ is set to be $0.005$  (resp. $0.002$) for the MNIST (resp. CIFAR-10) dataset  if {\bf HLU} is not applied, and it is set to be $0.002$  (resp. $0.001$) for the MNIST (resp. CIFAR-10) dataset otherwise. Besides, the $\eta_{lc}$ of FedCPSL is set as $2 \eta$ and $\eta_g$ is the average of the number of local steps of the local edge devices. We adopt  \eqref{eqn: prob_pseudo_labels} to update the pseudo labels $\nub_{i}$ for each round as \eqref{eqn: prob_pseudo_labels} has a closed-form solution \cite{FedSSL}. 
Finally, all algorithms stop when $100$ communication rounds are achieved.

\subsection{The effect of model personalization}

First, we examine the performance of the proposed FedCPSL algorithm when different ratios of the global model are considered in the personalized model. We conducted experiments with different choices of $\beta_i$ on both the MNIST and CIFAR-10 datasets. The numerical results are depicted in Fig. \ref{fig: FedCPSL_beta}. Note that no local momentum is used, i.e., $\gamma = 0$. As seen, the personalized model significantly improve the test accuracy of FedCPSL over the global model when faced with non-i.i.d. data. In particular, when $m = 0.1N$, smaller values of $\beta_i$ (using more of the global model as the personalized model) yield lower test accuracy and exhibit unstable convergence curves. Increasing $\beta_i$ can diminish the negative impacts of partial participation on the personalized models. This demonstrates that larger values of $\beta_i$ are preferred for highly non-i.i.d. datasets, and it is necessary to adopt model personalization instead of merely using the global model. With $\beta_i = 0.75$ for all edge devices, the proposed FedCPSL algorithm significantly outperforms the case where $\beta_i = 0.25$ in terms of both convergence speed and test accuracy. In addition, one can also see that $\beta_i = 1$ may not be the optimal choice as the test accuracy can be improved with the assistance of the global model. Similar results can be observed when $m = 0.2N$ on both the MNIST and CIFAR-10 datasets.
\begin{figure*}[h]
	\centering
	\subfigure[MNIST, $m=0.1N$]{\includegraphics[width=6.6cm]{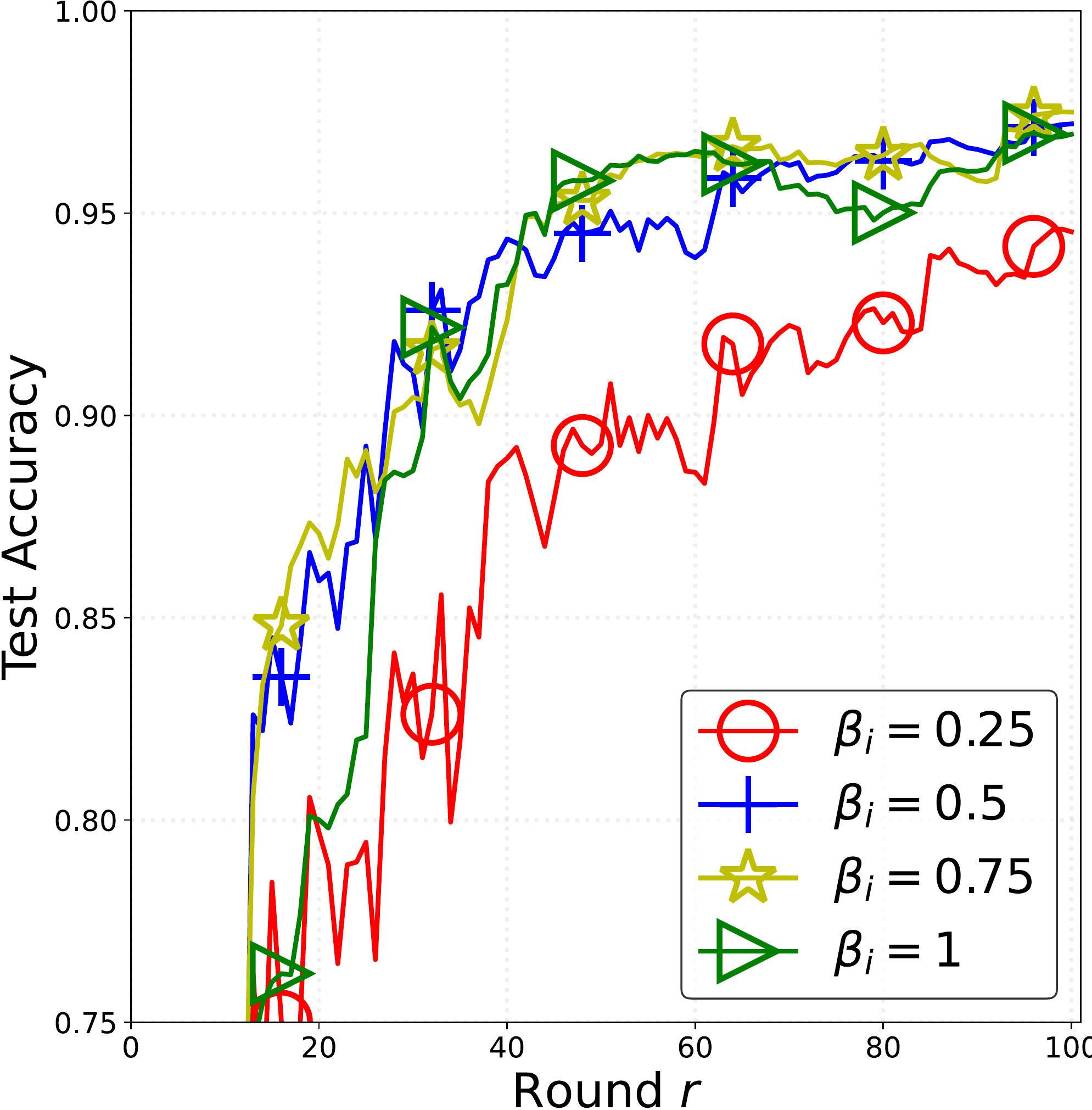}}	
	\subfigure[MNIST, $m=0.2N$]{\includegraphics[width=6.6cm]{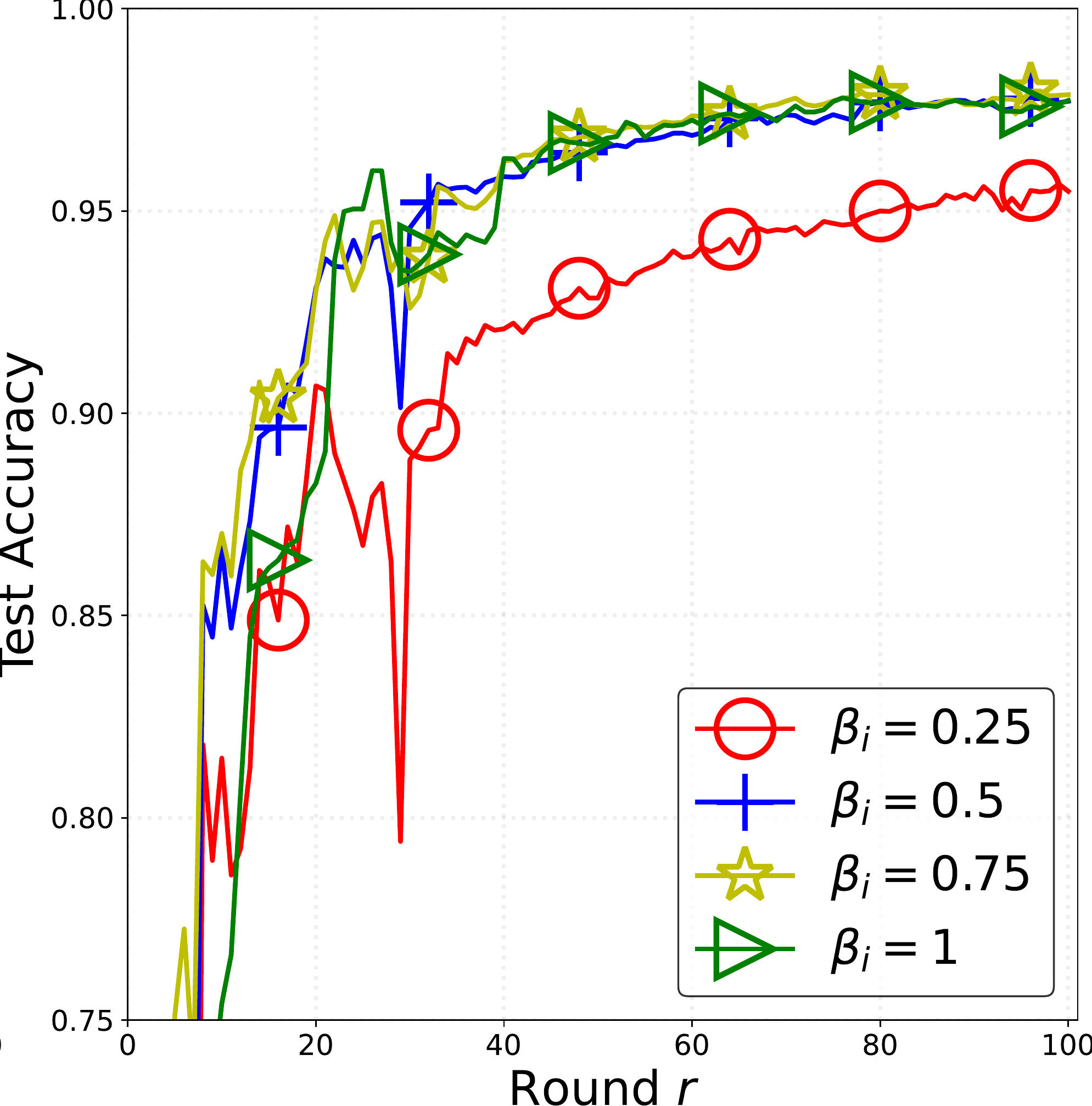}}
	\subfigure[CIFAR-10, $m=0.1N$]{\includegraphics[width=6.6cm]{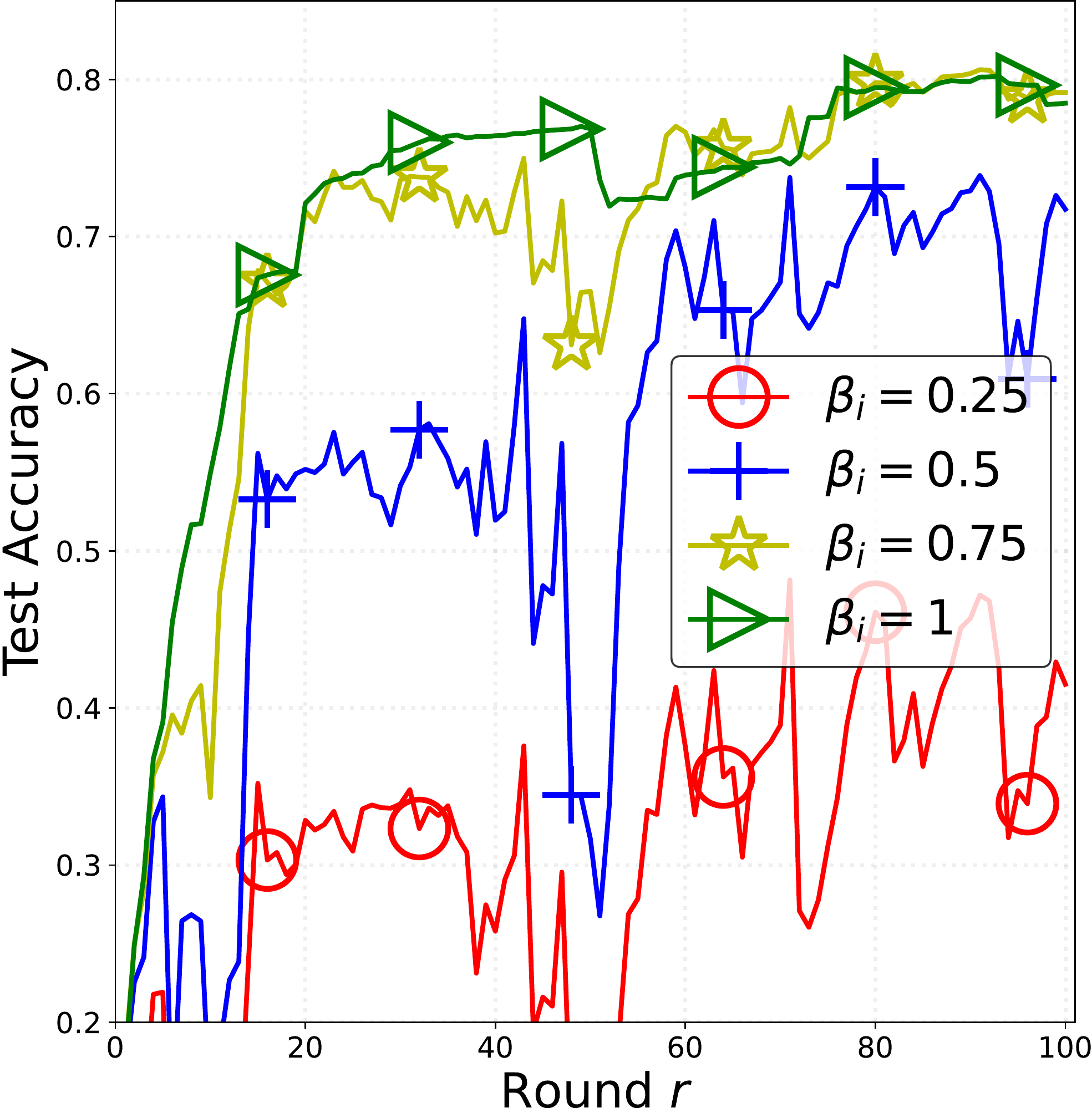}}
	\subfigure[CIFAR-10, $m=0.2N$]{\includegraphics[width=6.6cm]{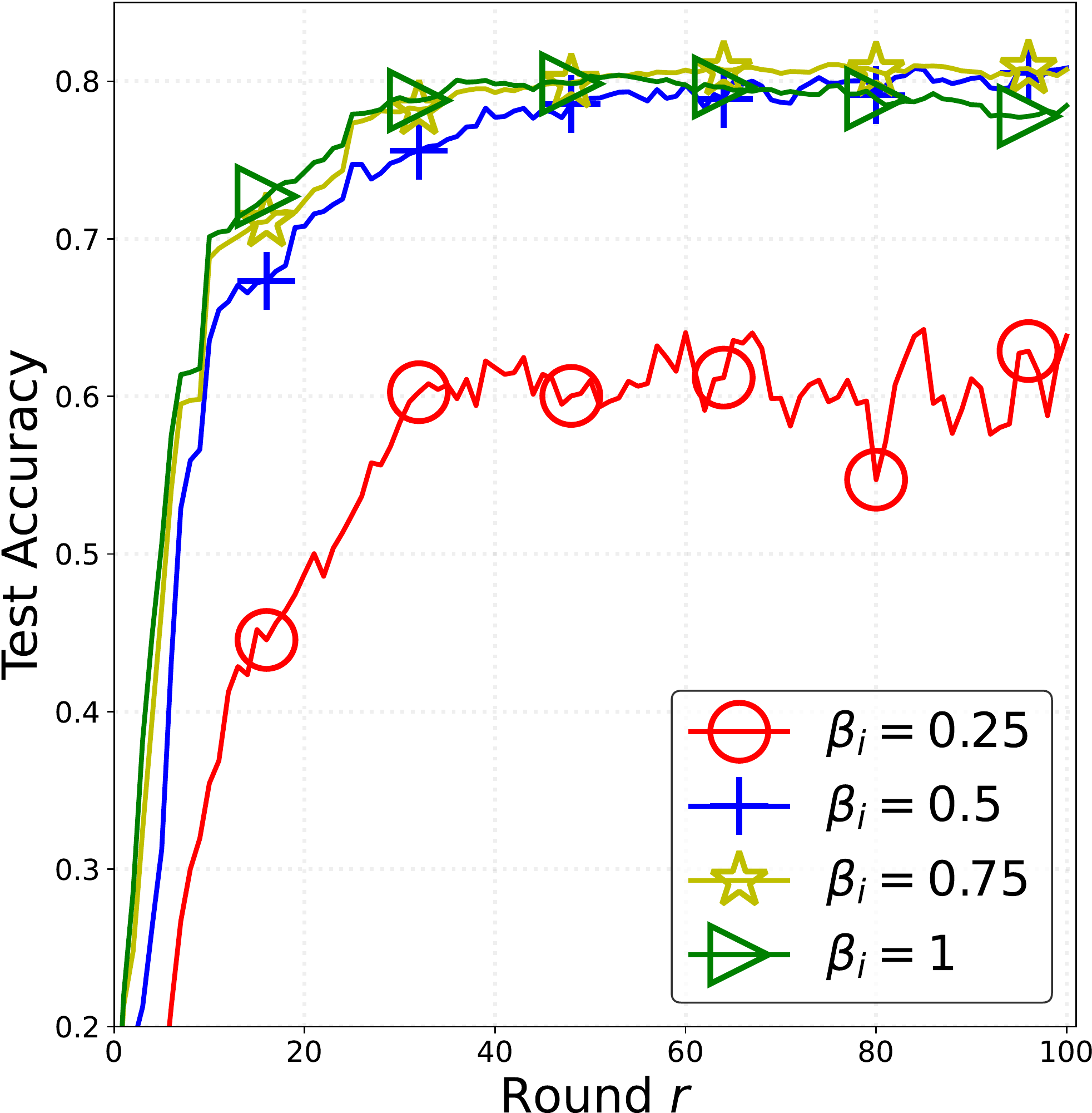}}
	\centering\caption{Test accuracy versus number of rounds of FedCPSL with different values of $\beta_i$ and $m$. Note that $\gamma$ is set as $0$ for all cases.}\label{fig: FedCPSL_beta}
\end{figure*}  

\begin{figure} [t!]
	\centering
	\subfigure[\scriptsize MNIST, $m=0.1N$]{\includegraphics[width=6.6cm]{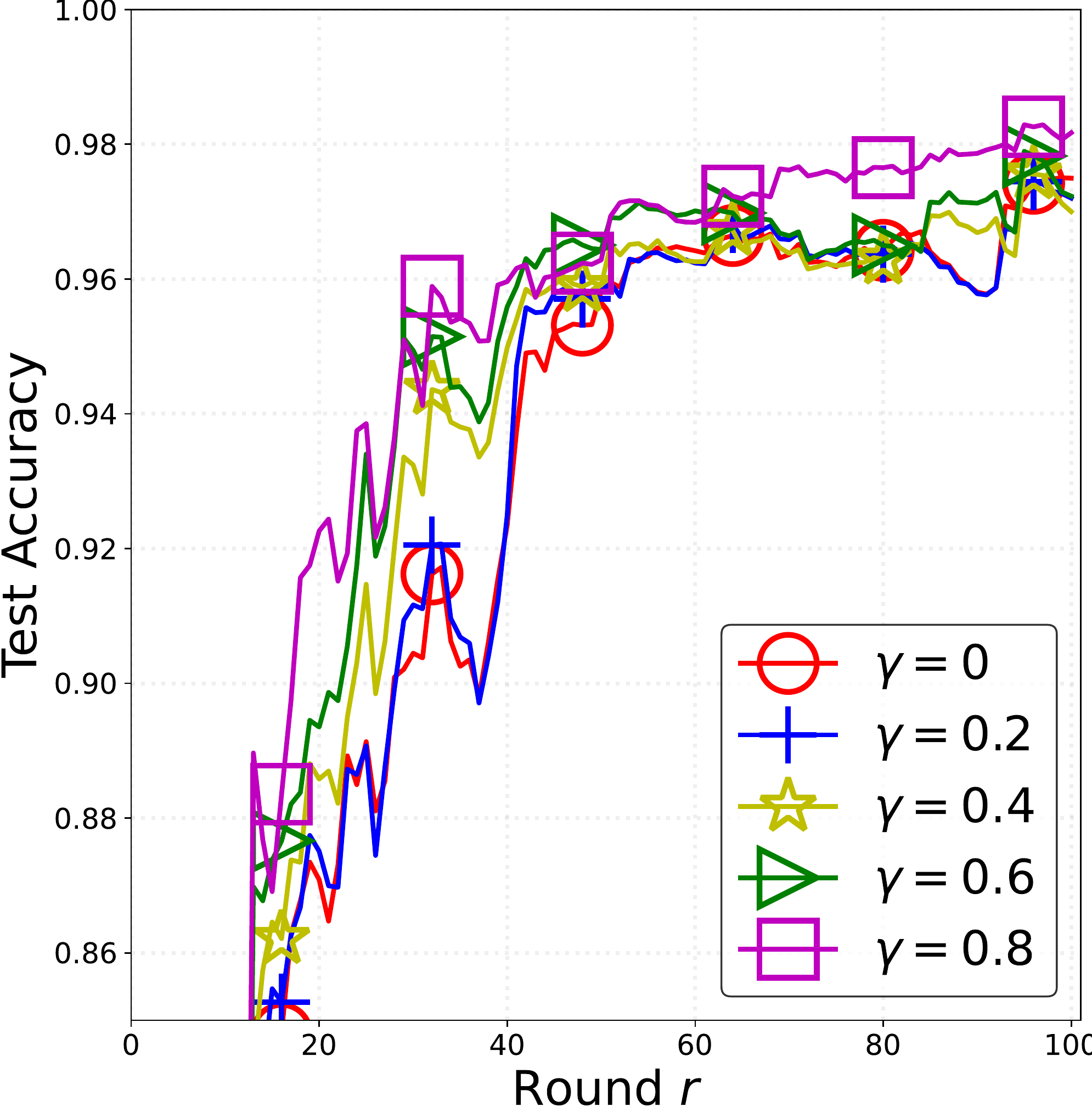}\label{fig: momentum_a}}	
	\subfigure[\scriptsize MNIST, $m=0.2N$]{\includegraphics[width=6.6cm]{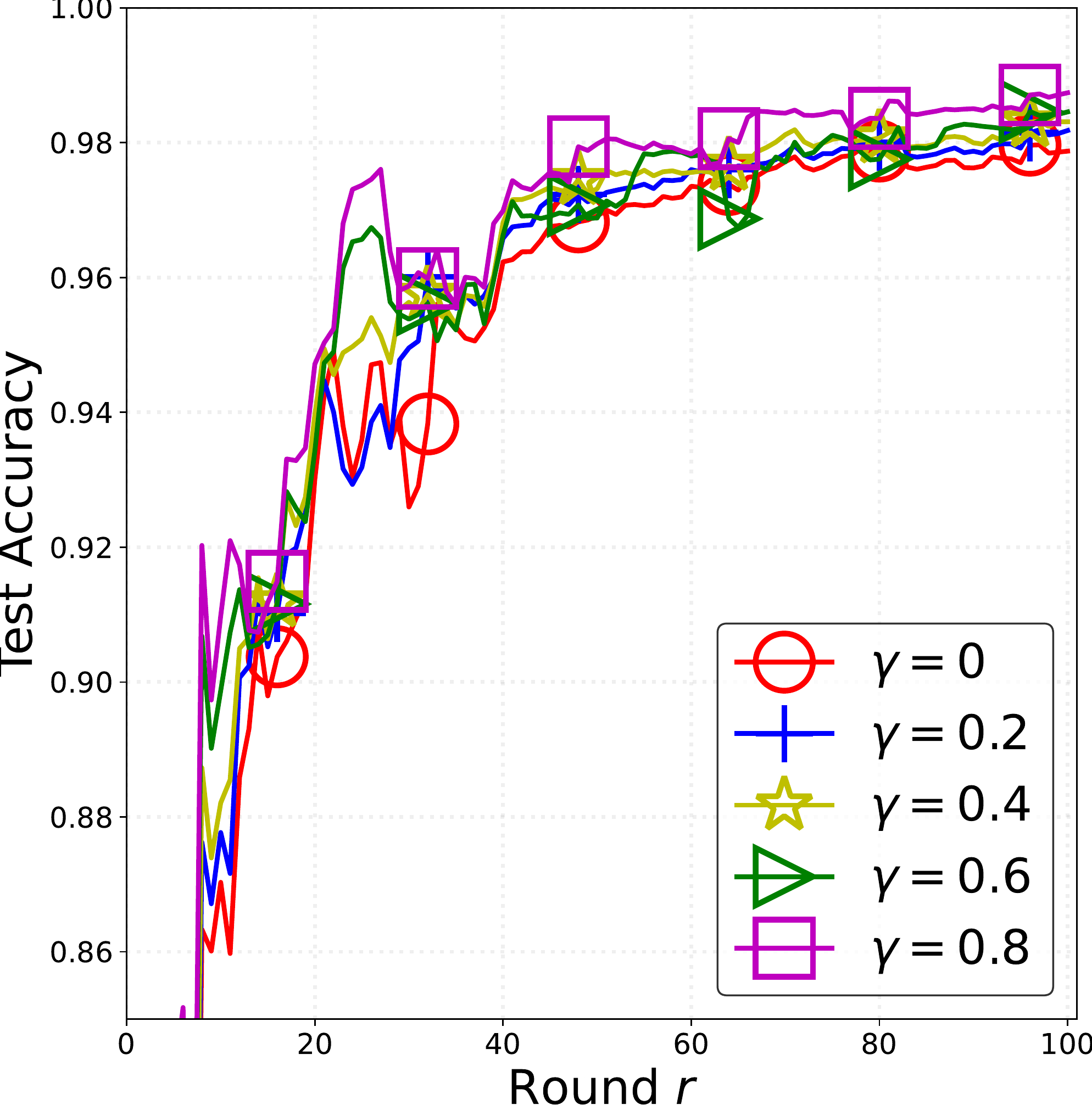} \label{fig: momentum_b}}
	\centering \caption{Test accuracy versus number of rounds of FedCPSL with different values of $\gamma$ and $m$. }\label{fig: FedCPSL_momentum}
\end{figure}  

\begin{figure}[t!]
	\centering
	\subfigure[\scriptsize MNIST]{\includegraphics[width=6.6cm]{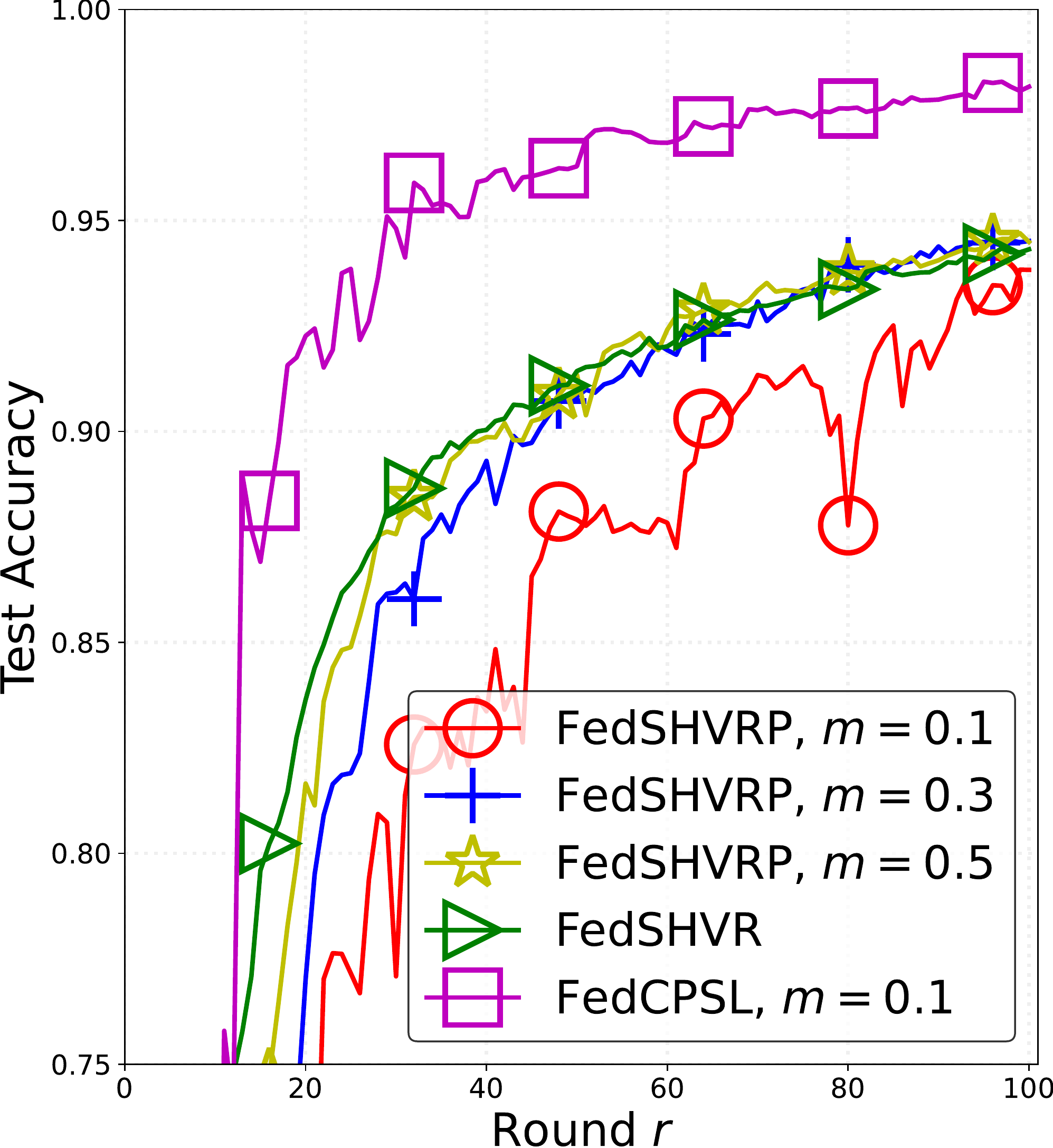}}
	\subfigure[\scriptsize CIFAR-10]{\includegraphics[width=6.6cm]{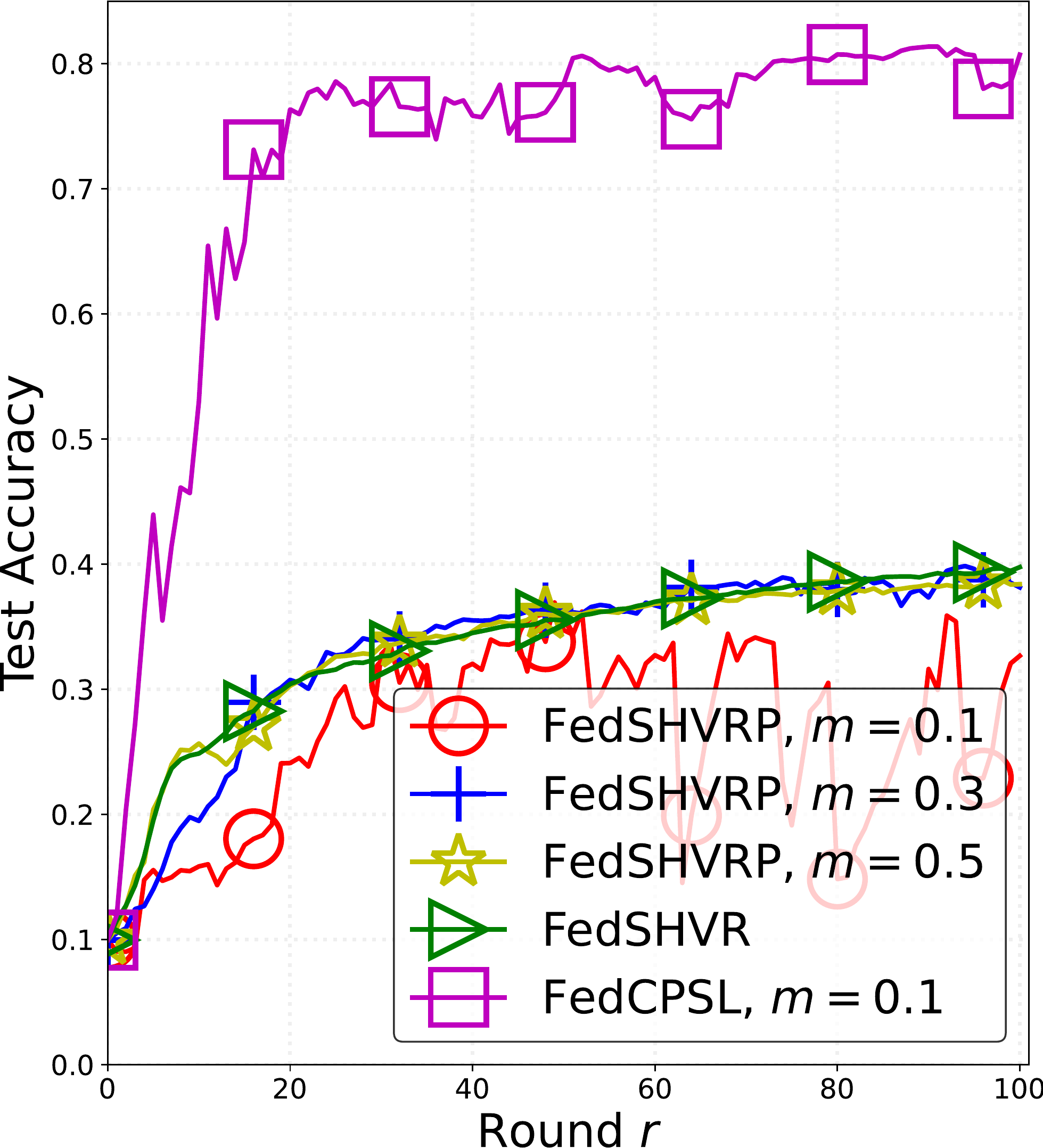}}
	\caption{Test accuracy versus number of rounds of FedCPSL, FedSHVRP and FedSHVR.}\label{fig: comp_FedSHVRP}
\end{figure}

\begin{figure}[t!]
	\centering
	\subfigure[\scriptsize MNIST, $m=0.1N$]{\includegraphics[width=6.6cm]{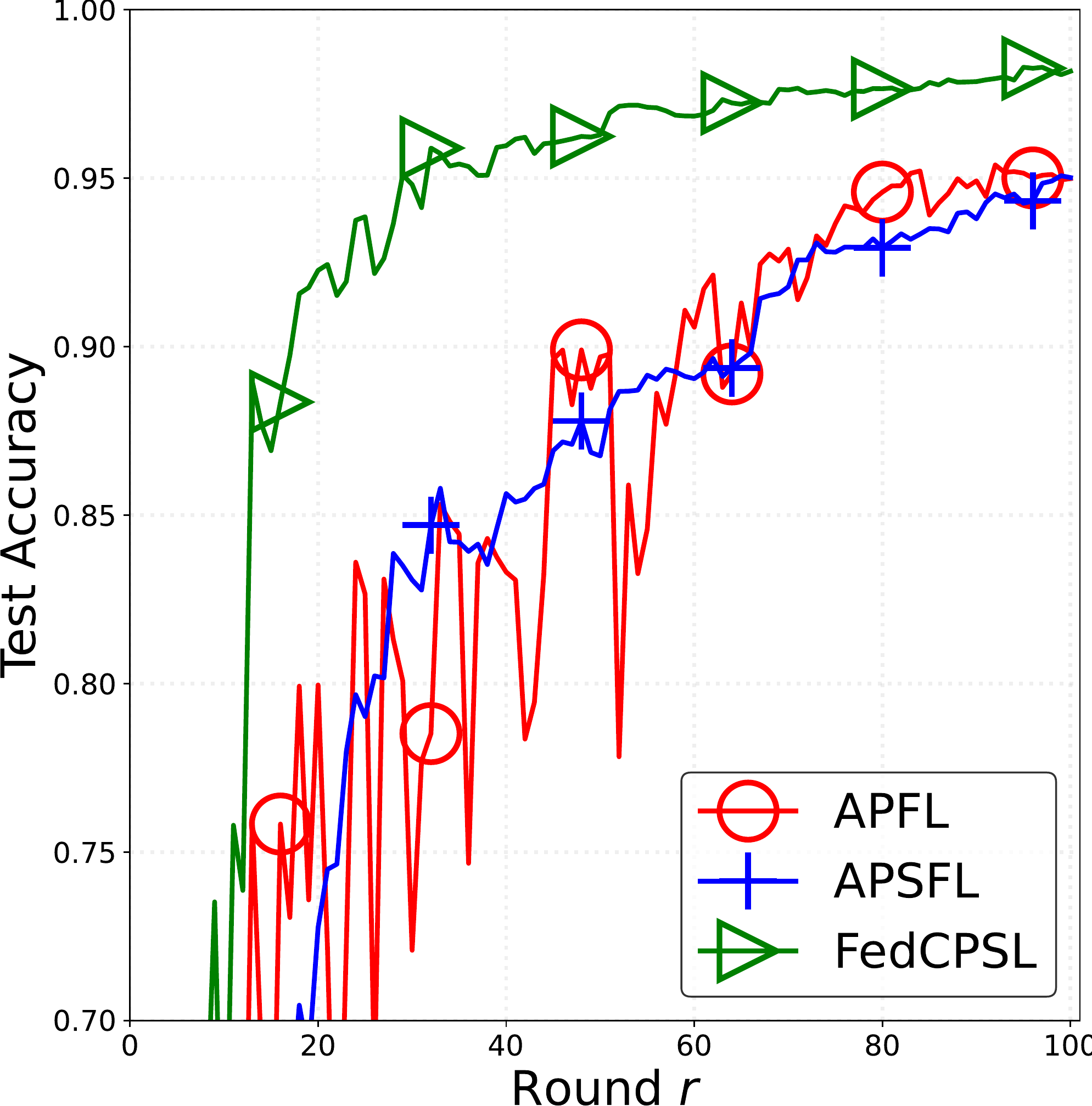} \label{fig: comp_mnist_a}}	
	\subfigure[\scriptsize MNIST, $m=0.2N$]{\includegraphics[width=6.6cm]{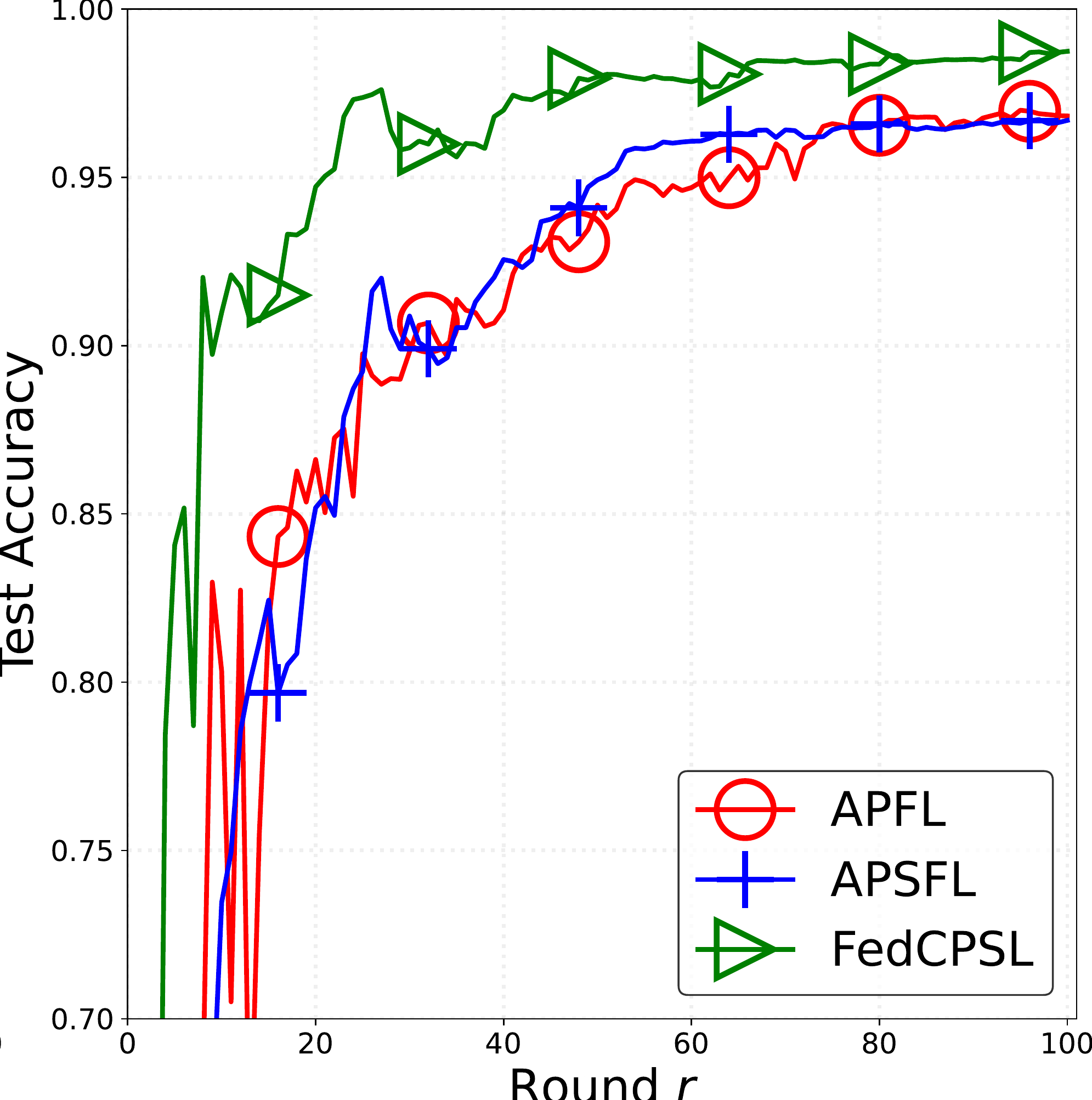} \label{fig: comp_mnist_b}}
	\subfigure[\scriptsize CIFAR-10, $m=0.1N$]{\includegraphics[width=6.6cm]{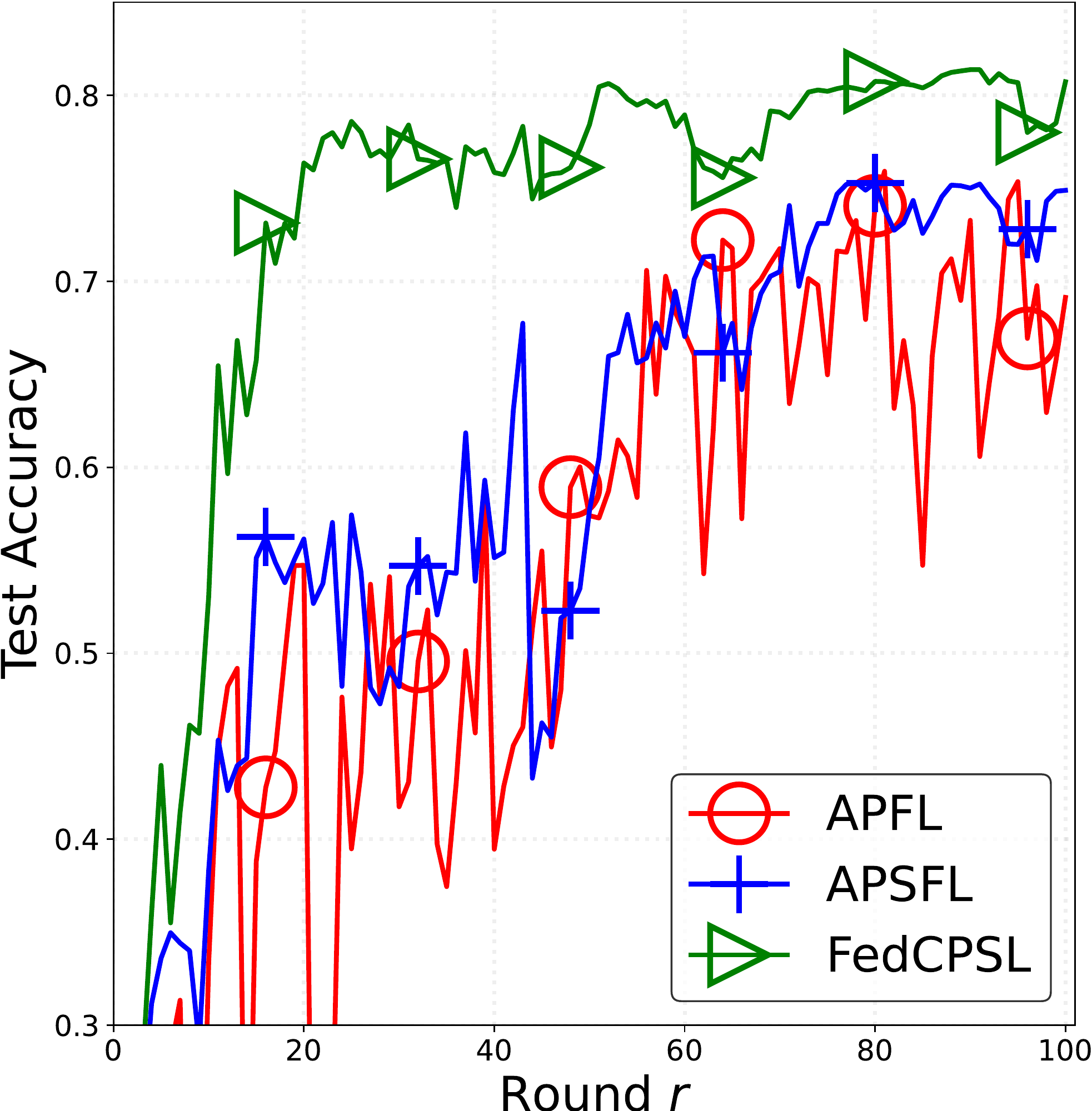} \label{fig: comp_cifar_c}}
	\subfigure[\scriptsize  CIFAR-10, $m=0.2N$]{\includegraphics[width=6.6cm]{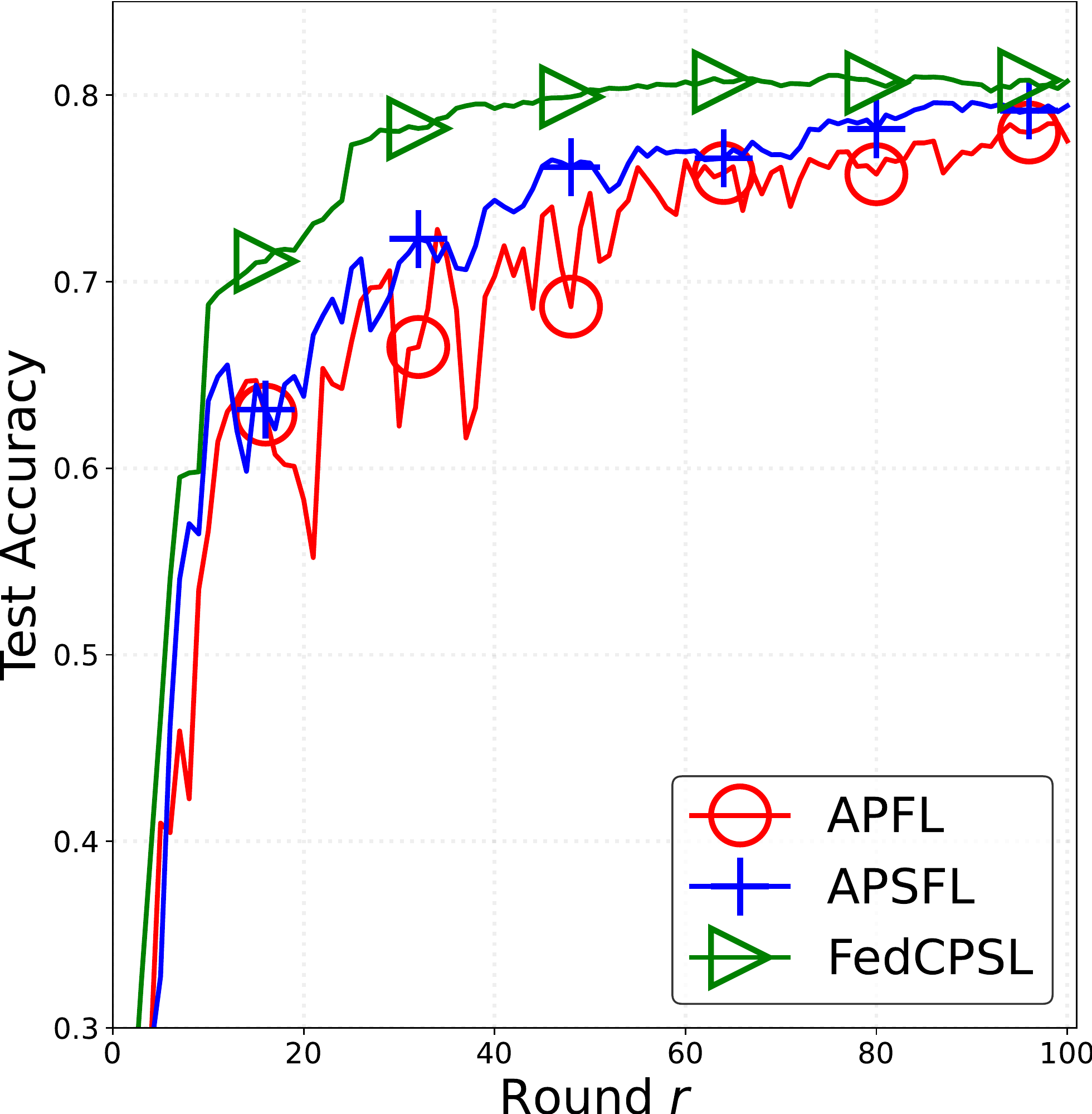}\label{fig: comp_cifar_d}}
	\caption{Test accuracy versus number of rounds of FedCPSL, APFL and APSFL.}\label{fig: comp_APFL}
\end{figure}

\begin{figure}[t!]
	\centering
	\subfigure[\scriptsize MNIST, $m=0.1N$]{\includegraphics[width=6.6cm]{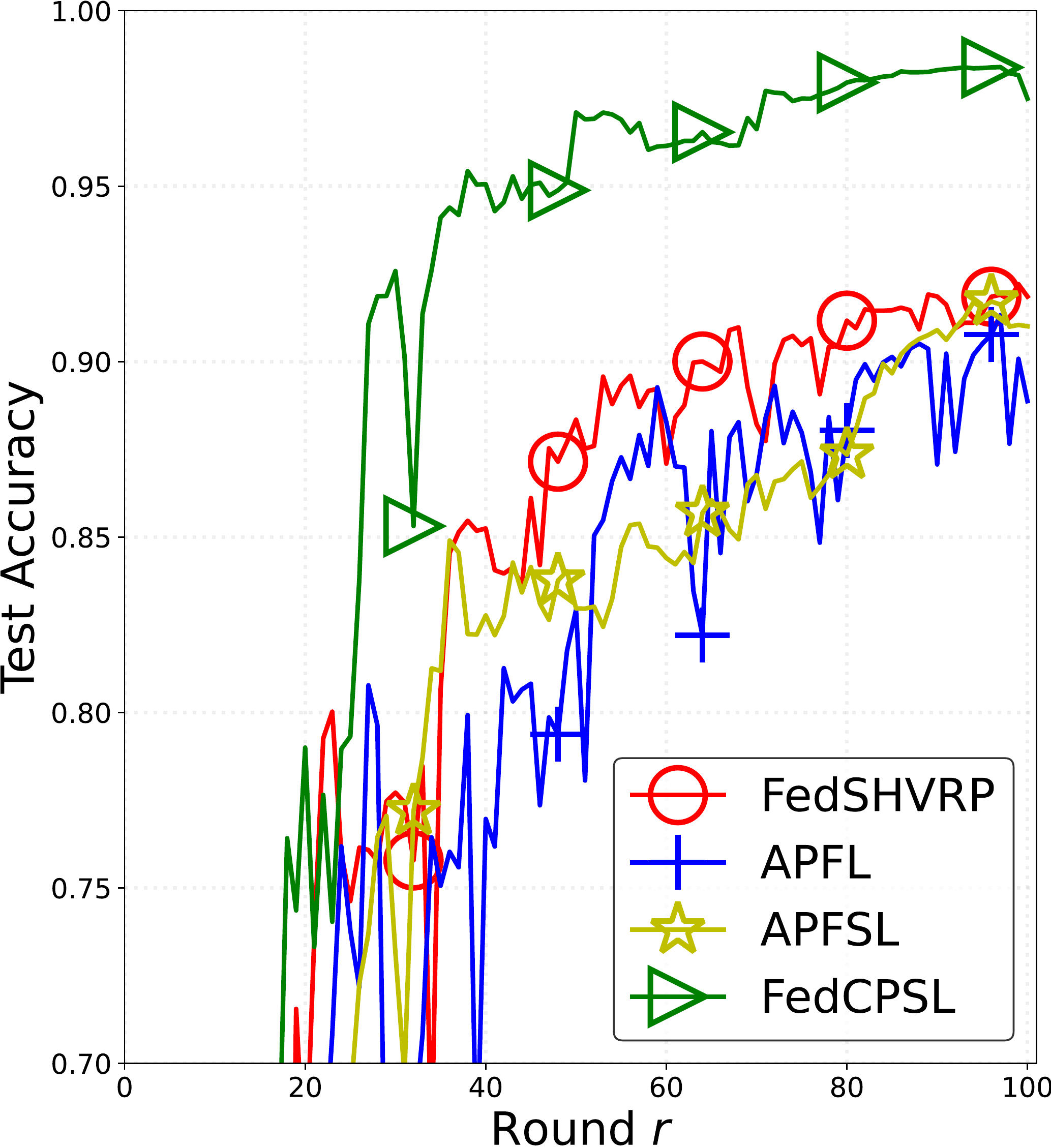} \label{fig: comp_mnist_hlu}}	
	\subfigure[\scriptsize CIFAR-10, $m=0.1N$]{\includegraphics[width=6.6cm]{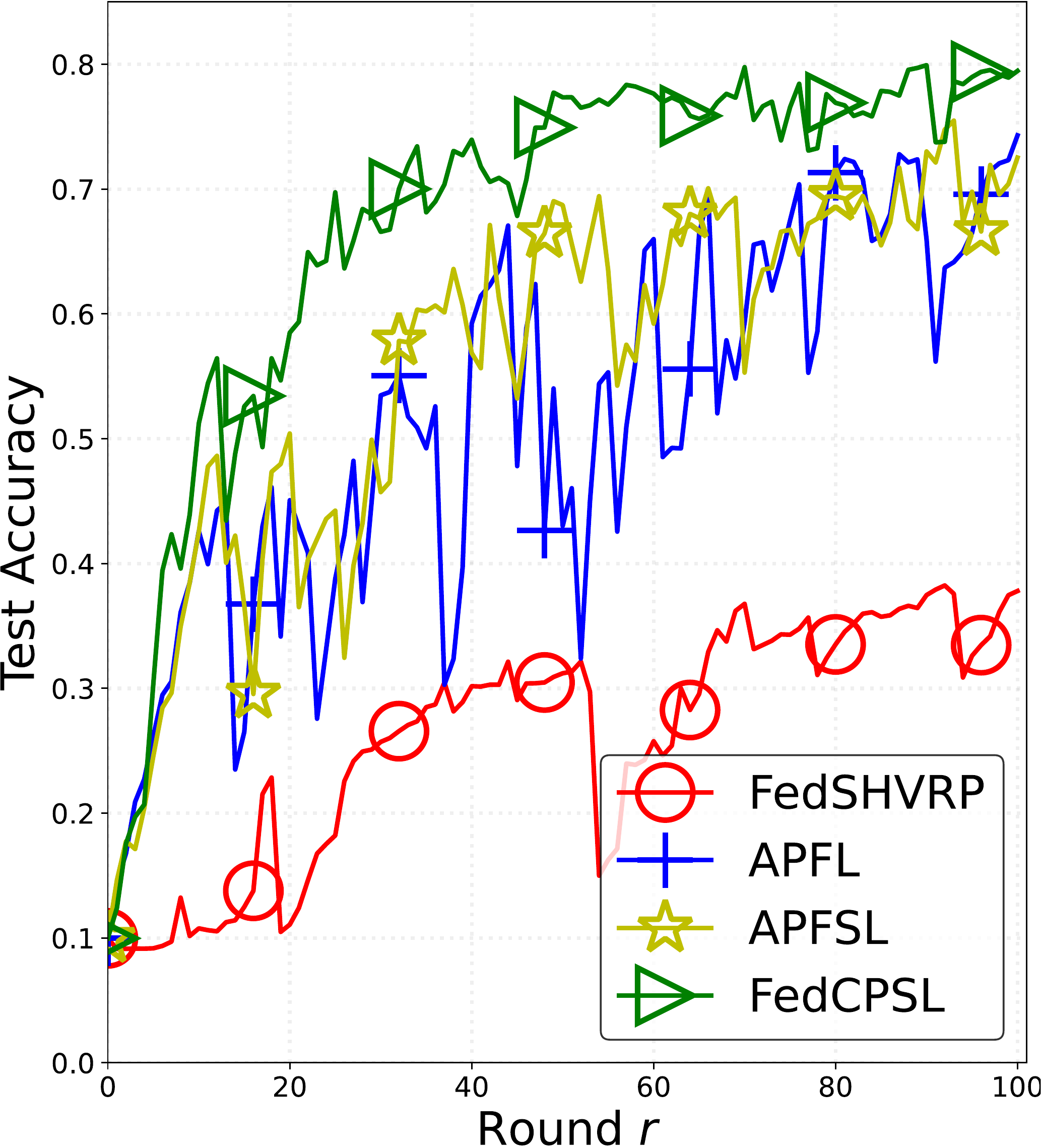} \label{fig: comp_cifar_hlu}}
	\caption{Test accuracy versus number of rounds of FedCPSL, FedSHVRP, APFL and APSFL under {\bf HLU}.}\label{fig: comp_hlu}
\end{figure}

\subsection{The effect of local momentum}
We further examine the effect of local momentum on the performance of the proposed FedCPSL algorithm by considering different values of $\gamma$. Fig. \ref{fig: FedCPSL_momentum} displays the corresponding test accuracy of FedCPSL versus the number of communication rounds on the MNIST dataset.  One can see from both Fig. \ref{fig: momentum_a} and Fig. \ref{fig: momentum_b} that a non-zero $\gamma$ can speed up the convergence of FedCPSL and FedCPSL with $\gamma = 0.8$ can quickly reach a higher test accuracy with fewer communication rounds than almost all other values of $\gamma$ under test. Such results corroborate our thoughts that local momentum could speed up the algorithm convergence and improve the performance of FedCPSL.

\subsection{Performance comparison }

In this subsection, we compare our proposed FedCPSL algorithm against the four aforementioned baseline algorithms on both the MNIST and CIFAR-10 datasets. Based on above results, we choose $\beta_i = 0.75, i \in [N]$ and $\gamma = 0.8$ for FedCPSL. The mix ratios used in APFL and APFSL are also set as $0.75$. Fig. \ref{fig: comp_FedSHVRP} displays the comparison results of FedCPSL against FedSHVRP and FedSHVR while Fig. \ref{fig: comp_APFL} displays the comparison results of FedCPSL against APFL and APSFL. 

First of all, one can see from Fig. \ref{fig: comp_FedSHVRP} that on both datasets the proposed FedCPSL algorithm with only $10\%$ of all edge devices being active performs significantly better than FedSHVRP with all values of $m$ under test. Although, with increased $m$, FedSHVRP obtains improved convergence speed and test performance, it still has a big performance gap with our proposed FedCPSL algorithm. Note that FedSHVR corresponds to FedSHVRP with $m = N$. The proposed FedCPSL algorithm is also quite effective in reducing the communication cost because even a much smaller value of $m$ ($m = 0.1N$) can yield better performance than FedSHVRP. Since both FedSHVRP and FedSHVR employ the global model as the personalized model, this again demonstrates the necessity and superiority of model personalization, especially in the presence of heterogeneous edge devices.

Then, as shown in Fig. \ref{fig: comp_APFL}, the proposed FedCPSL algorithm greatly outperforms APFL and APSFL on both the MNIST and CIFAR-10 datasets. On one hand, it can be seen from Fig. \ref{fig: comp_mnist_a} and Fig. \ref{fig: comp_cifar_c} that the proposed FedCPSL algorithm and APSFL yield a much more stable convergence and perform better than APFL. This demonstrates the advantage of semi-supervised FL which leverages both labeled and unlabeled data. Similar results can be observed in Fig. \ref{fig: comp_mnist_b} and Fig. \ref{fig: comp_cifar_d}. On the other hand, we also see that the proposed FedCPSL algorithm usually converges much faster to a high test accuracy than the others. In particular, as seen in Fig. \ref{fig: comp_mnist_a} (resp. Fig. \ref{fig: comp_cifar_c}), FedCPSL takes about 30 (resp. 21) rounds to achieve $95\%$ test accuracy, whereas the others need more than 80 rounds to obtain such performance. In summary, the number of communication rounds needed to reach a high accuracy is reduced by up to $60\%$. 

We further examine these algorithms under the scenario of {\bf HLU}. Fig. \ref{fig: comp_hlu} depicts the convergence performance of FedCPSL against the baseline algorithms when $m = 0.1N$ and {\bf HLU} is applied.  We can observe from Fig. \ref{fig: comp_hlu} that, when {\bf HLU} is considered, all algorithms under test would have a performance degradation and their convergence curves are less stable than that without {\bf HLU}. However, using {\bf HLU} has a much less influence on the convergence performance of our proposed FedCPSL and FedCPSL still performs much better than FedSHVRP, APFL and APFSL in terms of both convergence speed and application performance. This demonstrates the superior robustness of our proposed FedCPSL algorithm to {\bf HLU} over the others. Table \ref{table: test accuracy} shows the detailed test accuracy achieved by all the five algorithms under test. One can observe that the proposed FedCPSL algorithm performs the best and achieves the highest test accuracy for most cases. Moreover, there exists a considerable performance gap between FedCPSL and the others. One can also notice that, under {\bf HLU}, the APFL algorithm performs slightly better than FedCPSL in terms of test accuracy on the CIFAR-10 data with $m = 0.2N$. But FedCPSL converges much faster and it achieves that test accuracy using much less communication rounds.

\begin{table}[t!] 
	\centering 
	%	\scriptsize
	\caption{Test accuracy (\%) of the considered five algorithms on the Two datasets with two values of $m$. %Note that KM stands for K-means and all three
	}\vspace{-0.1cm}
	\setlength{\tabcolsep}{3.0mm}
	\label{table: test accuracy}
	\begin{tabular}{|c|c|c|c|c|}
		\hline    \rowcolor{gray!50}       
		Dataset      &     \tabincell{c}{ MNIST \\ ($m=0.1N$)  }      &       \tabincell{c}{ MNIST \\ ($m=0.2N$)  }     &         \tabincell{c}{ CIFAR-10 \\ ($m=0.1N$)  }       &    \tabincell{c}{CIFAR-10 \\ ($m=0.2N$)  }\\  \hline\hline 
		FedSHVRP     &     93.8  &      85.6       &     35.4  &   46.8  \\ \cline{1-5}
		FedSHVR       &    94.3     &   94.3   &        39.8   &  39.8  \\ \cline{1-5}
		APFL       &     94.9      &       96.8   &         69.1   &     77.6  \\ \cline{1-5}
		%			dist-kzc          & 13.8            &   20.4        &      48.5   &   33.4 \\ \cline{1-5}
		APSFL     &     95.1         &      96.7     &     74.9  &  79.5   \\ \cline{1-5}
		%			\tabincell{c}{FedMAvg ($m = 30$)}       &     84.6         &   52.9      &     58.1  &   47.6   \\ \cline{1-5}
		FedCPSL      &     \textbf{98.2}            &    \textbf{98.8}     &   \textbf{80.7}    &    \textbf{80.8}   \\ \cline{1-5}			     
		\hline \hline
		FedSHVRP ({\bf HLU})     &     91.8  &      91.8      &     37.8  &   38.3  \\ \cline{1-5}
		FedSHVR  ({\bf HLU})      &    73.0    &   73.0 &     31.2     &  31.2  \\ \cline{1-5}
		APFL  ({\bf HLU})      &     88.9      &       95.3   &         74.3   &      \textbf{81.1} \\ \cline{1-5}
		%			dist-kzc          & 13.8            &   20.4        &      48.5   &   33.4 \\ \cline{1-5}
		APSFL  ({\bf HLU})    &     91.0         &      95.1     &     72.5  &  78.7   \\ \cline{1-5}
		%			\tabincell{c}{FedMAvg ($m = 30$)}       &     84.6         &   52.9      &     58.1  &   47.6   \\ \cline{1-5}
		FedCPSL  ({\bf HLU})     &     \textbf{97.5}            &    \textbf{98.5}     &   \textbf{79.5}   &   80.3  \\ \cline{1-5}			     
		\hline
	\end{tabular}
	\vspace{-0.30cm}
\end{table}

\section{Conclusion}\label{sec: conclusion}
To address the challenges of label deficiency and device heterogeneity in wireless edge networks, we have investigated the PSSFL problem in this paper and proposed an efficient and effective algorithm, FedCPSL. Specifically, we rely on the technique of pseudo-labeling and interpolation based model personalization to model the PSSFL problem. The proposed FedCPSL algorithm incorporates the strategies of adaptive client variance reduction and normalized global aggregation to mitigate the adverse effects of heterogeneity on algorithm convergence. It also adopts the technique of local momentum to further boost the algorithm convergence. Theoretically, we have built the convergence property of FedCPSL, which shows that FedCPSL is resilient to the challenge of device heterogeneity and can converge in a sublinear manner. The experimental results have also demonstrated the superiority of FedCPSL over the existing methods with respect to both convergence speed and application performance.

% or


\appendices

\section{Proof of Theorem 1} \label{appdix: theorem1}

Before delving into the proof of Theorem 1, we define the virtual sequence $ \{(\wt \thetab_i^{r}, \wt \nub_i^r, \wt \cb_i^{r})\}$ by assuming that all edge devices are active at round $r$, i.e., $\forall i, 0 \leq t \leq Q_i^r - 1$,
\begin{small}
	\begin{subequations}
		\begin{align}
			& \wt \nub_i^{r+1} = \arg\min_{\substack{\nub_i \in \Vc_i}} ~f_i(\thetab^{r}, \nub_i)  \label{eq: virtual_vi},\\
			&\wt \thetab_i^{r, 0} = \thetab^{r}, \wt \mub_i^{r, 0} = \zerob, \label{eq: virtual_thetai}\\
			&\wt \mub_i^{r, t+1} = \gamma \wt \mub_i^{r, t} + g_i(\wt \thetab_i^{r,t}, \wt \nub_i^{r+1}) - \cb_i^r + \cb^r, \\
			& \wt \thetab_i^{r, t+1} =\wt \thetab_i^{r, t} - \eta \wt \mub_i^{r, t+1} ,\\
			& \wt \thetab_{i}^{r+1}=\wt \thetab_i^{r, Q_i^r}, \label{eqn: virtual_xi}\\
			&\wt \cb_i^{r+1} =\cb_i^r - \bigg(\cb^r + \frac{\wt \thetab_i^{r+1}-\thetab^{r} }{\eta \wt Q_i^r}\bigg). \label{eqn: def_virtual_dual}
		\end{align}
	\end{subequations}
\end{small}The following additional terms are introduced for ease of analysis.
\begin{small}
	\begin{align}
		&\Psi^r \triangleq \sum_{i=1}^{N}\omega_i\E\bigg[\sum_{t = 0}^{Q_i^r-1} \frac{(\bb_i^r)^t}{\wt Q_i^r} \|\wt \thetab_i^{r,t}-\thetab^r \|^2\bigg],\label{def: Psi_i}\\
		&\Xi^r \triangleq \sum_{i=1}^{N}\omega_i\E[\|\nabla_{\thetab} f_i(\thetab^r, \nub_i^{r}) - \cb_i^r\|^2], \label{def: Xi_i} \\
		&\Phi^r \triangleq \sum_{i=1}^{N}\omega_i\E[\|\wt \nub_i^{r+1}- \nub_i^r\|^2]. \label{def: Phi_i}
	\end{align}
\end{small}
We also build the following potential function
\begin{align} \label{eqn: def_potential}
	P^r = \E[f(\thetab^r, \nub^r)] +  \frac{24\wt \eta^2  N^2L}{m^2} \Xi^r. 
\end{align}where $\wt \eta \triangleq \eta \eta_g$. Then, we proceed to the proof of Theorem 1 with the help of the following three lemmas, while the detailed proofs of the lemmas refer to the supplementary materials.

\begin{Lemma} \label{lem: diff_f}
	For any round $r$, it holds that
	\begin{small}
		\begin{align}
			&\E[f(\thetab^{r+1}, \nub^{r+1})] - \E[f(\thetab^{r}, \nub^{r})] \leq  -  \bigg(\frac{\wt \eta}{2}-\frac{ 4\wt \eta^2 N L}{m}\bigg)\E[\| \nabla_{\thetab}f(\thetab^r, \nub^{r})\|^2]   +\bigg(\frac{\wt \eta L^2}{2} +  \frac{4 \wt \eta^2 NL^3}{m}\bigg)\Psi^r \notag \\
			&\qquad\qquad\qquad\qquad\qquad - \bigg(\bigg(\frac{\mu}{2} + \frac{1}{\eta_v} - \frac{3L}{2}\bigg)\frac{m}{N}- \frac{\wt \eta  L^2 }{2} -  \frac{4 \wt \eta^2 NL^3}{m}\bigg) \Phi^r + \frac{8 \wt \eta^2  NL}{m}\Xi^r + \frac{\wt \eta^2\sigma^2N L}{mS}.\label{lem: diff_f_bd}
		\end{align}
	\end{small}
\end{Lemma}

\begin{Lemma} \label{lem: opt_diff}
	For any round $r$, if $\wt \eta \leq \frac{m}{32NL} (1 + \frac{2N}{ m})^{-\frac{1}{2}}$, it holds that
	\begin{small}
		\begin{align}
			\Xi^{r+1} - \Xi^r \leq & \bigg(\frac{257}{256}+ \frac{ m}{2N} \bigg)\frac{m L^2 }{N}\Psi^r - \frac{m}{2N}\bigg(\frac{63}{64} + \frac{ m}{N}\bigg)\Xi^r + \frac{mL^2}{256N} \Phi^r \notag \\
			&+ \frac{ m}{256N}\E[\|\nabla_{\thetab} f(\thetab^r, \nub^{r}) \|^2] + \bigg(1 + \frac{m}{2N}\bigg)\frac{m\sigma^2}{NS}+  \frac{ m\sigma^2}{1024NS} .\label{lem: opt_diff_bd}
		\end{align}
	\end{small}
\end{Lemma}

\begin{Lemma} \label{lem: model_diff}
	Let $\ol Q = \max\limits_{i, r} \wt Q_i^r$, for any round, if $\eta \leq \frac{1}{\sqrt{5}L\ol Q}$, it holds that
	\begin{small}
		\begin{align}
			\Psi^r 
			\leq &  5\eta^2\ol Q^2\E[\|\nabla_{\thetab} f(\thetab^r, \nub^{r}) \|^2]+ \frac{5\eta^2\ol Q^2\sigma^2 }{4S} + 5 \eta^2 \ol Q^2L^2\Phi^r+ 10\eta^2 \ol Q^2\Xi^r,\label{lem: model_diff_bd}
		\end{align}
	\end{small}
\end{Lemma}

In particular, combining the results \eqref{lem: diff_f_bd} and \eqref{lem: opt_diff_bd} yields
\begin{small}
	\begin{align}
		P^{r+1} - P^r 
		\leq & -\bigg(\frac{\wt \eta}{2}-\frac{ 131\wt \eta^2 N L}{32m}\bigg)\E[\| \nabla_{\thetab}f(\thetab^r, \nub^{r})\|^2]  +\bigg(\frac{\wt \eta L^2}{2} +  \frac{4 \wt \eta^2 NL^3}{m}+\bigg(\frac{257}{32}+ \frac{4 m}{N} \bigg)\frac{3 \wt \eta^2 NL^3 }{ m}\bigg)\Psi^r  \notag \\
		&-  \bigg(\bigg(\frac{\mu}{2} + \frac{1}{\eta_v} - \frac{3L}{2}\bigg)\frac{m}{N}- \frac{\wt \eta  L^2 }{2} -  \frac{131 \wt \eta^2 NL^3 }{32m} \bigg)\Phi^r -\bigg(\frac{61}{16 }+ \frac{12 m}{N}\bigg)\frac{\wt \eta^2NL}{m}\Xi^r  + \frac{131\wt \eta^2N L\sigma^2}{128mS}  \notag \\
		&+\bigg(1 + \frac{ m}{2N}\bigg)\frac{24\wt \eta^2 NL\sigma^2}{ mS}. \label{thm1: bd1}
	\end{align}
\end{small}We multiply \eqref{lem: model_diff_bd} by $2\wt \eta L^2$ and then combine it with  \eqref{thm1: bd1} to obtain
\begin{small}
	\begin{align}
		&P^{r+1} \!-\! P^r \notag \\
		&\leq -  C_1\E[\| \nabla_{\thetab}f(\thetab^r, \nub^{r})\|^2] \!-\! C_2\Psi^r \!-\! C_3\Xi^r -  C_4\Phi^r + \bigg(\frac{259}{64} + \frac{ m}{N}\bigg)\frac{12\wt \eta^2 NL\sigma^2}{ mS} +\frac{5\eta^2 \wt \eta \ol Q^2L^2 \sigma^2}{2S}.\label{thm1: bd2}
	\end{align}
\end{small}where 
\begin{small}
	\begin{align}
		&C_1 \triangleq \frac{\wt \eta}{2}-\frac{ 131\wt \eta^2 N L}{32m} - 10\wt \eta \eta^2L^2\ol Q^2, \\
		&C_2 \triangleq \frac{3\wt \eta L^2}{2} - \frac{4 \wt \eta^2 NL^3}{m}-\bigg(\frac{257}{32}+ \frac{4 m}{N} \bigg)\frac{3 \wt \eta^2 NL^3 }{ m},\\
		&C_3 \triangleq \bigg(\frac{61}{16 }+ \frac{12 m}{N}\bigg)\frac{\wt \eta^2NL}{m} -20 \wt \eta \eta^2  L^2\ol Q^2, \\
		&C_4 \triangleq \bigg(\frac{\mu}{2} + \frac{1}{\eta_v} - \frac{L}{2}\bigg)\frac{m}{N} - \frac{\wt \eta  L^2 }{2} -  \frac{131 \wt \eta^2 NL^3 }{32m} -  10\wt \eta\eta^2 L^4\ol Q^2.
	\end{align}
\end{small}In order to get the result, we require the following four conditions
\begin{small}
	\begin{align}
		C_1 \geq  \frac{\wt \eta }{4}, ~C_2 \geq 0,~ C_3 \geq 0,~ C_4 \geq 0.
	\end{align}
\end{small}Then, we respectively show that these conditions hold. 

\underline{Condition I:  $C_1 \geq  \frac{\wt \eta }{4}$:} As $\wt \eta \leq \frac{m}{48NL}$, it suffices to have
\begin{small}
	\begin{align}
		\frac{\wt \eta}{4}-\frac{131 \wt \eta^2 N L}{32m}- 10\wt \eta \eta^2L^2\ol Q^2
		\geq & 	\frac{\wt \eta}{4}\bigg(1 - \frac{131}{8 \times 48}-40 \eta^2L^2\ol Q^2\bigg) \\
		\geq & \frac{\wt \eta}{4}\bigg(\frac{253}{8 \times 48}-40 \eta^2L^2\ol Q^2 \bigg) \\
		\geq & \frac{\wt \eta}{4}\bigg(\frac{40}{64}-40 \eta^2L^2Q^2\bigg) \geq 0,\label{thm1: c1}
	\end{align}
\end{small}which is equivalent to $\eta \leq \frac{1}{8 L\ol Q}$. Thus, the condition $C_1 \geq  \frac{\wt \eta }{4}$ is true. 

\underline{Condition II:  $C_2\geq  0$:} Similarly, we have
\begin{small}
	\begin{align}
		C_2 &= \frac{3\wt \eta L^2}{2}\bigg(1 - \frac{8 \wt \eta NL}{3m}-\bigg(\frac{257}{16}+ \frac{4m}{N} \bigg)\frac{\wt \eta NL }{m}\bigg) \\
		&\geq  \frac{3\wt \eta L^2}{2}\bigg(1 -\frac{23\wt \eta NL }{m}\bigg) \\
		&\geq  \frac{3\wt \eta L^2}{4}, \label{thm1: c2}
	\end{align}
\end{small}where \eqref{thm1: c2} follows because $\wt \eta \leq \frac{m}{48NL}$. 

\underline{Condition III:  $C_3\geq  0$:} It suffices to have
\begin{small}
	\begin{align}
		\bigg(\frac{61}{16 }+ \frac{12 m}{N}\bigg)\frac{\wt \eta^2NL}{m} -20 \wt \eta \eta^2  L^2\ol Q^2 
		\geq 	 2 \wt \eta \eta L \bigg( \frac{9\eta_g N}{5m} - 10\eta L \ol Q^2\bigg) + 12 \wt \eta^2 L
		\geq   \frac{3\wt \eta^2 LN}{m} +  12 \wt \eta^2 L,\label{thm1: c3}
	\end{align}
\end{small}where \eqref{thm1: c3} follows because $\eta \leq \frac{3\eta_g N}{100mL\ol Q^2}$. Thus, the condition $C_3 \geq 0$ is also true. 

\underline{Condition IV:  $C_4\geq  0$:} We have
\begin{small}
	\begin{align}
		C_4 &= \bigg(\frac{\mu}{2} + \frac{1}{\eta_v} - \frac{3L}{2}\bigg)\frac{m}{N} - \frac{\wt \eta  L^2 }{2} - \frac{131 \wt \eta^2 NL^3 }{32m} -  10\wt \eta\eta^2 L^4\ol Q^2 \\
		& \geq \frac{ mL}{2N} - \frac{\wt \eta  L^2 }{2} - \frac{\wt \eta  L^2 }{4} + \frac{\mu m}{2N}\\
		& \geq \frac{ 31mL}{64N}+ \frac{\mu m}{2N},
	\end{align}
\end{small}where the first inequality holds by \eqref{thm1: c1} and the fact that $\eta_v \leq \frac{1}{2L}$, the second inequality follows because $\wt \eta \leq \frac{m}{48NL}$.
Therefore, we have from \eqref{thm1: bd2} that
\begin{small}
	\begin{align}
		P^{r+1} - P^r \leq -\frac{\wt \eta}{4}\E[\| \nabla_{\thetab}f(\thetab^r, \nub^{r})\|^2] 
		+ \bigg(\frac{259}{64}  + \frac{ m}{N} + \frac{5\eta mL\ol Q^2}{24\eta_g N}\bigg)\frac{12\wt \eta^2 NL\sigma^2}{ mS},
	\end{align}
\end{small}which implies that
\begin{small}
	\begin{align}
		\E[\| \nabla_{\thetab}f(\thetab^r, \nub^{r})\|^2] \leq \frac{4(P^r - P^{r+1})}{\wt \eta} - \frac{31mL}{64N}\Phi^r + \bigg(\frac{259}{64} + \frac{ m}{N} + \frac{5\eta mL\ol Q^2}{24\eta_g N}\bigg)\frac{12\wt \eta NL\sigma^2}{ mS} . \label{thm1: bd3}
	\end{align}
\end{small}Summing \eqref{thm1: bd3} up from $r = 0$ to $R - 1$ and dividing it by $R$ yields
\begin{small}
	\begin{align}
		\frac{1}{R}\sum_{r = 0}^{R-1}\bigg(\E[\| \nabla_{\thetab}f(\thetab^r, \nub^{r})\|^2] +\frac{31L}{64}\sum_{i=1}^{N}\omega_i\E[\| \nub_i^{r+1} \!-\! \nub_i^r\|^2] \bigg)
		\leq  \frac{4(P^0 - \ul f)}{\wt \eta R}+ \bigg(\frac{259}{64} \!+\! \frac{ m}{N} \!+\! \frac{5\eta mL\ol Q^2}{24\eta_g N}\bigg)\frac{12\wt \eta NL\sigma^2}{mS}.
	\end{align}
\end{small}
This completes the proof. \hfill $\blacksquare$

\section{Poof of Lemma 1}
By Assumption 2, we have
\begin{small}
	\begin{align}
		&\E[f(\thetab^r, \nub^{r}) - f(\thetab^r, \nub^{r+1})] \notag\\
		& = \E\bigg[ \sum_{i=1}^{N}\omega_i\Ibb_{\Ac^r}^i(f_i(\thetab^r, \nub_i^{r}) - f_i(\thetab^r, \wt \nub_i^{r+1}))\bigg] \\
		& \geq  \E\bigg[\sum_{i=1}^{N}\omega_i\Ibb_{\Ac^r}^i\bigg(\langle \nabla_{\nub_i} f_i(\thetab^r, \wt \nub_{i}^{r+1}), \nub_{i}^r - \wt \nub_{i}^{r+1}\rangle +  \frac{\mu}{2}\|\wt \nub_i^{r+1} - \nub_i^r\|^2\bigg)\bigg] \label{lem1: bd1}\\
		& =  \E\bigg[\sum_{i=1}^{N}\omega_i\Ibb_{\Ac^r}^i\bigg(\langle \nabla_{\nub_i} f_i(\thetab^r, \wt \nub_{i}^{r+1}) \!-\! \nabla_{\nub_i} f_i(\thetab^r, \nub_{i}^{r}), \nub_{i}^r \!-\! \wt \nub_{i}^{r+1}\rangle +\langle \nabla_{\nub_i} f_i(\thetab^r, \nub_{i}^{r}), \nub_{i}^r \!-\! \wt \nub_{i}^{r+1}\rangle +  \frac{\mu}{2}\|\wt \nub_i^{r+1} \!-\! \nub_i^r\|^2\bigg)\bigg] \label{lem1: bd1_1}\\
		& \geq   \bigg(\frac{\mu}{2} + \frac{1}{\eta_v} - L\bigg)\frac{m}{N} \sum_{i=1}^{N}\omega_i\E[\|\wt \nub_i^{r+1} - \nub_i^r\|^2],\label{lem1: bd2}
	\end{align}
\end{small}where $\Ibb_{\Ac^r}^i$ is the indicator denoting whether $i \in \Ac^r$ or not; \eqref{lem1: bd1} follows by Assumption 2;\eqref{lem1: bd2} holds because $\Prob(i \in \Ac^r) = \frac{m}{N}$, (10) and the Cauchy-Schwarz inequality. Then, by Assumption 3,  $f_i(\thetab, \nub_i)$ is L-smooth with respect to $\thetab$, and thus we obtain
\begin{small}
	\begin{align}
		&\E[f(\thetab^{r+1}, \nub^{r+1})] - \E[f(\thetab^{r}, \nub^{r+1})] \notag\\
		& =  \sum_{i=1}^{N}\omega_i(\E[f_i(\thetab^{r+1}, \nub_i^{r+1})] - \E[f_i(\thetab^{r}, \nub_i^{r+1})])\notag \\
		& \leq  \sum_{i=1}^{N}\omega_i\bigg( \E[\langle \nabla_{\thetab}f_i(\thetab^r, \nub_i^{r+1}), \thetab^{r+1} - \thetab^r\rangle] + \frac{L}{2}\E[\|\thetab^{r+1} - \thetab^r\|^2]\bigg) \notag \\
		& =  \sum_{i=1}^{N}\omega_i\bigg(\E[\langle \nabla_{\thetab}f_i(\thetab^r, \nub_i^{r}), \thetab^{r+1} - \thetab^r\rangle]+ \frac{L}{2}\E[\|\thetab^{r+1} - \thetab^r\|^2] + \E[\langle \nabla_{\thetab}f_i(\thetab^r, \nub_i^{r+1}) - \nabla_{\thetab}f_i(\thetab^r, \nub_i^{r}), \thetab^{r+1} - \thetab^r\rangle] \bigg)\notag \\
		& \leq \sum_{i=1}^{N}\omega_i\bigg(  \E[\langle \nabla_{\thetab}f_i(\thetab^r, \nub_i^{r}), \thetab^{r+1} - \thetab^r\rangle] +  L\E[\|\thetab^{r+1} - \thetab^r\|^2] +\frac{1}{2L}\E[\| \nabla_{\thetab}f_i(\thetab^r, \nub_i^{r+1}) - \nabla_{\thetab}f_i(\thetab^r, \nub_i^{r})\|^2] \bigg) \label{lem1: bd3}\\
		& \leq   \E[\langle \nabla_{\thetab}f(\thetab^r, \nub^{r}), \thetab^{r+1} - \thetab^r\rangle] +\frac{L}{2}\sum_{i=1}^{N}\omega_i\E[\| \nub_i^{r+1} - \nub_i^r \|^2] +  L\E[\|\thetab^{r+1} - \thetab^r\|^2] \notag \\
		& = \frac{mL}{2N}\Phi^r + \E[\langle \nabla_{\thetab}f(\thetab^r, \nub^{r}), \thetab^{r+1} - \thetab^r\rangle] + L\E[\|\thetab^{r+1} - \thetab^r\|^2], \label{lem1: bd4}
	\end{align} 
\end{small}where \eqref{lem1: bd3} holds thanks to the Jensen's Inequality; \eqref{lem1: bd4} follows because $\Prob(i \in \Ac^r) = \frac{m}{N}$. The last two terms in the RHS of \eqref{lem1: bd2} can be further bounded using the following lemma.

\begin{Lemma} \label{lem: diff_x}
	For any round $r$, it holds that
	\begin{small}
		\begin{align}
			\E[\thetab^r - \thetab^{r+1}] &=  \wt \eta \sum_{i=1}^{N}\omega_i\E\bigg[\sum_{t = 0}^{Q_i^r-1} \frac{(\bb_i^r)^t}{\wt Q_i^r} \nabla_{\thetab} f_i(\wt \thetab_i^{r,t}, \wt \nub_i^{r+1}) \bigg],  \label{lem: diff_theta0_bd1}\\
			\E[\|\thetab^{r+1} - \thetab^r\|^2] &\leq  \frac{4 \wt \eta^2 NL^2}{m}\Psi^r+ \frac{8 \wt \eta^2 N}{m}\Xi^r + \frac{ 4\wt \eta^2 N}{m}\E[\|\nabla_{\thetab} f(\thetab^r, \nub^{r}) \|^2] + \frac{ 4\wt \eta^2 NL^2}{m} \Phi^r+ \frac{\wt \eta^2\sigma^2N}{mS}.\label{lem: diff_theta0_bd2}
		\end{align}
	\end{small}
\end{Lemma}
}
In particular, applying Lemma S.1 to \eqref{lem1: bd4} gives
\begin{small}
	\begin{align}
		\E[\langle \nabla_{\thetab}f(\thetab^r, \nub^{r}), \thetab^{r+1} - \thetab^r\rangle] 
		& = - \wt \eta\E\bigg[\bigg\langle \nabla_{\thetab}f(\thetab^r, \nub^{r}), \sum_{i=1}^{N}\omega_i\sum_{t = 0}^{Q_i^r-1}  \frac{(\bb_i^r)^t}{\wt Q_i^r}  \nabla_{\thetab} f_i(\wt \thetab_i^{r,t}, \wt \nub_i^{r+1})\bigg\rangle\bigg] \notag\\
		&= \frac{\wt \eta}{2}\E\bigg[\bigg\|\nabla_{\thetab}f(\thetab^r, \nub^{r}) - \sum_{i=1}^{N}\omega_i\sum_{t = 0}^{Q_i^r-1}  \frac{(\bb_i^r)^t}{\wt Q_i^r}  \nabla_{\thetab} f_i(\wt \thetab_i^{r,t}, \wt \nub_i^{r+1})\bigg\|^2\bigg] \notag \\
		&~~~ - \frac{\wt \eta}{2} \E\bigg[\bigg\|\sum_{i=1}^{N}\omega_i\sum_{t = 0}^{Q_i^r-1}  \frac{(\bb_i^r)^t}{\wt Q_i^r}  \nabla_{\thetab} f_i(\wt \thetab_i^{r,t}, \wt \nub_i^{r+1})\bigg\|^2\bigg] -  \frac{\wt \eta}{2}\E[\| \nabla_{\thetab}f(\thetab^r, \nub^{r})\|^2]  
		\label{lem1: bd5}\\
		& \leq -  \frac{\wt \eta}{2}\E[\| \nabla_{\thetab}f(\thetab^r, \nub^{r})\|^2]  +\frac{\wt \eta L^2}{2}\Psi^r + \frac{\wt \eta L^2}{2}  \Phi^r  \label{lem1: bd6}.
	\end{align}
\end{small}where \eqref{lem1: bd5} follows because $\langle \vb_1, \vb_2 \rangle = \frac{1}{2}\|\vb_1\|^2 + \frac{1}{2}\|\vb_2\|^2 - \frac{1}{2}\|\vb_1 - \vb_2\|^2, \forall \vb_1, \vb_2 \in \Rbb^n$; \eqref{lem1: bd6} holds by the convexity of $\|\cdot\|^2$ and Assumption 3. Substituting  \eqref{lem1: bd6} and \eqref{lem: diff_theta0_bd2} into \eqref{lem1: bd2} yields
\begin{small}
	\begin{align}
		\E[f(\thetab^{r+1}, \nub^{r+1})]-\E[f(\thetab^{r}, \nub^{r+1})]  
		&\leq  -  \bigg(\!\frac{\wt \eta}{2}-\frac{ 4\wt \eta^2 N L}{m}\bigg)\E[\| \nabla_{\thetab}f(\thetab^r, \nub^{r})\|^2] +\bigg(\!\frac{\wt \eta L^2}{2} +   \frac{4 \wt \eta^2 NL^3}{m}\!\bigg)\Psi^r\notag\\
		&~~~~+ \frac{8 \wt \eta^2  NL}{m}\Xi^r + \bigg( \frac{\wt \eta  L^2 }{2}  + \frac{mL}{2N} + \frac{4 \wt \eta^2 NL^3}{m}\!\bigg)\Phi^r + \frac{\wt \eta^2\sigma^2N L}{mS}. \label{lem1: bd7}
	\end{align} 
\end{small}Lastly, combining  \eqref{lem1: bd2} and \eqref{lem1: bd4} gives rise to
\begin{small}
	\begin{align}
		\E[f(\thetab^{r+1}, \nub^{r+1})] - \E[f(\thetab^{r}, \nub^{r})] &\leq   -  \bigg(\frac{\wt \eta}{2}-\frac{ 4\wt \eta^2 N L}{m}\bigg)\E[\| \nabla_{\thetab}f(\thetab^r, \nub^{r})\|^2]   +\bigg(\frac{\wt \eta L^2}{2} +  \frac{4 \wt \eta^2 NL^3}{m}\bigg)\Psi^r \notag\\
		&~~~~-  \bigg(\bigg(\frac{\mu}{2} + \frac{1}{\eta_v} - \frac{3L}{2}\bigg)\frac{m}{N}- \frac{\wt \eta  L^2 }{2} -  \frac{4 \wt \eta^2 NL^3}{m}\bigg) \Phi^r + \frac{8 \wt \eta^2  NL}{m}\Xi^r + \frac{\wt \eta^2\sigma^2N L}{mS}.  \label{lem1: bd8}
	\end{align}
\end{small}
\hfill $\blacksquare$

\section{Proof of Lemma  \ref{lem: opt_diff}}
First, we have
\begin{small}
	\begin{align}
		\E[\|\nabla_{\thetab} f_i(\thetab^{r+1}, \nub_i^{r+1}) - \cb_i^{r+1}\|^2] 
		& = \E[\|\nabla_{\thetab} f_i(\thetab^{r+1}, \nub_i^{r+1}) -\nabla_{\thetab} f_i(\thetab^{r}, \nub_i^{r+1}) + \nabla_{\thetab} f_i(\thetab^{r}, \nub_i^{r+1}) - \cb_i^{r+1}\|^2] \notag \\
		& \leq  \bigg(1 +\frac{1}{\epsilon} \bigg)L^2\E[\|\thetab^{r+1} - \thetab^{r}\|^2] + (1 + \epsilon) \E[\|\nabla_{\thetab} f_i(\thetab^{r}, \nub_i^{r+1}) - \cb_i^{r+1}\|^2]  \label{lem4: bd1}\\
		& =  \bigg(1 +\frac{1}{\epsilon} \bigg)L^2\E[\|\thetab^{r+1} - \thetab^{r}\|^2] + (1 + \epsilon)\frac{m}{N} \E[\|\nabla_{\thetab} f_i(\thetab^{r},\wt \nub_i^{r+1}) - \wt \cb_i^{r+1}\|^2] \notag \\
		&~~~~ + (1 + \epsilon) \bigg(1- \frac{m}{N}\bigg)\E[\|\nabla_{\thetab} f_i(\thetab^{r}, \nub_i^{r}) -  \cb_i^{r}\|^2],  \label{lem4: bd2}
	\end{align}
\end{small}where \eqref{lem4: bd1} holds by the Jensen's Inequality; \eqref{lem4: bd2} follows because $\Prob(i \in \Ac^r) = \frac{m}{N}$. Then, we proceed to bound $\E[\|\nabla_{\thetab} f_i(\thetab^{r},\wt \nub_i^{r+1}) - \wt \cb_i^{r+1}\|^2]$. By the definition of $\wt \cb_{i}^{r+1}$, we have
\begin{small}
	\begin{align}
		\E[\|\nabla_{\thetab} f_i(\thetab^{r},\wt \nub_i^{r+1}) - \wt \cb_i^{r+1}\|^2] 
		& = \E\bigg[\bigg\|\nabla_{\thetab} f_i(\thetab^{r},\wt \nub_i^{r+1})  - \sum_{t = 0}^{Q_i^r-1}\frac{(\bb_i^r)^t}{\wt Q_i^r}g_i(\wt \thetab_i^{r,t}, \wt \nub_i^{r+1})\bigg\|^2\bigg] \notag \\
		& \leq  \E\bigg[\bigg\|\nabla_{\thetab} f_i(\thetab^{r},\wt \nub_i^{r+1}) - \sum_{t = 0}^{Q_i^r-1}\frac{(\bb_i^r)^t}{\wt Q_i^r}\nabla_{\thetab} f_i(\wt \thetab_i^{r,t}, \wt \nub_i^{r+1})\bigg\|^2\bigg] \notag \\
		&~~~~ + \E\bigg[\bigg\| \sum_{t = 0}^{Q_i^r-1} \frac{(\bb_i^r)^t}{\wt Q_i^r}(\nabla_{\thetab} f_i(\wt \thetab_i^{r,t}, \wt \nub_i^{r+1}) - g_i(\wt \thetab_i^{r,t}, \wt \nub_i^{r+1}))\bigg\|^2\bigg]  \label{lem4: bd3}\\
		& \leq  \E\bigg[\sum_{t = 0}^{Q_i^r-1} \frac{(\bb_i^r)^t}{\wt Q_i^r}\|\nabla_{\thetab} f_i(\thetab^{r},\wt \nub_i^{r+1}) -\nabla_{\thetab} f_i(\wt \thetab_i^{r,t}, \wt \nub_i^{r+1})\|^2\bigg]  + \frac{\sigma^2}{S} \\
		& \leq    L^2 \E\bigg[\sum_{t = 0}^{Q_i^r-1} \frac{(\bb_i^r)^t}{\wt Q_i^r}\|\thetab^r - \wt \thetab_i^{r,t}\|^2\bigg]+\frac{\sigma^2}{S}, \label{lem4: bd4} 
	\end{align}
\end{small}where \eqref{lem4: bd3} follows because $\Var[\nu] = \E[\nu^2] - (\E[\nu])^2$, if $\nu$ is a random variable; \eqref{lem4: bd4} holds by Assumption 3, Assumption 4 and the independence of SGD over $t$. Substituting \eqref{lem4: bd4} into \eqref{lem4: bd2} yields\begin{small}
	\begin{align}
		&\E[\|\nabla_{\thetab} f_i(\thetab^{r+1}, \nub_i^{r+1}) - \cb_i^{r+1}\|^2] \notag \\
		 \leq& \bigg(1 + \frac{1}{\epsilon}\bigg) L^2\E[\|\thetab^{r+1} - \thetab^{r}\|^2] + \frac{(1 + \epsilon)m\sigma^2}{NS} \!+\! (1 \!+\! \epsilon)\bigg(1 \!-\! \frac{m}{N} \bigg)\E[\|\nabla_{\thetab} f_i(\thetab^{r}, \nub_i^{r}) \!-\! \cb_i^{r}\|^2] \notag\\
		& + (1 + \epsilon)\frac{m  L^2 }{N}  \E\bigg[\sum_{t = 0}^{Q_i^r-1} \frac{(\bb_i^r)^t}{\wt Q_i^r}\|\thetab^r - \wt \thetab_i^{r,t}\|^2\bigg]. \label{lem4: bd5}
	\end{align}
\end{small}Let us pick $\epsilon = \frac{m}{2N}$ and then \eqref{lem4: bd5} becomes
\begin{small}
	\begin{align}
		\E[\|\nabla_{\thetab} f_i(\thetab^{r+1}, \nub_i^{r+1}) - \cb_i^{r+1}\|^2] 
		& \leq \bigg(1 +\frac{2N}{m}\bigg)L^2 \E[\|\thetab^{r+1} - \thetab^{r}\|^2] + \bigg(1 + \frac{m}{2N}\bigg)\frac{m\sigma^2}{NS} \notag \\
		&~~~~ +\bigg(1- \frac{m}{2N}\bigg(1 + \frac{m}{N}\bigg)\bigg)\E[\|\nabla_{\thetab} f_i(\thetab^{r}, \nub_i^{r}) -  \cb_i^{r}\|^2]  \notag\\
		&~~~~+  \bigg(1 + \frac{m}{2N}\bigg)\frac{m L^2 }{N}  \E\bigg[\sum_{t = 0}^{Q_i^r-1} \frac{(\bb_i^r)^t}{\wt Q_i^r}\|\thetab^r - \wt \thetab_i^{r,t}\|^2\bigg], \label{lem4: bd6}
	\end{align}
\end{small}where \eqref{lem4: bd6} follows because
\begin{small}
	\begin{align}
		&	(1 + \epsilon)\bigg(1- \frac{m}{N} \bigg) 
		= 1- \frac{m}{2N}\bigg(1 + \frac{m}{N}\bigg), \\
		& 1 + \epsilon= 1 + \frac{m}{2N}, ~1 + \frac{1}{\epsilon} = 1 + \frac{2N}{m}.
	\end{align}
\end{small}Taking the average over the two sides of \eqref{lem4: bd6} with respect to all edge devices yields
\begin{small}
	\begin{align}
		\Xi^{r+1} - \Xi^r 
		& \leq  \bigg(1 +\frac{2N}{ m}\bigg)L^2 \E[\|\thetab^{r+1} - \thetab^{r}\|^2]+ \bigg(1 + \frac{ m}{2N}\bigg)\frac{m\sigma^2}{NS} \notag \\
		&~~~~ - \frac{ m}{2N}\bigg(1 + \frac{ m}{N}\bigg)\E[\|\nabla_{\thetab} f_i(\thetab^{r}, \nub_i^{r}) -  \cb_i^{r}\|^2] + \bigg(1 + \frac{ m}{2N}\bigg)\frac{mL^2 }{N} \Psi^r.  \label{lem4: bd7}
	\end{align}
\end{small}Lastly, substituting the results of Lemma S.1 into \eqref{lem4: bd7} yields
\begin{small}
	\begin{align}
		\Xi^{r+1} - \Xi^r 
		& \leq  \bigg(\bigg(1 + \frac{2N}{m}\bigg) \frac{4\wt \eta^2 NL^4}{m} + \bigg(1 + \frac{ m}{2N}\bigg)\frac{m L^2 }{N} \bigg)\Psi^r - \bigg(\frac{ m}{2N}\bigg(1 + \frac{ m}{N}\bigg) - \bigg(1 + \frac{2N}{ m}\bigg) \frac{8\wt \eta^2 NL^2}{m} \bigg)\Xi^r \notag \\
		&~~~~+\bigg(1 + \frac{2N}{ m}\bigg) \frac{4\wt \eta^2 NL^4}{m} \Phi^r+  \bigg(1 + \frac{2N}{m}\bigg)\frac{\wt \eta^2\sigma^2NL^2}{mS} + \bigg(1 + \frac{2N}{ m}\bigg) \frac{4\wt \eta^2 NL^2}{m}\E[\|\nabla_{\thetab} f(\thetab^r, \nub^{r}) \|^2] \notag \\
		&~~~~+\bigg(1 + \frac{ m}{2N}\bigg)\frac{m\sigma^2}{NS}. \label{lem4: bd8}
	\end{align}
\end{small}Since $\wt \eta \leq \frac{m}{32NL} (1 + \frac{2N}{ m})^{-\frac{1}{2}} $,
we have from  \eqref{lem4: bd8} that
\begin{small}
	\begin{align}
		\Xi^{r+1} - \Xi^r 
		& \leq  \bigg(\frac{257}{256}+ \frac{ m}{2N} \bigg)\frac{m L^2 }{N}\Psi^r - \frac{m}{2N}\bigg(\frac{63}{64} + \frac{ m}{N}\bigg)\Xi^r + \frac{mL^2}{256N} \Phi^r \notag\\
		&~~~~+ \frac{ m}{256N}\E[\|\nabla_{\thetab} f(\thetab^r, \nub^{r}) \|^2] +\bigg(1 + \frac{m}{2N}\bigg)\frac{m\sigma^2}{NS}+  \frac{ m\sigma^2}{1024NS}. 
	\end{align}
\end{small}\hfill $\blacksquare$

\section{Proof of Lemma \ref{lem: model_diff}}
By the definition of $\wt \thetab_i^{r,t}$, we have
\begin{small}
	\begin{align}
		\E[\|\wt \thetab_i^{r,t}\!-\!\thetab^r \|^2] & =  \eta^2	\E\bigg[\bigg\|\sum_{k = 0}^{t-1}(\bb_i^r)^{Q_i^r - t + k} (g_i(\wt \thetab_i^{r,k}, \wt \nub_i^{r+1})  + \cb^r - \cb_i^r)\bigg\|^2\bigg] \label{lem3: bd00}\\
		& =   \eta^2	\E\bigg[\bigg\|\sum_{k = 0}^{t-1} (\bb_i^r)^{Q_i^r - t + k} (\nabla_{\thetab} f_i(\wt \thetab_i^{r,k}, \wt \nub_i^{r+1})  + \cb^r - \cb_i^r)\bigg\|^2\bigg] \notag \\
		&~~~~ + \eta^2\E\bigg[\bigg\|\sum_{k = 0}^{t-1} (\bb_i^r)^{Q_i^r - t + k} (g_i(\wt \thetab_i^{r,k}, \wt \nub_i^{r+1}) - \nabla_{\thetab} f_i(\wt \thetab_i^{r,k}, \wt \nub_i^{r+1})  )\bigg\|^2\bigg]\label{lem3: bd0} \\
		& \leq   \eta^2 \E\bigg[\sum_{k = 0}^{t-1} (\bb_i^r)^{Q_i^r - t + k}\sum_{t = 0}^{Q_i^r-1} (\bb_i^r)^t	\|\nabla_{\thetab} f_i(\wt \thetab_i^{r,t}, \wt \nub_i^{r+1})  \!+\! \cb^r \!-\! \cb_i^r\|^2\bigg] + \frac{\eta^2\sigma^2}{S}\E\bigg[\sum_{k = 0}^{t-1} ((\bb_i^r)^{Q_i^r - t + k})^2\bigg],\label{lem3: bd1}
	\end{align}
\end{small}where \eqref{lem3: bd00} holds because
\begin{align*}
	\wt \thetab_{i}^{r, t} & =  \wt \thetab_{i}^{r, t-1} - \eta  \sum_{k = 0}^{t-1} \gamma^{t -1- k} (g_i(\wt \thetab_i^{r,k}, \wt \nub_i^{r+1})  + \cb^r - \cb_i^r) \notag\\
	& =  - \eta \sum_{s = 0}^{t-1}\sum_{k=0}^{s}\gamma^{s - k} (g_i(\wt \thetab_i^{r,k}, \wt \nub_i^{r+1})  + \cb^r - \cb_i^r)\\
	& =  -\eta \sum_{k=0}^{t-1}\sum_{s \geq k}^{t-1} \gamma^{s - k} (g_i(\wt \thetab_i^{r,k}, \wt \nub_i^{r+1})  + \cb^r - \cb_i^r)\\
	& =  -\eta \sum_{k=0}^{t-1}\frac{1- \gamma^{t- k}}{1-\gamma} (g_i(\wt \thetab_i^{r,k}, \wt \nub_i^{r+1})  + \cb^r - \cb_i^r) \\
	& =  - \eta \sum_{k=0}^{t-1}(\bb_i^r)^{Q_i^r - t +k}  (g_i(\wt \thetab_i^{r,k}, \wt \nub_i^{r+1})  + \cb^r - \cb_i^r) ;  \notag 
\end{align*}\eqref{lem3: bd0} follows because $\Var[\nu] = \E[\nu^2] - (\E[\nu])^2$, if $\nu$ is a random variable; \eqref{lem3: bd1} holds due to the independence of SGD over $t$ and Assumption 4. Furthermore, note that
\begin{small}
	\begin{align}
		\sum_{t = 0}^{Q_i^r - 1}\frac{(\bb_i^r)^t}{\wt Q_i^r}\sum_{k = 0}^{t-1} ((\bb_i^r)^{Q_i^r - t + k})^2 
		& \leq  \sum_{t = 0}^{Q_i^r - 1}\frac{(\bb_i^r)^t}{\wt Q_i^r}\sum_{k = 0}^{Q_i^r-2} ((\bb_i^r)^{k + 1})^2 \notag \\
		& =  \sum_{k = 0}^{Q_i^r-2} ((\bb_i^r)^{k + 1})^2 \notag \\
		& =  \|\bb_i^r\|^2 - ((\bb_i^r)^0)^2  \leq  \|\bb_i^r\|^2, \label{lem3: bd1_1}\\
		\sum_{t = 0}^{Q_i^r - 1}\frac{(\bb_i^r)^t}{\wt Q_i^r}\sum_{k = 0}^{t-1} (\bb_i^r)^{Q_i^r - t + k}& \leq  \sum_{t = 0}^{Q_i^r - 1}\frac{(\bb_i^r)^t}{\wt Q_i^r}\sum_{k = 0}^{Q_i^r-2} (\bb_i^r)^{ k+ 1} \notag \\
		& =  \sum_{k = 0}^{Q_i^r-2} (\bb_i^r)^{ k+ 1} \notag \\
		& =  \wt Q_i^r - (\bb_i^r)^0 \notag \\
		& \leq   \wt Q_i^r. \label{lem3: bd1_2}
	\end{align}
\end{small}Thus, taking average over the two sides of \eqref{lem3: bd1} with respect to $t$ gives rise to
\begin{small}
	\begin{align}
		&\E\bigg[\sum_{t = 0}^{Q_i^r-1} \frac{(\bb_i^r)^t}{\wt Q_i^r}\|\thetab^r \!-\! \wt \thetab_i^{r,t}\|^2\bigg] \notag \\
		 \leq&  (\eta \wt Q_i^r)^2\E\bigg[\sum_{t = 0}^{Q_i^r-1}\frac{(\bb_i^r)^t}{\wt Q_i^r}\|\nabla_{\thetab} f_i(\wt \thetab_i^{r,t}, \wt \nub_i^{r+1})  + \cb^r - \cb_i^r\|^2\bigg] + \frac{\eta^2\sigma^2 \|\bb_i^r\|^2}{S} \label{lem3: bd2}\\
		 \leq &  (\eta \wt Q_i^r)^2\bigg(4L^2\E\bigg[\sum_{t = 0}^{Q_i^r-1} \frac{(\bb_i^r)^t}{\wt Q_i^r}\|\thetab^r - \wt \thetab_i^{r,t}\|^2\bigg] + 4L^2\E[\|\wt \nub_i^{r+1} - \nub_i^r\|^2] + 4\Xi^r \notag \\
		& + 4\E[\|\nabla_{\thetab} f_i(\thetab^r, \nub_i^{r}) - \cb_i^r\|^2]+ 4 \E[\|\nabla_{\thetab} f(\thetab^r, \nub^{r})\|^2]\bigg) + \frac{\eta^2\sigma^2 \|\bb_i^r\|^2}{S}  \label{lem3: bd3}\\
		 \leq&   (\eta \wt Q_i^rL)^2\E\bigg[\sum_{t = 0}^{Q_i^r-1} \frac{(\bb_i^r)^t}{\wt Q_i^r}\|\thetab^r - \wt \thetab_i^{r,t}\|^2\bigg]+ 4(\eta \wt Q_i^r)^2 \Xi^r +  4(\eta \wt Q_i^rL)^2 \E[\|\wt \nub_i^{r+1} - \nub_i^r\|^2] \notag \\
		&+  \frac{\eta^2\sigma^2 \|\bb_i^r\|^2}{S} +  4(\eta \wt Q_i^r)^2 \E[\|\nabla_{\thetab} f(\thetab^r, \nub^{r}) \|^2] + 4(\eta \wt Q_i^r)^2 \E[\|\nabla_{\thetab} f_i(\thetab^r, \nub_i^{r}) - \cb_i^r\|^2],  \label{lem3: bd4}
	\end{align}
\end{small}where \eqref{lem3: bd2} follows by \eqref{lem3: bd1_1} and \eqref{lem3: bd1_2}; \eqref{lem3: bd3} holds by \eqref{lem2: bd8}. Rearranging the two sides of \eqref{lem3: bd2} yields
\begin{small}
	\begin{align}
	&	\E\bigg[\sum_{t = 0}^{Q_i^r-1} \frac{(\bb_i^r)^t}{\wt Q_i^r}\|\thetab^r \!- \!\wt \thetab_i^{r,t}\|^2\bigg] \notag \\
		%\notag \\
		 \leq&  \frac{1}{1-(\eta \wt Q_i^rL)^2 } \bigg(4(\eta \wt Q_i^r)^2 \E[\|\nabla_{\thetab} f(\thetab^r, \nub^{r}) \|^2] +4(\eta \wt Q_i^rL)^2 \E[\|\wt \nub_i^{r+1} - \nub_i^r\|^2] + 4(\eta \wt Q_i^r)^2 \Xi^r \notag \\
		&+4(\eta \wt Q_i^r)^2\E[\|\nabla_{\thetab} f_i(\thetab^r, \nub_i^{r}) - \cb_i^r\|^2]+  \frac{\eta^2\sigma^2 \|\bb_i^r\|^2}{S}\bigg) \\
		\leq  &5 (\eta \wt Q_i^r)^2 \E[\|\nabla_{\thetab} f(\thetab^r, \nub^{r}) \|^2] + 5 (\eta \wt Q_i^r)^2 \Xi^r +  \frac{5\eta^2\sigma^2 \|\bb_i^r\|^2}{4S}+5(\eta \wt Q_i^rL)^2 \E[\|\wt \nub_i^{r+1} - \nub_i^r\|^2]\notag \\
		& +5 (\eta \wt Q_i^r)^2\E[\|\nabla_{\thetab} f_i(\thetab^r, \nub_i^{r}) - \cb_i^r\|^2] \label{lem3: bd5},
	\end{align}
\end{small}where  \eqref{lem3: bd5} follows because $(\eta \wt Q_i^rL)^2  \leq \frac{1}{5}$.
Lastly, taking the average over the two sides of \eqref{lem3: bd5} with respect to all edge devices yields
\begin{small}
	\begin{align}
		\Psi^r 
		& \leq   5\eta^2\ol Q^2\E[\|\nabla_{\thetab} f(\thetab^r, \nub^{r}) \|^2]+ \frac{5\eta^2\ol Q^2\sigma^2 }{4S}  + 5 \eta^2 \ol Q^2L^2\Phi^r+ 10\eta^2 \ol Q^2\Xi^r. 
	\end{align}
\end{small}where $\ol Q = \max\limits_{i, r} \wt Q_i^r$.\hfill $\blacksquare$

\section{Proof of Lemma \ref{lem: diff_x}}
According to the update of $\thetab_0$ and $\thetab_{i}$, we have
\begin{small}
	\begin{align}
		\E[\thetab^r - \thetab^{r+1}]
		& =  \frac{N}{m}\eta_g\E\bigg[\sum_{i=1}^{N}\omega_i \Ibb_{\Ac^r}^{i}\frac{\thetab^r-\wt \thetab_i^{r+1}}{\wt Q_i^r} \bigg] \notag \\
		& = \eta_g \sum_{i=1}^{N}\omega_i\E\bigg[\frac{\thetab^r -\wt  \thetab_i^{r+1}}{\wt Q_i^r}\bigg] \\
		& =  \wt \eta \sum_{i=1}^{N}\omega_i\E\bigg[\sum_{t = 0}^{Q_i^r-1}  \frac{(\bb_i^r)^t}{\wt Q_i^r}(g_i(\wt \thetab_i^{r,t}, \wt \nub_i^{r+1}) + \cb^r - \cb_i^r)\bigg] \notag\\
		& = \wt \eta \sum_{i=1}^{N}\omega_i \E\bigg[\sum_{t = 0}^{Q_i^r-1}  \frac{(\bb_i^r)^t}{\wt Q_i^r}( \nabla_{\thetab} f_i(\wt \thetab_i^{r,t}, \wt \nub_i^{r+1})+ \cb^r - \cb_i^r) \bigg] \label{lem2: bd1} \\
		& = \wt \eta \sum_{i=1}^{N}\omega_i\E\bigg[\sum_{t = 0}^{Q_i^r-1} \frac{(\bb_i^r)^t}{\wt Q_i^r}  \nabla_{\thetab} f_i(\wt \thetab_i^{r,t}, \wt \nub_i^{r+1}) \bigg], \label{lem2: bd2}
	\end{align}
\end{small}where $\wt \eta \triangleq \eta\eta_g$;  \eqref{lem2: bd1} follows because $\Prob(i \in \Ac^r) = \frac{m}{N}$; \eqref{lem2: bd2} follows because $\cb^r = \sum_{i=1}^{N}\omega_i \cb_i^r$.
Next, we have
\begin{small}
	\begin{align}
	&	\E[\|\thetab^{r+1} - \thetab^r\|^2] \notag \\
		 = 	&\E\bigg[\bigg\|\eta_g\sum_{i=1}^{N}\omega_i \Ibb_{\Ac^r}^{i}\frac{N}{m}\frac{ \thetab^r-\wt \thetab_i^{r+1}}{\wt Q_i^r}\bigg\|^2\bigg] \notag \\
		 =&  \E\bigg[\bigg\|\wt \eta \sum_{i=1}^{N} \omega_i\Ibb_{\Ac^r}^{i}\frac{N}{m}\sum_{t = 0}^{Q_i^r-1} \frac{(\bb_i^r)^t}{\wt Q_i^r}(g_i(\wt \thetab_i^{r,t},\wt \nub_i^{r+1})  + \cb^r- \cb_i^r)\bigg\|^2\bigg] \notag \\
		 =& \wt \eta^2 \E\bigg[\bigg\| \sum_{i=1}^{N} \omega_i \Ibb_{\Ac^r}^{i}\frac{N}{m}\sum_{t = 0}^{Q_i^r-1}\frac{(\bb_i^r)^t}{\wt Q_i^r}(g_i(\wt \thetab_i^{r,t},\wt \nub_i^{r+1})  + \cb^r- \cb_i^r)\bigg\|^2\bigg]\notag \\
		 = &\wt \eta^2 \E\bigg[\bigg\| \sum_{i=1}^{N}\omega_i\Ibb_{\Ac^r}^{i}\frac{N}{m}\sum_{t = 0}^{Q_i^r-1} \frac{(\bb_i^r)^t}{\wt Q_i^r}(\nabla_{\thetab} f_i(\wt \thetab_i^{r,t},\wt \nub_i^{r+1})  + \cb^r- \cb_i^r)\bigg\|^2\bigg] \notag \\
		&+\wt \eta^2 \E\bigg[\bigg\| \sum_{i=1}^{N}\omega_i\Ibb_{\Ac^r}^{i}\frac{N}{m}\sum_{t = 0}^{Q_i^r-1}\frac{(\bb_i^r)^t}{\wt Q_i^r}(g_i(\wt \thetab_i^{r,t},\wt \nub_i^{r+1})  - \nabla_{\thetab} f_i(\wt \thetab_i^{r,t},\wt \nub_i^{r+1})  )\bigg\|^2\bigg] \label{lem2: bd3}\\
		=&  \wt \eta^2 \E\bigg[\bigg\| \sum_{i=1}^{N}\omega_i \Ibb_{\Ac^r}^{i}\frac{N}{m}\sum_{t = 0}^{Q_i^r-1}\frac{(\bb_i^r)^t}{\wt Q_i^r} (\nabla_{\thetab} f_i(\wt \thetab_i^{r,t},\wt \nub_i^{r+1})  + \cb^r- \cb_i^r)\bigg\|^2\bigg] \notag \\
		&+\wt \eta^2 \E\bigg[\sum_{i=1}^{N}\bigg(\omega_i\Ibb_{\Ac^r}^{i}\frac{N}{m}\bigg)^2\sum_{t = 0}^{Q_i^r-1}\bigg(\frac{(\bb_i^r)^t}{\wt Q_i^r}\bigg)^2\| g_i(\wt \thetab_i^{r,t},\wt \nub_i^{r+1})  - \nabla_{\thetab} f_i(\wt \thetab_i^{r,t},\wt \nub_i^{r+1})  \|^2\bigg] \label{lem2: bd4}\\
		 \leq &\frac{ \wt \eta^2 N}{m}\sum_{i=1}^{N}\omega_i\E\bigg[\sum_{t = 0}^{Q_i^r-1} \frac{(\bb_i^r)^t}{\wt Q_i^r} \|\nabla f_i(\wt \thetab_i^{r,t},\wt \nub_i^{r+1}) \!+\! \cb^r \!-\! \cb_i^r\|^2\bigg] \!+\! \frac{\wt \eta^2\sigma^2N}{mS} \sum_{i=1}^{N}\omega_i^2\E\bigg[\sum_{t = 0}^{Q_i^r-1}\bigg(\frac{(\bb_i^r)^t}{\wt Q_i^r}\bigg)^2\bigg] \label{lem2: bd5} \\
		 \leq & \frac{ \wt \eta^2 N}{m}\sum_{i=1}^{N}\omega_i\E\bigg[\sum_{t = 0}^{Q_i^r-1} \frac{(\bb_i^r)^t}{\wt Q_i^r} \|\nabla f_i(\wt \thetab_i^{r,t},\wt \nub_i^{r+1})  + \cb^r- \cb_i^r\|^2\bigg] + \frac{\wt \eta^2\sigma^2N}{mS},  \label{lem2: bd6}
	\end{align}
\end{small}where \eqref{lem2: bd3} follows because $\Var[\nu] = \E[\nu^2] - (\E[\nu])^2$, if $\nu$ is a random variable; \eqref{lem2: bd4} holds due to the independence of SGD over $i, t$; \eqref{lem2: bd5} follows by Assumption 4; \eqref{lem2: bd6} follows because 
\begin{small}
	\begin{align}
		\sum_{i=1}^{N}\omega_i^2	\E\bigg[\sum_{t = 0}^{Q_i^r-1}\bigg(\frac{(\bb_i^r)^t}{\wt Q_i^r}\bigg)^2\bigg] \leq 1.
	\end{align}
\end{small}Furthermore, note that
\begin{small}
	\begin{align}
		&\E\bigg[\sum_{t = 0}^{Q_i^r-1} \frac{(\bb_i^r)^t}{\wt Q_i^r} \|\nabla f_i(\wt \thetab_i^{r,t},\wt \nub_i^{r+1})  + \cb^r- \cb_i^r\|^2\bigg]\notag \\
		& =  \E\bigg[\sum_{t = 0}^{Q_i^r-1} \frac{(\bb_i^r)^t}{\wt Q_i^r} \|\nabla_{\thetab} f_i(\wt \thetab_i^{r,t}, \wt \nub_i^{r+1})- \nabla_{\thetab} f_i(\thetab^r, \nub_i^{r}) + \cb^r  - \nabla_{\thetab} f(\thetab^r, \nub^{r}) + \nabla_{\thetab}f(\thetab^r, \nub^{r}) - (\cb_i^r-\nabla_{\thetab} f_i(\thetab^r, \nub_i^{r}) )\|^2\bigg] \notag \\
		& \leq  \E\bigg[\sum_{t = 0}^{Q_i^r-1} \frac{(\bb_i^r)^t}{\wt Q_i^r} \bigg(4L^2\|\thetab^r \!-\! \wt \thetab_i^{r,t}\|^2 \!+\! 4L^2 \|\wt \nub_i^{r+1} \!-\! \nub_i^r\|^2 \!+\! 4\Xi^r \!+\! 4\|\nabla_{\thetab} f_i(\thetab^r, \nub_i^{r}) \!-\! \lambdab_i^r\|^2 \!+\! 4 \|\nabla_{\thetab} f(\thetab^r, \nub^{r})\|^2\bigg)\bigg] \label{lem2: bd7}  \\
		& =   4L^2\E\bigg[\sum_{t = 0}^{Q_i^r-1} \frac{(\bb_i^r)^t}{\wt Q_i^r}\|\thetab^r \!-\! \wt \thetab_i^{r,t}\|^2\bigg] \!+\! 4L^2\E[\|\wt \nub_i^{r+1} \!-\! \nub_i^r\|^2] \!+\! 4\Xi^r \!+\! 4\E[\|\nabla_{\thetab} f_i(\thetab^r, \nub_i^{r}) \!-\! \cb_i^r\|^2] \!+\! 4 \E[\|\nabla_{\thetab} f(\thetab^r, \nub^{r})\|^2],\label{lem2: bd8}
	\end{align}
\end{small}where  \eqref{lem2: bd7} follows by Assumption 3. As a result, we have from \eqref{lem2: bd6} that
\begin{small}
	\begin{align}
		\E[\|\thetab^{r+1} - \thetab^r\|^2] \leq  \frac{4 \wt \eta^2 NL^2}{m}\Psi^r+ \frac{8 \wt \eta^2 N}{m}\Xi^r + \frac{ 4\wt \eta^2 N}{m}\E[\|\nabla_{\thetab} f(\thetab^r, \nub^{r}) \|^2] + \frac{ 4\wt \eta^2 NL^2}{m} \Phi^r+ \frac{\wt \eta^2\sigma^2N}{mS}. 
	\end{align}
\end{small}
\hfill $\blacksquare$

\section{Proof of Theorem \ref{thm: localized model}}
\label{appdix: proof_theorem2}
Similar to the poof of Theorem 1, we start with defining the virtual sequences $\{\wt \thetab_{i, c}^r\}$ with by assuming that all edge devices are active at round $r$, i.e., $\forall i, 0 \leq t \leq Q_i^r - 1$,
\begin{small}
	\begin{align}
		&\wt \thetab_{i, lc}^{r, 0} = \thetab_{i, lc}^{r}, \wt \mub_{i, lc}^{r, 0} = \zerob, \label{eq: virtual_thetai_lc_r} \\
		&\wt \mub_{i, lc}^{r, t+1} = \gamma \mub_{i, lc}^{r, t} + G_i(\wt \thetab_{i, lc}^{r,t}, \wt \thetab_{i}^{r,t}, \wt \nub_i^{r+1}), \\
		& \wt \thetab_{i, lc}^{r, t+1} =\wt \thetab_{i, lc}^{r, t} - \eta \wt \mub_{i ,lc}^{r, t+1}, \\
		& \wt \thetab_{i, lc}^{r+1}=\wt \thetab_i^{r, Q_i^r}. \label{eqn: virtual_thetai_lc_r+}
	\end{align}
\end{small}Then we proceed with the help of the following Lemma. 

\begin{Lemma}\label{lem: local_round}
	For any round $r$, if $\eta_c \ol Q L_F \leq \frac{1}{2}$, it holds that
	\begin{small}
		\begin{align}
			&\E[F_i(\thetab_{i, lc}^{r+1}, \thetab^{r+1},  \nub_{i}^{r+1})] - \E[F_i(\thetab_{i, lc}^r,\thetab^{r},  \nub_{i}^{r})] \notag\\
			& \leq  - \! \frac{\eta_c \wt Q_i^rm}{2N} \E[\|\nabla_{\thetab_{i, lc}} F_i(\thetab_{i, lc}^r,\thetab^{r}, \wt \nub_{i}^{r+1})\|^2] \!+\! \frac{\eta_c^2L_F(\eta_c L_F + 1)(\wt Q_i^r)^2m\sigma^2}{2NS} \!+\! \frac{\eta_c \wt Q_i^rmL_F^2}{2N} \E\bigg[\sum_{t = 0}^{Q_i^r - 1} \frac{(\bb_i^r)^t}{\wt Q_i^r}\|\wt\thetab_i^{r, t} \!-\! \thetab^{r}\|^2\bigg] \notag \\
			&~~~~  +  (1-\beta_i) \E[ \langle \nabla_{\thetab} f_i(\thetab^r, \nub_{i}^{r}), \thetab^{r+1}- \thetab^r \rangle] + \frac{(1-\beta_i)(4\beta_i\wt LL_h \Gamma + 5L)}{8}\E[\|\thetab^{r+1} - \thetab^r\|^2] + \beta_i^2(1-\beta_i)(\wt L-\mu)\Gamma^2 \notag \\
			&~~~~+ \beta_i(1-\beta_i)\wt L \Gamma^2 +\frac{m\wt L^2 \beta_i^2\Gamma^2}{2NL} - \bigg(\frac{1}{\eta_v} + \frac{\mu}{2}- \frac{3L}{2} - \frac{2(1- \beta_i)mL}{N}\bigg) \E[\|\wt \nub_{i}^{r+1} - \nub_{i}\|^2]. \label{lem: one_round_thetai_lc_bd} 
		\end{align}
\end{small}
\end{Lemma}

Rearranging the two sides of \eqref{lem: one_round_thetai_lc_bd} and then taking its average over all edge devices yields
\begin{small}
	\begin{align}
		&\sum_{i=1}^{N}\omega_i \E[\|\nabla_{\thetab_{i, lc}} F_i(\thetab_{i, lc}^r,\thetab^{r}, \wt \nub_{i}^{r+1})\|^2] \notag \\
		& \leq  \frac{2N}{\eta_{c} \ul Q m} \sum_{i=1}^{N} \omega_i(\E[F_i(\thetab_{i, lc}^r,\thetab^{r},  \nub_{i}^{r})]-\E[F_i(\thetab_{i, lc}^{r+1}, \thetab^{r+1}, \nub_{i}^{r+1})] ) +  \frac{\eta_c L_F(\eta_c L_F + 1)\ol Q \sigma^2}{S} + L_F^2\Psi^r  \notag \\
		&~~~~ +  \frac{2N}{\eta_{c}\ul Qm} \E[ \langle \nabla_{\thetab} f(\thetab^r,  \nub^{r}), \thetab^{r+1}- \thetab^r \rangle] + \E[\|\thetab^{r+1} - \thetab^r\|^2] \sum_{i=1}^{N}\omega_i \frac{(1-\beta_i)(4\beta_i\wt LL_h \Gamma + 5L)N}{4\eta_{c} \wt Q_i^rm} \notag \\
		&~~~~+\Gamma^2\sum_{i=1}^{N}\omega_i\frac{2N\beta_i(1-\beta_i)(\beta_i(\wt L-\mu) + \wt L)}{\eta_{c}\wt Q_i^rm} +\frac{\wt L^2\Gamma^2}{\eta_{c} \ul Q L}\sum_{i=1}^{N}\omega_i \beta_i^2 - \bigg(\frac{1}{\eta_v} + \frac{\mu}{2}- \frac{3L}{2} - \frac{2(1- \ul \beta)mL}{N}\bigg) \frac{2N}{\eta_{c}\ul Qm}\Phi^r, \label{thm2: bd1} 
	\end{align}
\end{small}where $\ul \beta = \min\limits_i \beta_i$. Substituting \eqref{lem1: bd6} and \eqref{lem: diff_theta0_bd2} into \eqref{thm2: bd1} yields
\begin{small}
	\begin{align}
		&\sum_{i=1}^{N}\omega_i \E[\|\nabla_{\thetab_{i, lc}} F_i(\thetab_{i, lc}^r,\thetab^{r}, \wt \nub_{i}^{r+1})\|^2] \notag \\
		& \leq  \frac{2N}{\eta_{c} \ul Q m} \sum_{i=1}^{N} \omega_i(\E[F_i(\thetab_{i, lc}^r,\thetab^{r},  \nub_{i}^{r})]-\E[F_i( \thetab_{i, lc}^{r+1}, \thetab^{r+1},  \nub_{i}^{r+1})] ) +  \frac{\eta_c L_F(\eta_c L_F + 1)\ol Q \sigma^2}{S} \notag \\
		&~~~~ + \bigg(\frac{\wt \eta N L^2}{\eta_{c}\ul Q m} + L_F^2 + \frac{4 \wt \eta^2 N^2L^2D_0}{\eta_{c}\ul Qm^2}\bigg)\Psi^r - \frac{\wt \eta N}{\eta_{c}\ul Q m }\bigg(1 - \frac{ 4\wt \eta ND_0}{m} \bigg) \E[\| \nabla_{\thetab}f(\thetab^r,\nub^{r})\|^2]\notag \\
		&~~~~+\Gamma^2\sum_{i=1}^{N}\omega_i\frac{2N\beta_i(1-\beta_i)(\beta_i(\wt L-\mu) + \wt L)}{\eta_{c}\wt Q_i^rm} +\frac{\wt L^2\Gamma^2}{\eta_{c} \ul Q L}\sum_{i=1}^{N}\omega_i \beta_i^2 \notag\\
		&~~~~- \bigg(\frac{1}{\eta_v} + \frac{\mu}{2}- \frac{3L}{2} - \frac{2(1- \beta_i)mL}{N}- \frac{\wt \eta L^2}{2}-\frac{ 2\wt \eta^2 NL^2D_0}{m}\bigg) \frac{2N}{\eta_{c}\ul Qm}\Phi^r +\frac{8 \wt \eta^2 N^2D_0}{\eta_{c}\ul Qm^2}\Xi^r + \frac{\wt \eta^2N^2\sigma^2}{\eta_{c}\ul Qm^2S},\label{thm2: bd2} 
	\end{align}
\end{small}where $D_0 \triangleq \sum_{i=1}^{N}\omega_i \frac{(1-\beta_i)(4\beta_i\wt LL_h \Gamma + 5L)}{4}$. Furthermore, we conclude from Theorem 1 that 
\begin{small}
	\begin{align}
		P^{r+1} - P^r
		& \leq  -\frac{\wt \eta}{4}\E[\| \nabla_{\thetab}f(\thetab^r, \nub^{r})\|^2] - \frac{3\wt \eta L^2}{4} \Psi^r - \bigg(\frac{3\wt \eta^2 LN}{m} +  12 \wt \eta^2 L\bigg) \Xi^r \notag \\
		&~~~~ - \frac{ 31mL}{64N}\Phi^r   + \bigg(\frac{259}{64}  + \frac{ m}{N} + \frac{5\eta mL\ol Q^2}{24\eta_g N}\bigg)\frac{12\wt \eta^2 NL\sigma^2}{ mS}.\label{thm2: bd3} 
	\end{align}
\end{small}
Combining \eqref{thm2: bd2} and \eqref{thm2: bd3} gives rise to
\begin{small}
	\begin{align}
		&\sum_{i=1}^{N}\omega_i \E[\|\nabla_{\thetab_{i, lc}} F_i(\thetab_{i, lc}^r,\thetab^{r}, \wt \nub_{i}^{r+1})\|^2] \notag \\
		\leq&   \frac{2N}{\eta_{c} \ul Q m} \sum_{i=1}^{N} \omega_i(\E[F_i(\thetab_{i, lc}^r,\thetab^{r},  \nub_{i}^{r})]-\E[F_i( \thetab_{i, lc}^{r+1}, \thetab^{r+1},  \nub_{i}^{r+1})]) \notag \\
		& + \frac{4N }{\eta_{c}\ul Qm}\bigg(1 - \frac{ 4\wt \eta ND_0}{m} \bigg)(P^r - P^{r+1}) - D_1 \Psi^r - D_2 \Xi^r  - D_3 \Phi^r \notag \\
		&+  \frac{\eta_c L_F(\eta_c L_F + 1)\ol Q \sigma^2}{S} +\Gamma^2\sum_{i=1}^{N}\omega_i\frac{2N\beta_i(1-\beta_i)(\beta_i(\wt L-\mu) + \wt L)}{\eta_{c}\wt Q_i^rm} \notag \\
		& +\frac{\wt L^2\Gamma^2}{\eta_{c} \ul Q L}\sum_{i=1}^{N}\omega_i \beta_i^2 + \frac{\wt \eta^2N^2\sigma^2}{\eta_{c}\ul Qm^2S} + \bigg(\frac{259}{64} + \frac{ m}{N} + \frac{5\eta mL\ol Q^2}{24\eta_g N}\bigg)\frac{48\wt \eta^2 N^2L\sigma^2}{\eta_{c} \ul Q m^2S}\bigg(1 - \frac{ 4\wt \eta ND_0}{m} \bigg), \label{thm2: bd4}
	\end{align}
\end{small}where 
\begin{small}
	\begin{align}
		&D_1 \triangleq  \frac{3\wt \eta N L^2 }{\eta_{c}\ul Q m}\bigg(1 - \frac{ 4\wt \eta ND_0}{m} \bigg)-\frac{\wt \eta N L^2}{\eta_{c}\ul Qm} - L_F^2 - \frac{4 \wt \eta^2 N^2L^2D_0}{\eta_{c}\ul Qm^2}, \\
		&D_2 \triangleq \frac{4\wt \eta^2 NL}{\eta_c \ul Qm}\bigg(\frac{3 N}{m}+12\bigg)\bigg(1 - \frac{ 4\wt \eta ND_0}{m} \bigg)- \frac{8 \wt \eta^2 N^2D_0}{\eta_{c}\ul Qm^2}, \\
		&D_3 \triangleq \bigg(\frac{1}{\eta_v} + \frac{\mu}{2} + \frac{31mL}{32N}\bigg(1 - \frac{ 4\wt \eta ND_0}{m} \bigg)- \frac{3L}{2}- \frac{2(1-\ul \beta)mL}{N} - \frac{\wt \eta L^2}{2} -  \frac{ 2\wt \eta^2 NL^2D_0}{m}\bigg) \frac{2N}{\eta_{c}\ul Qm}. 
	\end{align}
\end{small}
We then proceed to prove that $D_1$, $D_2$ and $D_3$ all are nonnegative. 
First, as $\wt \eta \leq \frac{m}{48ND_0}$, it suffices to have
\begin{small}
	\begin{align}
		D_1 & \geq  \frac{11\wt \eta N L^2}{4\eta_{c} \ul Qm}  -  \frac{\wt \eta NL^2}{\eta_{c} \ul Qm} - L_F^2 - \frac{\wt \eta N L^2}{12\eta_{c} \ul Qm} \notag \\
		& = \frac{5\wt \eta NL^2}{3\eta_{c} \ul Qm}-L_F^2 \notag \\
		& \geq  \frac{2L_F^2}{3}, \label{thm2: bd5}
	\end{align}
\end{small}where \eqref{thm2: bd5} follows because $\eta_{c} \leq \frac{\wt \eta N L^2}{\ul Q m L_F^2}$. Second, for $D_2$,  it suffices to have
\begin{small}
	\begin{align}
		D_2 & \geq  \frac{11\wt \eta^2N L}{3\eta_c \ul Q m}\bigg(\frac{3 N}{m}+12 \bigg)- \frac{8 \wt \eta^2 N^2D_0}{\eta_{c}\ul Qm^2}\notag \\
		& = \frac{ \wt \eta^2 N^2}{\eta_c \ul Q m^2} \bigg(11L +  \frac{44 m}{N}- 8D_0 \bigg) \notag \\
		& \geq  0, \label{thm2: bd6}
	\end{align}
\end{small}where \eqref{thm2: bd6} follows because $11L +  \frac{44m}{N}- 8D_0 \geq 0$. Lastly, for $D_3$, it suffices to have
\begin{small}
	\begin{align}
		D_3 & \geq  \bigg(\frac{1}{\eta_v} + \frac{\mu}{2} + \frac{11 \times 31mL}{12 \times 32N}- \frac{7L}{2} - \frac{\wt \eta L^2}{2} -  \frac{ \wt \eta L^2}{24}\bigg) \frac{2N}{\eta_{c}\ul Qm} \notag \\
		& \geq \bigg(\frac{1}{\eta_v} + \frac{\mu}{2}  -\frac{7L}{2} -  \frac{ 13\wt \eta L^2}{24}\bigg) \frac{2N}{\eta_{c}\ul Qm} \notag \\
		& \geq \bigg(\frac{1}{\eta_v} + \frac{\mu}{2}  -\frac{7L}{2} -  \frac{ 13\wt \eta L^2}{24}\bigg) \frac{2N}{\eta_{c}\ul Qm} \notag \\
		& \geq \bigg(\frac{L}{2} + \frac{\mu}{2} -  \frac{L}{60}\bigg) \frac{2N}{\eta_{c}\ul Qm} \notag \\
		& \geq  46L^2 + 48\mu L, \label{thm2: bd7}
	\end{align}
\end{small}where \eqref{thm2: bd7} follows because $\eta_v \leq \frac{1}{4L}$, $\eta_{c} \leq \frac{\wt \eta N}{\ul Q m }$ and $\wt \eta \leq \frac{m}{48NL}$. Therefore, we have from 
\eqref{thm2: bd4} that
\begin{small}
	\begin{align}
		&\sum_{i=1}^{N}\omega_i \E[\|\nabla_{\thetab_{i, lc}} F_i(\thetab_{i, lc}^r,\thetab^{r}, \wt \nub_{i}^{r+1})\|^2] \notag \\
		 \leq & \frac{2N}{\eta_{c} \ul Q m} \sum_{i=1}^{N} \omega_i(\E[F_i(\thetab_{i, lc}^r,\thetab^{r},  \nub_{i}^{r})]-\E[F_i( \thetab_{i, lc}^{r+1}, \thetab^{r+1},  \nub_{i}^{r+1})]) + \frac{11N }{3\eta_{c}\ul Qm}(P^r - P^{r+1})\notag \\
		& - 46L^2 \Phi^r  +  \frac{\eta_c L_F(\eta_c L_F + 1)\ol Q \sigma^2}{S}  
		+\frac{\wt L^2\Gamma^2}{\eta_{c} \ul Q L}\sum_{i=1}^{N}\omega_i \beta_i^2 + \frac{\wt \eta^2N^2\sigma^2}{\eta_{c}\ul Qm^2S}\notag \\
		& +\Gamma^2\sum_{i=1}^{N}\omega_i\frac{2N\beta_i(1-\beta_i)(\beta_i(\wt L-\mu) + \wt L)}{\eta_{c}\wt Q_i^rm} + \bigg(\frac{259}{64} + \frac{m}{N} + \frac{5\eta mL\ol Q^2}{24\eta_g N}\bigg)\frac{44\wt \eta^2 N^2L\sigma^2}{\eta_{c} \ul Q m^2S},\label{thm2: bd8}
	\end{align}
\end{small}where \eqref{thm2: bd8} follows because $\wt \eta \leq \frac{m}{48ND_0}$. Furthermore, note that
\begin{align}
	&\E[\|\nabla_{\thetab_{i, lc}} F_i(\thetab_{i, lc}^r,\thetab^{r},  \nub_{i}^{r})\|^2] \leq  3 L^2\E[\|\wt \nub_{i}^{r+1} - \nub_{i}^r\|^2]	+ \frac{3}{2}\E[\|\nabla_{\thetab_{i, lc}} F_i(\thetab_{i, lc}^r,\thetab^{r}, \wt \nub_{i}^{r+1})\|^2]. 
\end{align}As a result, we obtain
\begin{small}
	\begin{align}
		&\sum_{i=1}^{N}\omega_i \E[\|\nabla_{\thetab_{i, lc}} F_i(\thetab_{i, lc}^r,\thetab^{r}, \nub_{i}^{r})\|^2] \notag \\
		 \leq & \frac{3N}{\eta_{c} \ul Q m} \sum_{i=1}^{N} \omega_i(\E[F_i(\thetab_{i, lc}^r,\thetab^{r},  \nub_{i}^{r})]-\E[F_i( \thetab_{i, lc}^{r+1}, \thetab^{r+1},  \nub_{i}^{r+1})]) \notag \\
		& + \frac{11N }{2\eta_{c}\ul Qm}(P^r - P^{r+1})   +  \frac{3\eta_c L_F(\eta_c L_F + 1)\ol Q \sigma^2}{2S}  
		+\frac{3\wt L^2 \Gamma^2}{2\eta_{c} \ul Q L} \sum_{i=1}^{N}\omega_i\beta_i^2	\notag \\
		&+ \frac{3\wt \eta^2N^2\sigma^2}{2\eta_{c}\ul Qm^2S}+ \frac{3\Gamma^2}{\eta_{c}\ul Q m}\sum_{i=1}^{N}\omega_i\beta_i(1-\beta_i)(\beta_i(\wt L-\mu) + \wt L)\notag\notag \\
		&+ \bigg(\frac{259}{64} + \frac{ m}{N} + \frac{5\eta mL\ol Q^2}{24\eta_g N}\bigg)\frac{66\wt \eta^2 N^2L\sigma^2}{\eta_{c} \ul Q m^2S}. \label{thm2: bd9}
	\end{align}
\end{small}Summing \eqref{thm2: bd9} up from $r = 0$ to $R - 1$ and then dividing it by $R$ yields
\begin{small}
	\begin{align}
		&\frac{1}{R}\sum_{r = 0}^{R- 1}\sum_{i=1}^{N}\omega_i \E[\|\nabla_{\thetab_{i, lc}} F_i(\thetab_{i, lc}^r,\thetab^{r}, \nub_{i}^{r})\|^2] \notag \\
		& \leq  \frac{3N}{\eta_{c} \ul Q m R} (f(\thetab^0, \nub_{i}^0)-\ul f) + \frac{11N }{2\eta_{c}\ul QmR}(P^0 - \ul f)  +  \frac{3\eta_c L_F(\eta_c L_F + 1)\ol Q \sigma^2}{2S}  
		+\frac{3\wt L^2 \Gamma^2}{2\eta_{c} \ul Q L} \sum_{i=1}^{N}\omega_i\beta_i^2 	\notag \\
		&~~~~ + \frac{3\wt \eta^2N^2\sigma^2}{2\eta_{c}\ul Qm^2S} + \frac{3N\Gamma^2}{\eta_{c}\ul Q m}\sum_{i=1}^{N}\omega_i\beta_i(1-\beta_i)(\beta_i(\wt L-\mu) + \wt L) + \bigg(\frac{259}{64}  + \frac{ m}{N} + \frac{5\eta mL\ol Q^2}{24\eta_g N}\bigg)\frac{66\wt \eta^2 N^2L\sigma^2}{\eta_{c} \ul Q m^2S}. 
	\end{align}
\end{small}\hfill $\blacksquare$

\section{Proof of Lemma \ref{lem: local_round}}
First, since $F_i(\thetab_{i, lc}, \thetab_{i}, \nub_i)$ is strongly convex with modulus $\mu$, we have
\begin{small}
	\begin{align}
		&\E[F_i(\thetab_{i, lc}^r, \thetab^r, \nub_i^r) ] -\E[F_i(\thetab_{i, lc}^r, \thetab^r,  \nub_i^{r+1})] \notag \\
		 = &\frac{m}{N}(\E[F_i(\thetab_{i, lc}^r, \thetab^r, \nub_i^r) ] -\E[F_i(\thetab_{i, lc}^r, \thetab^r, \wt \nub_i^{r+1})] )\notag \\
		 \geq&  \frac{m}{N}\E[\langle \nabla_{\nub} F_i(\thetab_{i, lc}^r, \thetab^r, \wt \nub_i^{r+1}), \nub_{i} - \wt \nub_{i}^{r+1}\rangle] +\frac{\mu m}{2N}\E[\|\wt \nub_{i}^{r+1} - \nub_{i}^{r}\|^2]. \label{lem5: bd1}
	\end{align}
\end{small}Note that
\begin{small}
	\begin{align}
		\nabla_{\nub} F_i(\thetab_{i, lc}^r,\thetab^{r}, \wt \nub_{i}^{r+1}) &= \nabla_{\nub} L_i(\beta_ih(\thetab_{i, lc}^{r} )+ (1-\beta_i) h(\thetab^r) , \wt \nub_{i}^{r+1}), \notag\\
		\nabla_{\nub} f_i(\thetab^r, \wt \nub_{i}^{r+1}) 
		&= \nabla_{\nub} L_i( h(\thetab^r) , \wt \nub_{i}^{r+1}), \label{lem5: bd2}
	\end{align}
\end{small}As a result, we have
\begin{small}
	\begin{align}
		&\E[\langle \nabla_{\nub} F_i(\thetab_{i, lc}^r, \thetab^r, \wt \nub_i^{r+1}), \nub_{i} - \wt 	\nub_{i}^{r+1}\rangle]\notag\\ 
		& = \E[\langle \nabla_{\nub} F_i(\thetab_{i, lc}^r, \thetab^r, \wt \nub_i^{r+1}) -  \nabla_{\nub} F_i(\thetab_{i, lc}^r, \thetab^r, \nub_i^{r}) , \nub_{i} - \wt \nub_{i}^{r+1}\rangle] \notag \\
		&~~~~+\E[\langle \nabla_{\nub} F_i(\thetab_{i, lc}^r, \thetab^r, \nub_i^{r}) -  \nabla_{\nub} f_i(\thetab^r, \nub_i^{r}) , \nub_{i} - \wt \nub_{i}^{r+1}\rangle] +\E[\langle \nabla_{\nub} f_i(\thetab^r, \nub_i^{r}) , \nub_{i} - \wt \nub_{i}^{r+1}\rangle]  \notag \\
		& \geq  - \E[\|\nabla_{\nub} F_i(\thetab_{i, lc}^r, \thetab^r, \wt \nub_i^{r+1})-\nabla_{\nub} F_i(\thetab_{i, lc}^r, \thetab^r,  \nub_i^{r})\|\|\wt \nub_{i}^{r+1} - \nub_{i}\|] \notag \\
		&~~~~ - \E[\|\nabla_{\nub} F_i(\thetab_{i, lc}^r, \thetab^r, \nub_i^{r})-\nabla_{\nub} f_i(\thetab^r, \nub_i^{r})\|\|\wt \nub_{i}^{r+1} - \nub_{i}\|] +\E[\langle \nabla_{\nub} f_i(\thetab^r, \nub_i^{r}) , \nub_{i} - \wt \nub_{i}^{r+1}\rangle] \label{lem5: bd2_1} \\
		&\geq  \bigg(\frac{1}{\eta_v}-L\bigg) \E[\|\wt \nub_{i}^{r+1} - \nub_{i}\|^2]  - \wt L\beta_i \E[\|h(\thetab_{i, lc}^r) - h(\thetab_{i}^r)\|\|\wt \nub_{i}^{r+1} - \nub_{i}^r\|] \label{lem5: bd2_2}\\
		& \geq  \bigg(\frac{1}{\eta_v}- \frac{3L}{2}\bigg) \E[\|\wt \nub_{i}^{r+1} - \nub_{i}\|^2]-\frac{\wt L^2 \beta_i^2\Gamma^2}{2L},  \label{lem5: bd3}
	\end{align}
\end{small}where \eqref{lem5: bd2_1} follows by the Cauchy-Schwarz inequality;  \eqref{lem5: bd2_1} follows by Assumption 3; \eqref{lem5: bd3} holds by the Young's inequality. Substituting \eqref{lem5: bd3} into \eqref{lem5: bd2} yields
\begin{small}
	\begin{align}
		\E[F_i(\thetab_{i, lc}^r, \thetab^r,\nub_i^{r+1})]- \E[F_i(\thetab_{i, lc}^r, \thetab^r, \nub_i^r)]  \leq - \bigg(\frac{1}{\eta_v} + \frac{\mu}{2}- \frac{3L}{2}\bigg)\frac{m}{N} \E[\|\wt \nub_{i}^{r+1} - \nub_{i}\|^2]+\frac{m\wt L^2 \beta_i^2\Gamma^2}{2NL}. \label{lem5: bd3_1}
	\end{align}
\end{small}
Second, according to Assumption 3, we have
\begin{small}
	\begin{align}
		&\E[F_i( \thetab_{i, lc}^{r+1}, \thetab^{r+1}, \nub_{i}^{r+1})] - \E[F_i(\thetab_{i, lc}^r,\thetab^{r},  \nub_{i}^{r+1})] \notag \\
		 \leq& 	\E[ \langle \nabla_{\thetab_{i, lc}} F_i(\thetab_{i, lc}^r,\thetab^{r},  \nub_{i}^{r+1}),\thetab_{i, lc}^{r+1}- \thetab_{i, lc}^r \rangle] + \frac{L_F}{2}\E[\| \thetab_{i, lc}^{r+1}- \thetab_{i, lc}^r\|^2] \notag \\
		&+	\E[ \langle \nabla_{\thetab} F_i(\thetab_{i, lc}^r,\thetab^{r},  \nub_{i}^{r+1}),\thetab^{r+1}- \thetab^r \rangle]+ \frac{L_F}{2}\E[\|\thetab^{r+1}- \thetab^r\|^2] \notag \\
		 =& \frac{m}{N}\E[ \langle \nabla_{\thetab_{i, lc}} F_i(\thetab_{i, lc}^r,\thetab^{r},  \wt \nub_{i}^{r+1}),\wt \thetab_{i, lc}^{r+1}- \thetab_{i, lc}^r \rangle] +	\E[ \langle \nabla_{\thetab} F_i(\thetab_{i, lc}^r,\thetab^{r},  \nub_{i}^{r+1}),\thetab^{r+1}- \thetab^r \rangle] \notag \\
		&+ \frac{mL_F}{2N}\E[\| \wt \thetab_{i, lc}^{r+1}- \thetab_{i, lc}^r\|^2] + \frac{L_F}{2}\E[\|\thetab^{r+1}- \thetab^r\|^2]. \label{lem5: bd4} 
	\end{align}
\end{small}We proceed to bound the terms in the RHS of \eqref{lem5: bd4} with the following Lemma.

\begin{Lemma} \label{lem: local_some}
	For any round $r$, it holds that
	\begin{small}
		\begin{align}
			\E[\|\wt \thetab_{i, lc}^{r+1} -  \thetab_{i, lc}^{r}\|^2] 
			& \leq  \eta_c^2(\wt Q_i^r)^2 \E\bigg[\bigg\|\sum_{t=0}^{Q_i^r-1}\frac{(\bb_i^r)^t}{\wt Q_i^r}\nabla_{\thetab_{i, lc}}F_i(\wt \thetab_{i, lc}^{r,t}, \wt \thetab_i^{r, t},\wt \nub_{i}^{r+1})\bigg\|^2\bigg] +\frac{ \eta_c^2(\wt Q_i^r)^2\sigma^2}{S}, \\
			\E[ \langle \nabla_{\thetab_{i, lc}} F_i(\thetab_{i, lc}^r,\thetab^{r}, \wt \nub_{i}^{r+1}),\wt \thetab_{i, lc}^{r+1}- \thetab_{i, lc}^r \rangle] 
			& \leq  - \frac{\eta_{c} \wt Q_i^r}{2} \E[\|\nabla_{\thetab_{i, lc}} F_i(\thetab_{i, lc}^r,\thetab^{r}, \wt \nub_{i}^{r+1})\|^2]+ \frac{\eta_c^3 L_F^2(\wt Q_i^r)^2\sigma^2}{2S} \notag \\
			&~~~~ - \frac{\eta_{c} \wt Q_i^r(1 - \eta_c^2(\wt Q_i^r)^2L_F^2)}{2} \E\bigg[\bigg\|\sum_{t - 0}^{Q_i^r - 1}\frac{(\bb_i^r)^t}{\wt Q_i^r} \nabla_{\thetab_{i, lc}} F_i(\wt \thetab_{i, lc}^{r,t}, \wt \thetab_i^{r, t}, \wt \nub_{i}^{r+1})\bigg\|^2\bigg] \notag \\
			&~~~ +\frac{\eta_c \wt Q_i^rL_F^2}{2} \E\bigg[\sum_{t = 0}^{Q_i^r - 1} \frac{(\bb_i^r)^t}{\wt Q_i^r}\|\wt\thetab_i^{r, t} - \thetab^{r}\|^2\bigg], \\
			\E[ \langle \nabla_{\thetab} F_i(\thetab_{i, lc}^r,\thetab^{r},  \nub_{i}^{r+1}),\thetab^{r+1}- \thetab^r \rangle]
			& \leq  \beta_i^2(1-\beta_i)(\wt L-\mu)\Gamma^2+ (1-\beta_i) \E[ \langle \nabla_{\thetab} f_i(\thetab^r, \nub_{i}^{r}), \thetab^{r+1}- \thetab^r \rangle]  	 \notag \\
			&~~~~ +\frac{2(1-\beta_i)mL}{N}\E[\|\wt \nub_{i}^{r+1} - \nub_i^r\|^2] + \beta_i(1-\beta_i)\wt L \Gamma^2 \notag \\
			&~~~~ + \frac{(1-\beta_i)(4\beta_i\wt LL_h \Gamma + L)}{8}\E[\|\thetab^{r+1} - \thetab^r\|^2]. 
		\end{align}
	\end{small}
\end{Lemma}

Applying Lemma \ref{lem: local_some} to \eqref{lem5: bd4} yields
\begin{small}
	\begin{align}
		&\E[F_i(\thetab_{i, lc}^{r+1}, \thetab^{r+1}, \nub_{i}^{r+1})] - \E[F_i(\thetab_{i, lc}^r,\thetab^{r}, \nub_{i}^{r+1})] \notag \\
		& \leq  - \frac{\eta_c \wt Q_i^r m}{2N} \E[\|\nabla_{\thetab_{i, lc}} F_i(\thetab_{i, lc}^r,\thetab^{r}, \wt \nub_{i}^{r+1})\|^2] + \frac{\eta_c^2L_F(\eta_c L_F + 1)(\wt Q_i^r)^2m\sigma^2}{2NS} \notag \\
		&~~~~ - \frac{\eta_c \wt Q_i^rm(1 - \eta_c\wt Q_i^rL_F - \eta_c^2(\wt Q_i^r)^2L_F^2)}{2N}  \E\bigg[\bigg\|\sum_{t - 0}^{Q_i^r - 1}\frac{(\bb_i^r)^t}{\wt Q_i^r} \nabla_{\thetab_{i, lc}} F_i(\wt \thetab_{i, lc}^{r,t}, \wt \thetab_i^{r, t}, \wt \nub_{i}^{r+1})\bigg\|^2\bigg] \notag \\
		&~~~~ +\frac{\eta_c \wt Q_i^rmL_F^2}{2N} \E\bigg[\sum_{t = 0}^{Q_i^r - 1} \frac{(\bb_i^r)^t}{\wt Q_i^r}\|\wt\thetab_i^{r, t} - \thetab^{r}\|^2\bigg] + \frac{2(1-\beta_i)mL}{N} \E[\|\wt \nub_{i}^{r+1} - \nub_i^r\|^2]\notag \\
		&~~~~ +  (1-\beta_i) \E[ \langle \nabla_{\thetab} f_i(\thetab^r,  \nub_{i}^{r}), \thetab^{r+1}- \thetab^r \rangle] + \frac{(1-\beta_i)(4\beta_i\wt LL_h \Gamma + L) + 4L_F}{8}\E[\|\thetab^{r+1} - \thetab^r\|^2] \notag \\
		&~~~~ +\beta_i^2(1-\beta_i)(\wt L-\mu)\Gamma^2+ \beta_i(1-\beta_i)\wt L \Gamma^2	\notag \\
		& \leq  - \frac{\eta_c \wt Q_i^rm}{2N} \E[\|\nabla_{\thetab_{i, lc}} F_i(\thetab_{i, lc}^r,\thetab^{r}, \wt \nub_{i}^{r+1})\|^2] + \frac{\eta_c^2L_F(\eta_c L_F + 1)(\wt Q_i^r)^2m\sigma^2}{2NS} \notag \\
		&~~~~ +\frac{\eta_c \wt Q_i^r m L_F^2}{2N} \E\bigg[\sum_{t = 0}^{Q_i^r - 1} \frac{(\bb_i^r)^t}{\wt Q_i^r}\|\wt\thetab_i^{r, t} - \thetab^{r}\|^2\bigg] + \frac{2(1-\beta_i)mL}{N} \E[\|\wt \nub_{i}^{r+1} - \nub_i^r\|^2]\notag \\
		&~~~~ +  (1-\beta_i) \E[ \langle \nabla_{\thetab} f_i(\thetab^r,  \nub_{i}^{r}), \thetab^{r+1}- \thetab^r \rangle] + \frac{(1-\beta_i)(4\beta_i\wt LL_h \Gamma + 5L)}{8}\E[\|\thetab^{r+1} - \thetab^r\|^2] \notag \\
		&~~~~ +\beta_i^2(1-\beta_i)(\wt L-\mu)\Gamma^2+ \beta_i(1-\beta_i)\wt L \Gamma^2, \label{lem5: bd5} 
	\end{align}
\end{small}where \eqref{lem5: bd5} follows because $\eta_c \ol Q L_F \leq \frac{1}{2}$, and $L_F \leq (1-\beta_i)L$. We combine \eqref{lem5: bd3_1} and \eqref{lem5: bd5} to obtain
\begin{small}
	\begin{align}
		&\E[F_i(\thetab_{i, lc}^{r+1}, \thetab^{r+1},  \nub_{i}^{r+1})] - \E[F_i(\thetab_{i, lc}^r,\thetab^{r},  \nub_{i}^{r})] \notag \\
		& \leq  - \frac{\eta_c \wt Q_i^rm}{2N} \E[\|\nabla_{\thetab_{i, lc}} F_i(\thetab_{i, lc}^r,\thetab^{r}, \wt \nub_{i}^{r+1})\|^2] + \frac{\eta_c^2L_F(\eta_c L_F + 1)(\wt Q_i^r)^2m\sigma^2}{2NS} \notag \\
		&~~~~ +\frac{\eta_c \wt Q_i^rmL_F^2}{2N} \E\bigg[\sum_{t = 0}^{Q_i^r - 1} \frac{(\bb_i^r)^t}{\wt Q_i^r}\|\wt\thetab_i^{r, t} - \thetab^{r}\|^2\bigg] +  (1-\beta_i) \E[ \langle \nabla_{\thetab} f_i(\thetab^r, \nub_{i}^{r}), \thetab^{r+1}- \thetab^r \rangle] \notag \\
		&~~~~ + \frac{(1-\beta_i)(4\beta_i\wt LL_h \Gamma + 5L)}{8}\E[\|\thetab^{r+1} - \thetab^r\|^2] + \beta_i^2(1-\beta_i)(\wt L-\mu)\Gamma^2+ \beta_i(1-\beta_i)\wt L \Gamma^2 +\frac{m\wt L^2 \beta_i^2\Gamma^2}{2NL}\notag \\
		&~~~~ - \bigg(\frac{1}{\eta_v} + \frac{\mu}{2}- \frac{3L}{2} - \frac{2(1- \beta_i)mL}{N}\bigg) \E[\|\wt \nub_{i}^{r+1} - \nub_{i}\|^2]. \label{lem5: bd6} 
	\end{align}
\end{small}\hfill $\blacksquare$

\section{Proof of Lemma \ref{lem: local_some}}
First, by the definition of $\wt \thetab_{i, lc}^{r+1}$, we have
\begin{small}
	\begin{align}
		\E[\wt \thetab_{i, lc}^{r+1}- \thetab_{i, lc}^r] 
		& =  -\eta_{c}\wt Q_i^r	\E\bigg[ \sum_{t - 0}^{Q_i^r - 1} \frac{(\bb_i^r)^t}{\wt Q_i^r}G_i(\wt \thetab_{i, lc}^{r,t}, \wt \thetab_i^{r, t}, \wt \nub_{i}^{r+1})\bigg] \label{lem6: bd1}\\
		& =  - \eta_{c} \wt Q_i^r \E\bigg[ \sum_{t - 0}^{Q_i^r - 1}  \frac{(\bb_i^r)^t}{\wt Q_i^r}\nabla_{\thetab_{i, lc}} F_i(\wt \thetab_{i, lc}^{r,t}, \wt \thetab_i^{r, t},\wt  \nub_{i}^{r+1})\bigg], \label{lem6: bd2}
	\end{align}
\end{small}where \eqref{lem6: bd1} holds by following the spirit as (21). Then, we get
\begin{small}
	\begin{align}
		&\E[ \langle \nabla_{\thetab_{i, lc}} F_i(\thetab_{i, lc}^r,\thetab^{r}, \wt \nub_{i}^{r+1}),\wt \thetab_{i, lc}^{r+1}- \thetab_{i, lc}^r \rangle] \notag \\
		& = -\eta_{c}  \wt Q_i^r \E\bigg[\bigg\langle \nabla_{\thetab_{i, lc}} F_i(\thetab_{i, lc}^r,\thetab^{r}, \wt \nub_{i}^{r+1}),\sum_{t - 0}^{Q_i^r - 1}  \frac{(\bb_i^r)^t}{\wt Q_i^r}\nabla_{\thetab_{i, lc}} F_i(\wt \thetab_{i, lc}^{r,t}, \wt \thetab_i^{r, t}, \wt \nub_{i}^{r+1}) \bigg\rangle\bigg]	\notag \\
		& =  - \frac{\eta_{c} \wt Q_i^r}{2} \E[\|\nabla_{\thetab_{i, lc}} F_i(\thetab_{i, lc}^r,\thetab^{r}, \wt \nub_{i}^{r+1})\|^2]  - \frac{\eta_{c} \wt Q_i^r}{2} \E\bigg[\bigg\|\sum_{t - 0}^{Q_i^r - 1}  \frac{(\bb_i^r)^t}{\wt Q_i^r}\nabla_{\thetab_{i, lc}} F_i(\wt \thetab_{i, lc}^{r,t}, \wt \thetab_i^{r, t}, \wt \nub_{i}^{r+1})\bigg\|^2\bigg] \notag \\
		&~~~~ + \frac{\eta_{c} \wt Q_i^r}{2} \E\bigg[\bigg\|\nabla_{\thetab_{i, lc}} F_i(\thetab_{i, lc}^r,\thetab^{r}, \wt \nub_{i}^{r+1}) -\sum_{t - 0}^{Q_i^r - 1}  \frac{(\bb_i^r)^t}{\wt Q_i^r}\nabla_{\thetab_{i, lc}} F_i(\wt \thetab_{i, lc}^{r,t}, \wt \thetab_i^{r, t}, \wt \nub_{i}^{r+1}) \bigg\|^2\bigg]  \label{lem6: bd3}\\
		& \leq  - \frac{\eta_{c} \wt Q_i^r}{2} \E[\|\nabla_{\thetab_{i, lc}} F_i(\thetab_{i, lc}^r,\thetab^{r}, \wt \nub_{i}^{r+1})\|^2] - \frac{\eta_{c} \wt Q_i^r}{2} \E\bigg[\bigg\|\sum_{t - 0}^{Q_i^r - 1}  \frac{(\bb_i^r)^t}{\wt Q_i^r}\nabla_{\thetab_{i, lc}} F_i(\wt \thetab_{i, lc}^{r,t}, \wt \thetab_i^{r, t}, \wt \nub_{i}^{r+1})\bigg\|^2\bigg] \notag \\
		&~~~~ + \frac{\eta_{c} \wt Q_i^r}{2}\E\bigg[\sum_{t = 0}^{Q_i^r - 1} \frac{(\bb_i^r)^t}{\wt Q_i^r}\|\nabla_{\thetab_{i, lc}} F_i(\thetab_{i, lc}^r,\thetab^{r}, \wt \nub_{i}^{r+1}) - \nabla_{\thetab_{i, lc}} F_i(\wt \thetab_{i, lc}^{r,t}, \wt\thetab_i^{r, t}, \wt \nub_{i}^{r+1})\|^2\bigg]  \label{lem6: bd4}\\
		& \leq - \frac{\eta_{c} \wt Q_i^r}{2} \E[\|\nabla_{\thetab_{i, lc}} F_i(\thetab_{i, lc}^r,\thetab^{r}, \wt \nub_{i}^{r+1})\|^2] - \frac{\eta_{c} \wt Q_i^r}{2} \E\bigg[\bigg\|\sum_{t - 0}^{Q_i^r - 1} \frac{(\bb_i^r)^t}{\wt Q_i^r}\nabla_{\thetab_{i, lc}} F_i(\wt \thetab_{i, lc}^{r,t}, \wt \thetab_i^{r, t}, \wt \nub_{i}^{r+1})\bigg\|^2\bigg] \notag \\
		&~~~~ + \frac{\eta_{c} L_F^2\wt Q_i^r}{2}\E\bigg[\sum_{t = 0}^{Q_i^r - 1}\frac{(\bb_i^r)^t}{\wt Q_i^r}(\|\wt \thetab_{i, lc}^{r,t} -  \thetab_{i, lc}^{r}\|^2 + \|\wt\thetab_i^{r, t} - \thetab^{r}\|^2)\bigg], \label{lem6: bd5}
	\end{align}
\end{small}where \eqref{lem6: bd3} follows because $\langle \vb_1, \vb_2 \rangle = \frac{1}{2}\|\vb_1\|^2 + \frac{1}{2}\|\vb_2\|^2 - \frac{1}{2}\|\vb_1 - \vb_2\|^2, \forall \vb_1, \vb_2 \in \Rbb^n$; \eqref{lem6: bd4} follows by the convexity of $\|\cdot\|^2$; \eqref{lem6: bd5} holds by Assumption 3. Furthermore, note that
\begin{small}
	\begin{align}
		\E[\|\wt \thetab_{i, lc}^{r+1} -  \thetab_{i, lc}^{r}\|^2] 
		& =   \eta_{c}^2(\wt Q_i^r)^2\E\bigg[\bigg\|\sum_{t=0}^{Q_i^r-1}\frac{(\bb_i^r)^t}{\wt Q_i^r} G_i(\wt \thetab_{i, lc}^{r,t}, \wt \thetab_i^{r, t},\wt \nub_{i}^{r+1})\bigg\|^2\bigg] \notag \\
		& \leq  \eta_{c}^2 (\wt Q_i^r)^2\E\bigg[\bigg\|\sum_{t=0}^{Q_i^r-1}\frac{(\bb_i^r)^t}{\wt Q_i^r}\nabla_{\thetab_{i, lc}}F_i(\wt \thetab_{i, lc}^{r,t}, \wt \thetab_i^{r, t},\wt \nub_{i}^{r+1})\bigg\|^2\bigg] +\frac{ \eta_{c}^2(\wt Q_i^r)^2\sigma^2}{S}, \label{lem6: bd6}
	\end{align}
\end{small}where \eqref{lem6: bd6} is obtained by following the same spirit as \eqref{lem2: bd6}. Thus, we substitute \eqref{lem6: bd6} into \eqref{lem6: bd5} to have
\begin{small}
	\begin{align}
		&\E[ \langle \nabla_{\thetab_{i, lc}} F_i(\thetab_{i, lc}^r,\thetab^{r}, \wt \nub_{i}^{r+1}),\wt \thetab_{i, lc}^{r+1}- \thetab_{i, lc}^r \rangle] \notag \\
		& \leq  - \frac{\eta_{c} \wt Q_i^r}{2} \E[\|\nabla_{\thetab_{i, lc}} F_i(\thetab_{i, lc}^r,\thetab^{r}, \wt \nub_{i}^{r+1})\|^2]+ \frac{\eta_c^3 L_F^2(\wt Q_i^r)^2\sigma^2}{2S} \notag \\
		&~~~~ - \frac{\eta_{c} \wt Q_i^r(1 - \eta_c^2(\wt Q_i^r)^2L_F^2)}{2} \E\bigg[\bigg\|\sum_{t - 0}^{Q_i^r - 1}\frac{(\bb_i^r)^t}{\wt Q_i^r} \nabla_{\thetab_{i, lc}} F_i(\wt \thetab_{i, lc}^{r,t}, \wt \thetab_i^{r, t}, \wt \nub_{i}^{r+1})\bigg\|^2\bigg] \notag \\
		&~~~~ +\frac{\eta_c L_F^2\wt Q_i^r}{2} \E\bigg[\sum_{t = 0}^{Q_i^r - 1} \frac{(\bb_i^r)^t}{\wt Q_i^r}\|\wt\thetab_i^{r, t} - \thetab^{r}\|^2\bigg].\label{lem6: bd7}
	\end{align}
\end{small}
Second, by the definition of $F_i$ and $L_i$, we have
\begin{small}
	\begin{align}
		\nabla_{\thetab} F_i(\thetab_{i, lc}^r,\thetab^{r},  \nub_{i}^{r+1}) 
		& = (1-\beta_i) \nabla h(\thetab^r) \nabla L_i(\beta_i h(\thetab_{i, lc}^{r} )+ (1-\beta_i) h(\thetab^r) , \nub_{i}^{r+1}), \\
		\nabla_{\thetab} f_i(\thetab^r, \nub_{i}^{r+1}) 
		& =  \nabla h(\thetab^r) \nabla L_i( h(\thetab^r) , \nub_{i}^{r+1}).
	\end{align}
\end{small}
Then, we obtain
\begin{small}
	\begin{align}
		&\E[ \langle \nabla_{\thetab} F_i(\thetab_{i, lc}^r,\thetab^{r},  \nub_{i}^{r+1}),\thetab^{r+1}- \thetab^r \rangle] \notag \\
		 =&  \E[ \langle \nabla_{\thetab} F_i(\thetab_{i, lc}^r,\thetab^{r},  \nub_{i}^{r+1}) - (1-\beta_i)\nabla_{\thetab} f_i(\thetab^r,  \nub_{i}^{r+1}), \thetab^{r+1}- \thetab^r \rangle] + (1-\beta_i)\E[ \langle \nabla_{\thetab} f_i(\thetab^r, \nub_{i}^{r+1}), \thetab^{r+1}- \thetab^r \rangle] \notag \\
		 =&  (1-\beta_i) \E[ \langle \nabla L_i(\beta_ih(\thetab_{i, lc}^{r} )+ (1-\beta_i) h(\thetab^r),  \nub_{i}^{r+1})- \nabla L_i( h(\thetab^r, \nub_{i}^{r+1}),\nabla h(\thetab^r)^\top (\thetab^{r+1}- \thetab^r )\rangle] \notag \\
		& + (1-\beta_i) \E[ \langle \nabla_{\thetab} f_i(\thetab^r, \nub_{i}^{r+1}), \thetab^{r+1}- \thetab^r \rangle] \notag \\
		= & (1-\beta_i) \E[ \langle \nabla L_i(\zb_i^r , \nub_{i}^{r+1})- \nabla L_i( h(\thetab^r) , \nub_{i}^{r+1}), \nabla h(\thetab^r)^\top (\thetab^{r+1}- \thetab^r )\rangle] \notag \\
		& + (1-\beta_i) \E[ \langle \nabla_{\thetab} f_i(\thetab^r, \nub_{i}^{r+1}), \thetab^{r+1}- \thetab^r \rangle]  \\
		= & (1-\beta_i) \E[ \langle \nabla L_i(\zb_i^r , \nub_{i}^{r+1})- \nabla L_i( h(\thetab^r) , \nub_{i}^{r+1}), -(\zb_i^r -h(\thetab^r)) +(\zb_i^r -h(\thetab^r)) + \nabla h(\thetab^r)^\top (\thetab^{r+1}- \thetab^r )\rangle] \notag \\
		&+ (1-\beta_i) \E[ \langle \nabla_{\thetab} f_i(\thetab^r,  \nub_{i}^{r+1}), \thetab^{r+1}- \thetab^r \rangle] \notag \\
		 \leq&  -(1-\beta_i)\mu\E[\|\zb_i^r -h(\thetab^r)\|^2] + (1-\beta_i) \E[ \langle \nabla_{\thetab} f_i(\thetab^r, \nub_{i}^{r+1}), \thetab^{r+1}- \thetab^r \rangle] \notag \\
		& + (1-\beta_i) \E[ \langle \nabla L_i(\zb_i^r ,  \nub_{i}^{r+1})- \nabla L_i( h(\thetab^r) ,  \nub_{i}^{r+1}), \zb_i^r -h(\thetab^r) + \nabla h(\thetab^r)^\top (\thetab^{r+1}- \thetab^r )\rangle]\label{lem6: bd8} \\
		 \leq & -(1-\beta_i)\mu\E[\|\zb_i^r -h(\thetab^r)\|^2] + (1-\beta_i) \E[ \langle \nabla_{\thetab} f_i(\thetab^r, \nub_{i}^{r+1}), \thetab^{r+1}- \thetab^r \rangle] \notag \\
		& + (1-\beta_i)\wt L \E[\|\zb_i^r -h(\thetab^r)\| \|\zb_i^r -h(\thetab^r) + \nabla h(\thetab^r)^\top (\thetab^{r+1}- \thetab^r )\|], \label{lem6: bd9} 
	\end{align}
\end{small}where $\zb_i^r \triangleq \beta_ih(\thetab_{i, lc}^{r} )+ (1-\beta_i) h(\thetab^r)$; \eqref{lem6: bd8} follows by the strong convexity of $L_i(\cdot, \nub_{i})$, i.e.,
\begin{small}
	\begin{align}
		&\langle \nabla L_i(\zb_i^r , \nub_{i}^{r+1})- \nabla L_i( h(\thetab^r) ,  \nub_{i}^{r+1}), \zb_i^r -h(\thetab^r) \rangle \geq  \mu\|\zb_i^r -h(\thetab^r)\|^2; 
	\end{align}
\end{small}\eqref{lem6: bd9} follows by the Cauchy-Schwarz inequality and Assumption 3.  We proceed to bound $\|\zb_i^r -h(\thetab^r) + \nabla h(\thetab^r)^\top (\thetab^{r+1}- \thetab^r )\|$ by
\begin{small}
	\begin{align}
		&\|\zb_i^r -h(\thetab^r) + \nabla h(\thetab^r)^\top (\thetab^{r+1}- \thetab^r )\| \notag \\
		& = 	\|\beta_i(h(\thetab_{i, lc}^r) -h(\thetab^r)) + \nabla h(\thetab^r)^\top (\thetab^{r+1}- \thetab^r )\| \notag \\
		& \leq  \beta_i\|h(\thetab_{i, lc}^r) -h(\thetab^r) \|+ \|\nabla h(\thetab^r)^\top (\thetab^{r+1}- \thetab^r )\| \notag \\
		& \leq  \beta_i\|h(\thetab_{i, lc}^r) -h(\thetab^r) \| + \bigg\|h(\thetab^r) - h(\thetab^{r+1}) + \frac{L_h}{2}\|\thetab^{r+1} - \thetab^r\|^2\bigg\| \label{lem6: bd10} \\
		& \leq   \beta_i\|h(\thetab_{i, lc}^r) -h(\thetab^r) \|  + \|h(\thetab^r) - h(\thetab^{r+1})\| + \frac{L_h}{2}\|\thetab^{r+1} - \thetab^r\|^2, \label{lem6: bd11}
	\end{align}
\end{small}where \eqref{lem6: bd10} follows by Assumption 3. Substituting \eqref{lem6: bd11} into \eqref{lem6: bd9} gives rise to
\begin{small}
	\begin{align}
		&	\E[ \langle \nabla_{\thetab} F_i(\thetab_{i, lc}^r,\thetab^{r},  \nub_{i}^{r+1}),\thetab^{r+1}- \thetab^r \rangle] \notag \\
		& \leq \beta_i^2(1-\beta_i)(\wt L-\mu)\E[\|h(\thetab_{i, lc}) -h(\thetab^r)\|^2]+ (1-\beta_i) \E[ \langle \nabla_{\thetab} f_i(\thetab^r, \nub_{i}^{r+1}), \thetab^{r+1}- \thetab^r \rangle] \notag \\
		&~~~~+ \beta_i(1-\beta_i)\wt L \E[\|h(\thetab_{i, lc}) -h(\thetab^r)\|
		\|h(\thetab^r) - h(\thetab^{r+1})\|]	\notag \\
		&~~~~ + \frac{\beta_i(1-\beta_i)\wt L L_h}{2}\E[\|h(\thetab_{i, lc}) -h(\thetab^r)\|\|\thetab^{r+1} - \thetab^r\|^2] \notag \\
		& \leq  \beta_i^2(1-\beta_i)(\wt L-\mu)\E[\|h(\thetab_{i, lc}) -h(\thetab^r)\|^2]+ (1-\beta_i) \E[ \langle \nabla_{\thetab} f_i(\thetab^r, \nub_{i}^{r+1}), \thetab^{r+1}- \thetab^r \rangle] \notag \\
		&~~~~+ \beta_i(1-\beta_i) \wt L \E[\|h(\thetab_{i, lc}) -h(\thetab^r)\|
		\|h(\thetab^r) - h(\thetab^{r+1})\|]	\notag \\
		&~~~~ + \frac{\beta_i(1-\beta_i)\wt LL_h}{2}\E[\|h(\thetab_{i, lc}) -h(\thetab^r)\|\|\thetab^{r+1} - \thetab^r\|^2] \notag \\
		& \leq  \beta_i^2(1-\beta_i)(\wt L-\mu)\Gamma^2+ (1-\beta_i) \E[ \langle \nabla_{\thetab} f_i(\thetab^r, \nub_{i}^{r+1}), \thetab^{r+1}- \thetab^r \rangle] + \beta_i(1-\beta_i)\wt L \Gamma^2 \notag \\
		&~~~~ + \frac{\beta_i(1-\beta_i)\wt LL_h \Gamma}{2}\E[\|\thetab^{r+1} - \thetab^r\|^2] \label{lem6: bd12}\\
		& = \beta_i^2(1 \!-\! \beta_i)(\wt L \!-\! \mu)\Gamma^2 \!+\! (1 \!-\! \beta_i) \E[ \langle \nabla_{\thetab} f_i(\thetab^r, \nub_{i}^{r}), \thetab^{r+1} \!-\! \thetab^r \rangle]  \!+\! (1 \!-\! \beta_i) \E[ \langle \nabla_{\thetab} f_i(\thetab^r,  \nub_{i}^{r+1}) - \nabla_{\thetab} f_i(\thetab^r,  \nub_{i}^{r}), \thetab^{r+1} \!-\! \thetab^r \rangle] \notag \\
		&~~~~+ \beta_i(1-\beta_i)\wt L \Gamma^2	 + \frac{\beta_i(1-\beta_i)\wt LL_h \Gamma}{2}\E[\|\thetab^{r+1} - \thetab^r\|^2] \notag \\
		& \leq  \beta_i^2(1-\beta_i)(\wt L-\mu)\Gamma^2+ (1-\beta_i) \E[ \langle \nabla_{\thetab} f_i(\thetab^r, \nub_{i}^{r}), \thetab^{r+1}- \thetab^r \rangle] +\frac{2(1-\beta_i)mL}{N}\E[\|\wt \nub_{i}^{r+1} - \nub_i^r\|^2] 	 \notag \\
		&~~~~  + \beta_i(1-\beta_i)\wt L \Gamma^2 + \frac{(1-\beta_i)(4\beta_i\wt LL_h \Gamma + L)}{8}\E[\|\thetab^{r+1} - \thetab^r\|^2],  \label{lem6: bd13}
	\end{align}
\end{small}where \eqref{lem6: bd12} follows by Assumption 5; \eqref{lem6: bd13} holds by the Cauchy-Schwarz inequality, i,e., $\langle \db_1, \db_2\rangle \leq \frac{2}{L} \|\db_1\|^2 + \frac{L}{8}\|\db_2\|^2, \forall \db_1, \db_2$, and $\Prob(i \in \Ac^r) = \frac{m}{N}$.\hfill $\blacksquare$

\bibliographystyle{IEEEtran}
\bibliography{refs20,refs10}

\begin{thebibliography}{10}
\providecommand{\url}[1]{#1}
\csname url@samestyle\endcsname
\providecommand{\newblock}{\relax}
\providecommand{\bibinfo}[2]{#2}
\providecommand{\BIBentrySTDinterwordspacing}{\spaceskip=0pt\relax}
\providecommand{\BIBentryALTinterwordstretchfactor}{4}
\providecommand{\BIBentryALTinterwordspacing}{\spaceskip=\fontdimen2\font plus
\BIBentryALTinterwordstretchfactor\fontdimen3\font minus
  \fontdimen4\font\relax}
\providecommand{\BIBforeignlanguage}[2]{{%
\expandafter\ifx\csname l@#1\endcsname\relax
\typeout{** WARNING: IEEEtran.bst: No hyphenation pattern has been}%
\typeout{** loaded for the language `#1'. Using the pattern for}%
\typeout{** the default language instead.}%
\else
\language=\csname l@#1\endcsname
\fi
#2}}
\providecommand{\BIBdecl}{\relax}
\BIBdecl

\bibitem{Tariq2020WC}
F.~Tariq, M.~R.~A. Khandaker, K.-K. Wong, M.~A. Imran, M.~Bennis, and
  M.~Debbah, ``A speculative study on {6G},'' \emph{IEEE Wireless Commun.},
  vol.~27, no.~4, pp. 118--125, Aug. 2020.

\bibitem{Tong2022WC}
W.~Tong and G.~Y. Li, ``Nine challenges in artificial intelligence and wireless
  communications for {6G},'' \emph{IEEE Wireless Commun.}, pp. 1--10,
  10.1109/MWC.006.2100543 2022.

\bibitem{Letaief2019CM}
K.~B. Letaief, W.~Chen, Y.~Shi, J.~Zhang, and Y.-J.~A. Zhang, ``The roadmap to
  {6G: AI} empowered wireless networks,'' \emph{IEEE Commun. Mag.}, vol.~57,
  no.~8, pp. 84--90, Aug. 2019.

\bibitem{Nguyen2021CST}
D.~C. Nguyen, M.~Ding, P.~N. Pathirana, A.~Seneviratne, J.~Li, and
  H.~Vincent~Poor, ``Federated learning for {Internet} of things: A
  comprehensive survey,'' \emph{IEEE Commun. Surv. Tutorials}, vol.~23, no.~3,
  pp. 1622--1658, Apr. 2021.

\bibitem{Khan2021CST}
L.~U. Khan, W.~Saad, Z.~Han, E.~Hossain, and C.~S. Hong, ``Federated learning
  for {Internet} of things: Recent advances, taxonomy, and open challenges,''
  \emph{IEEE Commun. Surv. Tutorials}, vol.~23, no.~3, pp. 1759--1799, Jun.
  2021.

\bibitem{Yang2020TCOM}
H.~H. Yang, Z.~Liu, T.~Q.~S. Quek, and H.~V. Poor, ``Scheduling policies for
  federated learning in wireless networks,'' \emph{IEEE Trans. Commun.},
  vol.~68, no.~1, pp. 317--333, Jan. 2020.

\bibitem{ChenTWC2021}
M.~Chen, Z.~Yang, W.~Saad, C.~Yin, H.~V. Poor, and S.~Cui, ``A joint learning
  and communications framework for federated learning over wireless networks,''
  \emph{IEEE Trans. Wireless Commun.}, vol.~20, no.~1, pp. 269--283, Jan. 2021.

\bibitem{ZhuTWC2020}
G.~Zhu, Y.~Wang, and K.~Huang, ``Broadband analog aggregation for low-latency
  federated edge learning,'' \emph{IEEE Trans. Wireless Commun.}, vol.~19,
  no.~1, pp. 491--506, Oct. 2020.

\bibitem{Elbir2020CL}
A.~M. Elbir and S.~Coleri, ``Federated learning for hybrid beamforming in
  mm-wave massive {MIMO},'' \emph{IEEE Commun. Lett.}, vol.~24, no.~12, pp.
  2795--2799, Dec. 2020.

\bibitem{Zheng2022TSP}
X.~Zheng and V.~Lau, ``Federated online deep learning for csit and csir
  estimation of fdd multi-user massive mimo{} systems,'' \emph{IEEE Trans.
  Signal Process.}, vol.~70, pp. 2253--2266, Apr. 2022.

\bibitem{Elbir2022TWC}
A.~M. Elbir and S.~Coleri, ``Federated learning for channel estimation in
  conventional and {RIS}-assisted massive {MIMO},'' \emph{IEEE Trans. Wireless
  Commun.}, vol.~21, no.~6, pp. 4255--4268, Jun. 2022.

\bibitem{FedAvg_noniid_2019}
X.~Li, K.~Huang, W.~Yang, S.~Wang, and Z.~Zhang, ``On the convergence of
  {FedAvg} on non-iid data,'' in \emph{Proc. ICLR}, Addis Ababa, ETHIOPIA, Apr.
  26 - May 1, 2020, pp. 1--11.

\bibitem{shiTWC2022}
D.~Shi, L.~Li, M.~Wu, M.~Shu, R.~Yu, M.~Pan, and Z.~Han, ``To talk or to work:
  Dynamic batch sizes assisted time efficient federated learning over future
  mobile edge devices,'' \emph{IEEE Trans. Wireless Commun.}, vol.~21, no.~12,
  pp. 11\,038--11\,050, Dec. 2022.

\bibitem{millsTPDS2021}
J.~Mills, J.~Hu, and G.~Min, ``Multi-task federated learning for personalised
  deep neural networks in edge computing,'' \emph{IEEE Trans. Parallel Distrib.
  Syst.}, vol.~33, no.~3, pp. 630--641, Mar. 2022.

\bibitem{ZhaoTWC2022}
Z.~Zhao, C.~Feng, W.~Hong, J.~Jiang, C.~Jia, T.~Q.~S. Quek, and M.~Peng,
  ``Federated learning with {Non-IID} data in wireless networks,'' \emph{IEEE
  Trans. Wireless Commun.}, vol.~21, no.~3, pp. 1927--1942, Mar. 2022.

\bibitem{Zhang2021IoT}
C.~Zhang, Y.~Zhu, C.~Markos, S.~Yu, and J.~J. Yu, ``Towards crowdsourced
  transportation mode identification: A semi-supervised federated learning
  approach,'' \emph{IEEE Internet Things J.}, pp. 1--1, 2021.

\bibitem{Zhang2022IoT}
Z.~Zhang, S.~Ma, Z.~Yang, Z.~Xiong, J.~Kang, Y.~Wu, K.~Zhang, and D.~Niyato,
  ``Robust semi-supervised federated learning for images automatic recognition
  in internet of drones,'' \emph{IEEE Internet Things J.}, pp. 1--1, 2022.

\bibitem{FedSSL}
\BIBentryALTinterwordspacing
Z.~Wang, X.~Wang, R.~Sun, and T.-H. Chang, ``Federated semi-supervised learning
  with class distribution mismatch,'' 2021. [Online]. Available:
  \url{https://arxiv.org/abs/2111.00010}
\BIBentrySTDinterwordspacing

\bibitem{Ma2021JSAC}
Q.~Ma, Y.~Xu, H.~Xu, Z.~Jiang, L.~Huang, and H.~Huang, ``{FedSA}: A
  semi-asynchronous federated learning mechanism in heterogeneous edge
  computing,'' \emph{IEEE J. Sel. Areas Commun.}, vol.~39, no.~12, pp.
  3654--3672, Otc. 2021.

\bibitem{Zhao2022TWC}
Z.~Zhao, C.~Feng, W.~Hong, J.~Jiang, C.~Jia, T.~Q.~S. Quek, and M.~Peng,
  ``Federated learning with {Non-IID} data in wireless networks,'' \emph{IEEE
  Trans. Wireless Commun.}, vol.~21, no.~3, pp. 1927--1942, Sep. 2022.

\bibitem{SCAFFOLD_2020}
S.~P. Karimireddy, S.~Kale, M.~Mohri, S.~Reddi, S.~Stich, and A.~T. Suresh,
  ``Scaffold: Stochastic controlled averaging for federated learning,'' in
  \emph{Proc. ICML}, Jul. 13-18 2020, pp. 5132--5143.

\bibitem{FedMAJ_2020}
S.~Wang and T.-H. Chang, ``Federated matrix factorization: Algorithm design and
  application to data clustering,'' \emph{IEEE Trans. Signal Process.},
  vol.~70, pp. 1625--1640, Feb. 2022.

\bibitem{APFL_2020}
Y.~Deng, M.~M. Kamani, and M.~Mahdavi, ``Adaptive personalized federated
  learning,'' \emph{arXiv preprint arXiv:2003.13461}, 2020.

\bibitem{PerFedAvg_2020}
A.~Fallah, A.~Mokhtari, and A.~Ozdaglar, ``Personalized federated learning with
  theoretical guarantees: A model-agnostic meta-learning approach,'' in
  \emph{Proc. NeuIPS}, Vancouver, Canada, Dec. 6-12 2020, pp. 1--6.

\bibitem{FedAMP_2021}
Y.~Huang, L.~Chu, Z.~Zhou, L.Wang, J.~Liu, J.~Pei, and Y.~Zhang, ``Personalized
  cross-silo federated learning on non-iid data,'' in \emph{Proc. AAAI}, Dec.
  6-12 2021, pp. 7865--7873.

\bibitem{FedToE2022}
Y.~Wang, Y.~Xu, Q.~Shi, and T.-H. Chang, ``Quantized federated learning under
  transmission delay and outage constraints,'' \emph{IEEE J. Sel. Areas
  Commun.}, vol.~40, no.~1, pp. 323--341, Jan. 2022.

\bibitem{AsyncFedOpt_2020}
\BIBentryALTinterwordspacing
C.~Xie, S.~Koyejo, and I.~Gupta, ``Asynchronous federated optimization,'' 2020.
  [Online]. Available: \url{https://arxiv.org/abs/1903.03934}
\BIBentrySTDinterwordspacing

\bibitem{LocalSGD_2019}
S.~U. Stich, ``Local {SGD} converges fast and communicates little,'' in
  \emph{Proc. ICLR}, New Orleans, LA, USA, May 6 - May 9 2019, pp. 1--5.

\bibitem{FLANP_2022}
A.~Reisizadeh, I.~Tziotis, H.~Hassani, A.~Mokhtari, and R.~Pedarsani,
  ``Straggler-resilient federated learning: {Leveraging} the interplay between
  statistical accuracy and system heterogeneity,'' \emph{IEEE Journal on
  Selected Areas in Information Theory}, vol. 3, No. 2, pp. 197--205, June
  2022.

\bibitem{lin2021semifed}
H.~Lin, J.~Lou, L.~Xiong, and C.~Shahabi, ``Semi{Fed}: Semi-supervised
  federated learning with consistency and pseudo-labeling,'' \emph{arXiv
  preprint arXiv:2108.09412}, 2021.

\bibitem{FlexMatch_2021}
B.~Zhang, Y.~Wang, W.~Hou, H.~Wu, J.~Wang, M.~Okumura, and T.~Shinozaki,
  ``Flexmatch: Boosting semi-supervised learning with curriculum pseudo
  labeling,'' in \emph{Proc. NeuIPS}, California, USA, Dec. 6-14 2021, pp.
  1--12.

\bibitem{FedMatch_2021}
W.~Jeong, J.~Yoon, E.~Yang, and S.~J. Hwang, ``Federated semi-supervised
  learning with inter-client consistency \& disjoint learning,'' in \emph{Proc.
  ICLR}, May 3-7 2021, pp. 1--6.

\bibitem{FedSemL2022}
\BIBentryALTinterwordspacing
A.~Albaseer, M.~Abdallah, A.~Al-Fuqaha, A.~Erbad, and O.~A. Dobre,
  ``Semi-supervised federated learning over heterogeneous wireless iot edge
  networks: Framework and algorithms,'' 2022. [Online]. Available:
  \url{https://doi.org/10.36227/techrxiv.19317632}
\BIBentrySTDinterwordspacing

\bibitem{SSFL_TMI2022}
Y.~Zhu, Y.~Liu, J.~J.~Q. Yu, and X.~Yuan, ``Semi-supervised federated learning
  for travel mode identification from gps trajectories,'' \emph{IIEEE Trans.
  Intell. Transp. Syst.}, vol.~23, no.~3, pp. 2380--2391, 2022.

\bibitem{FedProx_2018}
T.~Li, A.~K. Sahu, M.~Sanjabi, M.~Zaheer, A.~Talwalkar, and V.~Smith,
  ``Federated optimization in heterogeneous networks,'' in \emph{Proc. MLSys},
  Austin, TX, USA, Mar. 2-4, 2020, pp. 1--12.

\bibitem{FedDyn_2021}
D.~A.~E. Acar, Y.~Zhao, R.~Matas, M.~Mattina, P.~Whatmough, and V.~Saligrama,
  ``Federated learning based on dynamic regularization,'' in \emph{Proc. ICLR},
  May 3-7 2021, pp. 1--6.

\bibitem{Ditto_2021}
T.~Li, S.~Hu, A.~Beirami, and V.~Smith, ``Ditto: Fair and robust federated
  learning through personalization,'' in \emph{Proc. ICML}, Virtual Conference,
  Jul. 18-24 2021, pp. 1--10.

\bibitem{pFedMe_2020}
C.~T. Dinh, N.~H. Tran, and T.~D. Nguyen, ``Personalized federated learning
  with moreau envelopes,'' in \emph{Proc. NeuIPS}, Vancouver, Canada, Dec. 6-12
  2020, pp. 1--12.

\bibitem{DRFL_2020}
Y.~Deng, M.~M. Kamani, and M.~Mahdavi, ``Distributionally robust federated
  averaging,'' in \emph{Proc. NeuIPS}, Vancouver, Canada, Dec. 6-12 2020, p.
  15111–15122.

\bibitem{DRFLM_2022}
B.~Wu, Z.~Liang, Y.~Han, Y.~Bian, P.~Zhao, and J.~Huang, ``Drflm:
  Distributionally robust federated learning with inter-client noise via local
  mixup,'' \emph{arXiv preprint arXiv:2204.07742}, 2022.

\bibitem{FedACL_2019}
M.~Mohri, G.~Sivek, and A.~T. Suresh, ``Agnostic federated learning,'' in
  \emph{Proc. ICML}, Long Beach, CA, USA, Jun. 9-15, 2019, pp. 4615--4625.

\bibitem{CE-DRDL_2022}
M.~Zecchin, M.~Kountouris, and D.~Gesbert, ``Communication-efficient
  distributionally robust decentralized learning,'' \emph{arXiv preprint
  arXiv:2205.15614}, 2022.

\bibitem{bettini2021personalized}
C.~Bettini, G.~Civitarese, and R.~Presotto, ``Personalized semi-supervised
  federated learning for human activity recognition,'' \emph{arXiv preprint
  arXiv:2104.08094}, 2021.

\bibitem{Yu2021TMC}
H.~Yu, Z.~Chen, X.~Zhang, X.~Chen, F.~Zhuang, H.~Xiong, and X.~Cheng,
  ``Fed{HAR}: Semi-supervised online learning for personalized federated human
  activity recognition,'' \emph{IEEE Trans. Mob. Comput.}, pp. 1--1, Dec. 2021.

\bibitem{tashakori2022semipfl}
A.~Tashakori, W.~Zhang, Z.~J. Wang, and P.~Servati, ``Semi{PFL}: Personalized
  semi-supervised federated learning framework for edge intelligence,''
  \emph{arXiv preprint arXiv:2203.08176}, 2022.

\bibitem{Shuai_FedMA_2021}
S.~Wang and T.-H. Chang, ``Demystifying model averaging for
  communication-efficient federated matrix factorization,'' in \emph{Proc. IEEE
  ICASSP}, Toronto, Ontario, Canada, June 6-11, 2021, pp. 1--5.

\bibitem{CE_DDNN_2017}
H.~B. McMahan, E.~Moore, D.~Ramage, S.~Hampson, and B.~A. Areas,
  ``Communication-efficient learning of deep networks from decentralized
  data,'' in \emph{Proc. ICML}, Sydney, Australia, Aug. 6-11, 2017, pp. 1--10.

\bibitem{EDDL_DModelAvg_2018}
M.~Kamp, L.~Adilova, J.~Sicking, F.~Huger, P.~Schlicht, T.~Wirtz, and
  S.~Wrobel, ``Efficient decentralized deep learning by dynamic model
  averaging,'' in \emph{Proc. ECML PKDD}, Dublin, Ireland, Sept. 10-14, 2018,
  pp. 393--409.

\bibitem{Parallel_RSGD_2019}
H.~Yu, S.~Yang, and S.~Zhu, ``Parallel restarted {SGD} with faster convergence
  and less communication: Demystifying why model averaging works,'' in
  \emph{Proc. AAAI}, Honolulu, Hawaii, USA, Jan. 27-Feb. 1, 2019, pp.
  5693--5700.

\bibitem{QFL_Outage_2022}
Y.~Wang, Y.~Xu, Q.~Shi, and T.-H. Chang, ``Quantized federated learning under
  transmission delay and outage constraints,'' \emph{IEEE Journal on Selected
  Areas in Communications}, vol.~40, pp. 323 -- 341, Jan. 2022.

\bibitem{FedNova_2020}
J.~Wang, Q.~Liu, H.~Liang, G.~Joshi, and H.~V. Poor, ``Tackling the objective
  inconsistency problem in heterogeneous federated optimization,'' \emph{arXiv
  preprint arXiv:2007.07481}, 2020.

\bibitem{ClusterFL2020}
F.~Sattler, K.-R. Müller, and W.~Samek, ``Clustered federated learning:
  Model-agnostic distributed multitask optimization under privacy
  constraints,'' \emph{IEEE Transactions on Neural Networks and Learning
  Systems}, vol.~32, pp. 3710--3722, 08 2020.

\bibitem{UserFL2021}
M.~Mestoukirdi, M.~Zecchin, D.~Gesbert, Q.~Li, and N.~Gresset, ``User-centric
  federated learning,'' in \emph{WCDI 2021, GLOBECOM Workshop on Wireless
  Communication for Distributed Intelligence}, Madrid, Spain, 2021.

\bibitem{website_MNIST}
\BIBentryALTinterwordspacing
Y.~LeCun, C.~Cortes, and C.~Burges, ``The mnist database.'' [Online].
  Available: \url{http://yann.lecun.com/exdb/mnist/}
\BIBentrySTDinterwordspacing

\bibitem{CIFAR10_Krizhevsky09}
A.~Krizhevsky and G.~Hinton, ``Learning multiple layers of features from tiny
  images,'' \emph{Master's thesis, Department of Computer Science, University
  of Toronto}, 2009.

\end{thebibliography}

\end{document}